\documentclass[a4paper,11pt]{article}
\pdfoutput=1 

\usepackage{jheppub} 
\usepackage[mathscr]{euscript}
\usepackage[T1]{fontenc} 

 \usepackage{mathtools}
\usepackage{url}

 \usepackage{musicography}
\usepackage{empheq}
  \usepackage{mathrsfs}
 
\usepackage{bm}
\usepackage{verbatim}
 
 \usepackage{makecell}
 \usepackage{float}
 \usepackage{hyperref}

 \usepackage[makeroom]{cancel}
 
 \newcommand{\bv}{ \begin{verbatim}}

    \newcommand{\bz}{{ \bar z}}
  
    \newcommand{\Split}{{ \mathsf{Split}}}

\newcommand{\be}{\begin{equation}}
\newcommand{\ee}{\end{equation}}
\newcommand{\bpm}{\begin{pmatrix}}
\newcommand{\epm}{\end{pmatrix}}

\newcommand{\EV}[1]{\langle #1 \rangle}
\newcommand{\beqn}{\begin{eqnarray}}
\newcommand{\eeqn}{\end{eqnarray}}

\newcommand{\zb}{\bar{z}}

\usepackage{slashed}

\newcommand{\cX}{\mathcal X}

\newcommand{\cO}{\mathcal O}
\newcommand{\cV}{\mathcal V}

\newcommand{\cW}{\mathcal W}

\newcommand{\cH}{\mathcal H}
\newcommand{\cK}{\mathcal K}

\newcommand{\cI}{\mathcal I}
\newcommand{\cN}{\mathcal N}

\newcommand{\sO}{\mathscr {O}}
\newcommand{\ap}{\alpha'{}}
 
\newcommand{\lk}{\mathsf k}

\newcommand{\sfz}{{\sf z}} 
\newcommand{\sfzb}{ \bar{\sf z }}

\newcommand{\bi}{{\bar i}}
\newcommand{\bj}{{\bar j}}

\newcommand{\ta}{{\tilde a}}
\newcommand{\tb}{{\tilde b}}
\newcommand{\tc}{{\tilde c}}

\newcommand{\p}{\partial}

\DeclareMathOperator{\Tr}{Tr}

\author[\natural]{Hongliang Jiang }

 \affiliation[\natural{}]{Centre for Theoretical Physics, Department of Physics and Astronomy,
Queen Mary University of London, Mile End Road, E1 4NS,  UK}

 \emailAdd{h.jiang@qmul.ac.uk}

\preprint{QMUL-PH-21-42}

\title{\boldmath\huge Celestial OPEs  and  $  w_{1+\infty}$ algebra from worldsheet in string theory }

\abstract{Celestial     operator product expansions (OPEs)    arise  from the collinear limit of scattering amplitudes and play a vital role in celestial holography. In this paper, we derive the celestial OPEs of massless fields   in string theory from the worldsheet.  By studying the worldsheet OPEs of vertex operators in worldsheet CFT and further examining their behaviors  in the collinear limit, we find that new vertex operators for the massless fields in string theory  are generated and become dominant in the collinear limit. Mellin transforming to the conformal basis yields exactly the celestial OPEs in  celestial CFT.
We also derive the celestial OPEs from the collinear factorization of string amplitudes and the results derived in these two different methods are in perfect agreement with each other. Our  final formulae of celestial OPEs are applicable to general dimensions, corresponding to Einstein-Yang-Mills theory supplemented   by   some possible higher derivative interactions. Specializing to 4D, we reproduce  all the celestial OPEs for gluon  and graviton    in the literature. 
We consider   various  string theories, including the open and closed bosonic string, as well as the closed superstring theory with   $\cN=1$ and $\cN=2$  worldsheet supersymmetry. In the case of $\cN=2$ string,  we also  derive  all the $\overline{SL (2,\mathbb R)}$ descendant contributions  in  the celestial OPE; the soft sector of such   OPE just yields the $w_{1+\infty}$ algebra after  rewriting in terms of chiral modes.
Our stringy derivation of celestial OPEs thus initiates the first step towards the realization of celestial holography in string theory.

}

\begin{document} 
\maketitle
\flushbottom
\allowdisplaybreaks

 \section{Introduction}
 
 The quest for quantum gravity is one of the most fundamental questions in theoretical physics. Although the quantization of quantum gravity is notoriously hard,    string theory has provided us  the framework for studying   quantum gravity, at least perturbatively. Moreover, string theory enables us to learn some profound   aspects of quantum gravity, in particular the holographic nature.  Although the holographic principle was first proposed  based on black hole entropy \cite{tHooft:1993dmi,Susskind:1994vu},   such an idea was very vague until a concrete model  was realized in string theory \cite{Maldacena:1997re}. In string theory, such a holographic duality, now known as the AdS/CFT correspondence,     naturally arises from the duality between open and closed strings. 
  After more than two decades of intensive study, the AdS/CFT correspondence has been tested very precisely and   has  also taught us even more profound aspects of quantum gravity, like entanglement.  Considering the  very beautiful and   successful story of AdS/CFT correspondence, it is natural to wonder whether we can  study   quantum gravity  beyond AdS. One   interesting and natural generalization is flat spacetime,   where the boundary  is null and even non-smooth. This turns out to be very difficult  and very little progress was made in the past.  
On the other hand, the last two decades also witnessed   fruitful achievements in scattering amplitude  both computationally and conceptually.  
In particular, the idea of   on-shellness, locality, unitarity and causality  has been playing a crucial role.
Interestingly, scattering amplitudes  are just the observables of quantum field theories  in  flat spacetime. Bridging the  ideas from two seemingly unrelated areas together, a promising program towards flat holography, called celestial holography, starts to emerge in recent years \cite{Cheung:2016iub,Pasterski:2016qvg}.  The precursor of celestial holography comes from   noticing the equivalence between asymptotic symmetries and soft theorems \cite{Strominger:2013jfa,He:2014laa}. Since then   various  interesting progress has been made. See      \cite{Raclariu:2021zjz,Pasterski:2021rjz} 
for recent reviews.  In spite, the study of celestial holography has been mostly focusing on  symmetries   and using the bottom-up approach. 
 This is drastically  different from the AdS/CFT correspondence which has various  top-down concrete realizations in string theory. It is then natural to ask whether we can also find a   concrete realization for  celestial holography in string theory.~\footnote{~For previous studies of celestial holography related to (ambitwistor) string theory, see  \cite{Stieberger:2018edy}    for  celestial   string amplitudes, and   \cite{Adamo:2019ipt}   \cite{Casali:2020uvr} for  conformally soft theorems and    celestial double copy in ambitwistor string theory.}
This is particularly important for several reasons. First of all, the celestial amplitudes seem to be very UV sensitive and it is only well-defined in theories, like string theory, where the UV behavior is soft enough. 
 Secondly, a stringy construction may give rise to an exact model  of  celestial holography and thus can be used to test various salient ideas there. In  AdS/CFT, the remarkable agreement  between $\cN=4$ SYM and type IIB string on $AdS_5\times S^5$  just provides a good justification on this point.

 The goal of this paper is to initiate the first step towards the  stringy  realization of celestial holography. More specifically, we will derive the celestial operator production expansion (OPE) from string theory. Although we have not been able to construct an explicit model for celestial holography in string theory, it turns out that the worldsheet of string theory already implies something nontrivial about celestial holography. In particular, we can derive the celestial OPEs from the worldsheet OPEs. The lack of an exact model and thus the model independence of our derivation   show that the relation between the two kinds of  OPEs   is    universal.
 
The celestial OPEs  characterize  the behavior of two   operators in the coincident limit in celestial CFT (CCFT).  They can be obtained from the   collinear limit  of scattering amplitude  by performing the Mellin transformation. Since the collinear factorization is a  universal property of scattering amplitude, the celestial OPE is supposed to also play an important role in CCFT.  
Moreover, the soft sector of celestial OPEs also encodes the underlying symmetry of CCFT and   the bulk scattering amplitude. In particular, starting from celestial OPEs, an infinite dimensional holographic symmetry algebra has been discovered recently \cite{Guevara:2021abz,Strominger:2021lvk}.~\footnote{ It is more precise to refer to this algebra as holographic chiral algebra as   this algebra governs   the chiral subsector of soft particles with positive helicity  only  and all the symmetries  are generated by chiral currents. }   In the case of gravity, this symmetry algebra is just the $w_{1+\infty}$ algebra \cite{Strominger:2021lvk}. 
The supersymmetric extension and the infinite Ward identities associated with this algebra were studied in   \cite{Jiang:2021ovh}.   See \cite{Banerjee:2020kaa,Guevara:2021tvr} for other aspects of celestial OPEs.

 The celestial OPEs can be obtained in several methods. The most direct way is  to consider  the  collinear limit  of scattering amplitude  and then  perform the Mellin transformation \cite{Fan:2019emx,Fotopoulos:2020bqj,Jiang:2021xzy}. Alternatively, one can bootstrap celestial OPEs using conformal symmetry as well as the input from soft theorems \cite{Pate:2019lpp}. Using these methods, the general formula of celestial OPEs for spinning massless particles with cubic interactions in 4D has been recently derived in \cite{Jiang:2021ovh,Himwich:2021dau}. 
 
 In this paper, we offer another derivation of celestial OPEs from  the string worldsheet perspective.~\footnote{  The celestial OPEs in 4D  can also be derived from ambitwistor string   \cite{ambiOPE}.}
 The general strategy is as follows.   
 In celestial holography, the  celestial amplitudes are defined as the Mellin transformation of momentum space scattering amplitude \footnote{ In this paper, we only consider    massless fields, which just correspond to the  gluon  and graviton/dilaton/Kalb-Ramond field in string theory. }
 and can be regarded as   the correlation functions of celestial  operators $\EV{\cO_{\Delta_1}(x_1)\cO_{\Delta_2}(x_2)\cdots}$ in  some putative CCFT on the celestial sphere  living at boundary null infinity \cite{Pasterski:2016qvg,Pasterski:2017kqt}.  
 This is very similar to  the scattering amplitudes in string theory which are computed by  the correlator of vertex operators in worldsheet CFT.  One can make the relation more precise by introducing the so-called conformal vertex operators, which are defined as the   Mellin transformation of the standard vertex operators.  The celestial string amplitude can then be alternatively regarded as the correlator of conformal vertex operators $\EV{\cV_{\Delta_1}(x_1)\cV_{\Delta_2}(x_2)\cdots}$  in worldsheet CFT.  The fact that the celestial amplitude can be computed in two different ways thus suggests a map between   worldsheet CFT and CCFT, and correspondingly a map between  conformal vertex operators  $\cV_{\Delta}$ and  celestial operators $\cO_{\Delta}$. Then the derivation of   OPEs of celestial operators, namely  $\lim_{x_1\to x_2}\cO_{\Delta_1}(x_1)\cO_{\Delta_2}(x_2)$, boils down to the computation of OPEs of conformal vertex operators  $\lim_{x_1\to x_2} \cV_{\Delta_1}(x_1)\cV_{\Delta_2}(x_2)$.  
It turns out that the latter can be obtained by first computing  worldsheet OPEs of  two vertex operators and then performing the Mellin transformation.~\footnote{ In principle, one may compute the worldsheet OPE of two conformal vertex operators directly, and then take the collinear limit. We will not pursue this approach in this paper, and leave it to the future.  } 
More precisely, the    vertex operator  is given by   some  operator  $ V_p( z)$ in the worldsheet  CFT  integrating   over the  worldsheet.~\footnote{ Here to be explicit, we only discuss the integrated form of vertex operators in closed string. The discussions for open string and the unintegrated form of vertex operators are the same.   }
In order to compute the celestial OPE, namely $\lim_{x_1\to x_2} \cV_{\Delta_1}(x_1)\cV_{\Delta_2}(x_2)$, we find it is sufficient  to compute the worldsheet OPE $\lim_{z_1\to z_2}  V_{p_1}( z_1)V_{p_2}( z_2)$. Explicit computation shows that worldsheet OPE in the collinear limit $p_1\cdot p_2\to 0$, which is  equivalent to  the celestial  coincident  limit  $x_1\to x_2$, localizes to a delta-function $\delta^2(z_1-z_2)$ on the worldsheet, and thus produces another vertex operator  after integrating over the worldsheet. 
 This thus fulfills our derivation of    celestial  OPEs from the  worldsheet perspective. And the computation essentially reduces to computing OPEs of free fields on the worldsheet.

 We  will consider  various string theories, including open and closed bosonic string,  and closed superstring with $\cN=1$ and $\cN=2$ worldsheet supersymmetry. For bosonic string, we  compute  the worldsheet OPE of two vertex operators for massless gluons in open string   and obtain the vertex operators for various open string fields including tachyon, gluon, etc.  In spite, we show that in the collinear limit, only the gluon  in the OPE is dominant and the  rest can be ignored, up to some subtle boundary contact terms. The boundary contact terms arise when the two  vertex operators in the OPE hit     another vertex operator.  They can partially be attributed to the remnant  contribution from tachyon.   
On the other hand, the gluon contribution in the   worldsheet OPE of two gluon vertex operators comes with a delta-function in the collinear limit. Doing a worldsheet integral and performing the Mellin transformation thus give us the   celestial OPEs for gluons.
The same derivation is also done for    massless  fields in the closed string, namely graviton, Kalb-Ramond two-form field and dilaton, and  we obtain the celestial OPEs for these fields.  We also   discuss the   OPEs between open   and closed string massless fields. This is a bit different; at tree level the open string vertex operator sits only at the boundary of the disk, while the closed string vertex operator sits in  the interior of the disk.  Nevertheless, we manage   deriving the corresponding celestial OPEs, namely the fusion of gluon and graviton to  another gluon. However, it is known that graviton can also appear in the celestial OPE of two gluons \cite{Pate:2019lpp}. The derivation of this OPE from the   open-closed string setup is not clear, as one needs to produce a graviton vertex operator in the interior of the disk from two gluon vertex operators sitting on the boundary of the disk.
We sidestep this problem by considering heterotic string where one can  realize  both  gluon and graviton/dilaton/Kalb-Ramond field  from the closed string. Using similar techniques in the bosonic string, we are able to derive  all the celestial OPEs involving     gluon and graviton/dilaton/Kalb-Ramond field, including the fusion of two gluons into one graviton. We also discuss the OPEs of NS-NS massless fields in type I and type IIA/IIB string theory. 
  
 To further corroborate our celestial OPEs, we study     string amplitudes in the collinear limit  and then perform the Mellin transformation, which offers another derivation of    celestial OPEs. It turns out that   the three-point string amplitude is  just enough to determine the celestial OPEs. The final results agree exactly with those  derived from worldsheet. 
  Our derivation is for string theory in critical dimensional flat spacetime, namely 26 dimensions for bosonic string and 10 dimensions for superstring. Nevertheless, the final formulae of celestial OPEs are supposed to be applicable to general dimensions, corresponding   to Einstein-Yang-Mills theory  with some possible higher derivative corrections.   Specializing to four dimensional spacetime, we recover all the gluon and graviton OPEs obtained in the literature \cite{Fan:2019emx,Pate:2019lpp,Jiang:2021ovh,Himwich:2021dau}.
 
 Last but not least, we also generalize our discussions to the $\cN=2$ string theory \cite{Ooguri:1991fp,Ooguri:1990ww,Marcus:1992wi}. The   $\cN=2$ string theory has critical dimension four and is consistent in (2,2) signature instead of Minkowski signature. The simplest version of $\cN=2$ theory has a massless degree of freedom and the low energy effective action for $\cN=2$ string is   described by some kind of self-dual gravity. Following the same approach in   other string theories, we study the worldsheet OPE of two vertex operators in $\cN=2$ string theory and derive the corresponding celestial OPE. 
 The derivation is very similar to other string theories with less supersymmetry. However,  now we can go   further beyond the previous derivations. It turns out that in the spacetime with (2,2) signature, we can even derive all the $\overline{SL (2,\mathbb R)}$  descendants  in  the OPE. We find that this essentially comes from the momentum conservation, and the  fact  that in (2,2) signature we 
 can vary two celestial coordinates  independently and thus
have  more freedom to realize the collinear limit. 
 These two features,  $\overline{SL (2,\mathbb R)}$  descendants  and independent  celestial coordinates, are just the crucial ingredients in the derivation of $w_{1+\infty}$ symmetry. Indeed, focusing on the soft sector of our OPE with descendants and then performing the mode expansion, we recover the $w_{1+\infty}$ algebra  \cite{Guevara:2021abz,Strominger:2021lvk}.  We thus provide  a stringy derivation of the $w_{1+\infty}$ symmetry, although it is   indirect. \footnote{ Note that  the  $w_{1+\infty}$ algebra         appeared before in   self-dual gravity and $\cN=2$ string, but in a different way  \cite{Boyer:1985aj,Ooguri:1991fp}. }
  In a more direct derivation, one should be able to construct the chiral generators in $w_{1+\infty}$ algebra directly from $\cN=2$  string worldsheet. 
 We will make some comments   and defer the direct construction to the future. 
 
This paper is organized as follows. In section~\ref{prelim}, we will first discuss the kinematics in general dimensions in terms of celestial variables, and   then review the vertex operators in  bosonic string theory. A map between conformal vertex operators and celestial operators will also be discussed. 
     In section~\ref{OPEfromworldsheet}, we will  derive the celestial OPEs from the  worldsheet OPEs in bosonic string theory. 
      In section~\ref{superstringOPE}, we will generalize the worldsheet derivation of celestial OPEs to superstring. 
 In section~\ref{OPEfromfactorization}, we will derive the celestial OPEs from the collinear factorization of string amplitude. The specialization of celestial OPEs to 4D will be presented.   
 In section~\ref{N2stringOPE}, we will derive the celestial OPE with descendants in $\cN=2$ string theory, and then discuss the resulting $w_{1+\infty}$ algebra.
  In section~\ref{Conclusion}, we will conclude and discuss possible future directions. 
The paper also has two appendices. 
 In  appendix~\ref{appOPE}   we will   collect all the  computation details of worldsheet OPEs of vertex operators in various string theories. In appendix~\ref{appOpenClosed}, we will derive the celestial OPE between gluon and graviton in the open-closed string setup from both the worldsheet  perspective and the amplitude approach.

  \bigskip
 \noindent{\bf Notation:} The spacetime dimension is $D+2$ and the corresponding celestial sphere is $D$-dimensional. We will use  $\mu, \nu,\cdots=0, 1, \cdots, D, D+1$ for spacetime indices and    $a,b,c,\cdots=1, \cdots, D$ for celestial sphere indices. The spacetime metric is $\eta^{\mu\nu}=\text{diag}(-1, +1, \cdots,+1)$ and the celestial sphere metric is $\delta^{ab}$, hence we will raise or lower the position of celestial sphere indices freely. Repeated indices are summed over.
We use $x_i^a$ to label the $a$-th coordinate of  the $i$-th particle/operator. 
We use  $y$ and $z,\bar z$ for the  open and closed string worldsheet coordinates, respectively, and $\sfz,\sfzb$  for the celestial sphere coordinates in four dimensional spacetime.  The polarizations are denoted as $\zeta, \xi, e, \varepsilon$, while $\epsilon$ always refers to infinitesimal quantity. 
%
%
%
%
%
%
%
%
 
\section{Preliminary}    \label{prelim}
 
 In this preliminary section, we will introduce some tools and background knowledge that will   be used in the later sections. We will first introduce the kinematics of massless fields in terms of celestial sphere variables in general dimensions. Then we will review the vertex operators in open and closed bosonic string theory. Finally we will introduce the notion of conformal vertex operators and their relation with celestial operators. 
  
  \subsection{Kinematics   in general dimension}
  We consider $D+2$-dimensional spacetime with   $D$-dimensional celestial sphere at null infinity.  A null momentum  $k$ can be parametrized as  \cite{Kapec:2017gsg}
    \be\label{CSmomentum}
k^\mu(\omega_k, x)=\eta \omega_k \hat k^\mu(x)~, \qquad
   \hat k^\mu(x) =\Big( \frac{1+(x)^2}{2}~, x^a,\frac{1-(x)^2}{2}  \Big)~, \qquad
n^\mu =\Big( -1, 0^a,  1  \Big)~.
  \ee
  where $\omega_k>0$, $\eta=\pm 1$ labels out-going/in-coming particles and $(x)^2=\sum_a x^a x^a$.   We  also introduce the following basis of   polarization vectors
    \be\label{polarizationBasis}
  \varepsilon_a^\mu (x)\equiv   \varepsilon_a^\mu (k)= \p_a    \hat k^\mu(x)=(x^a, \delta ^{ab}, -x^a)~, \qquad  \mu=(0,b, D+1)~.
  \ee
  These $D$  polarization vectors transform under the vector representation of the little group $SO(D)$ for massless particles. And we have
      \be\label{identitynke}
  n\cdot n =\hat k\cdot \hat k=  \varepsilon_a \cdot n = \varepsilon_a  \cdot \hat k=0~, \qquad
  n \cdot \hat k=1~, \qquad
    \varepsilon_a \cdot     \varepsilon_b =\delta_{ab}~.
  \ee
  So the vectors $\{\frac{n+\hat k}{\sqrt 2}, \varepsilon_a,\frac{n-\hat k}{\sqrt 2}\}$ form a complete orthonormal basis. The little group rotates polarization vectors $\varepsilon_a$ but leaves $n, \hat k$ invariant. 

For different momenta, we further have the following identities 
  \be\label{kenDiffpts}
      \hat k (x_i)\cdot    \hat k (x_j)\equiv  \hat k _i \cdot    \hat k _j=-\frac12 ( x_{ij})^2~, \qquad
         \hat k_i \cdot     \varepsilon_a(x_j)= x_{ij}^a~, \qquad
 \varepsilon_a(x_i) \cdot   \varepsilon_b(x_j)=\delta^{ab}~,
  \ee
  where $x_{ij}=x_i -x_j$. 
 
  Taking the product of two polarizations, we get $D^2$  two-index tensors 
 \be\label{varepsab}
 \varepsilon_{ab}^{\mu \nu} =\varepsilon_a^\mu \varepsilon^\nu_b~.
 \ee 
  It is reducible under the little group $SO(D)$ and can be decomposed into symmetric traceless, anti-symmetric and singlet representation: 
  \be
  \bm D\otimes \bm D=\bm{ \frac{(D+2)(D-1)}{2}} +\bm{\frac{D(D-1)}{2} }+\bm 1~.
  \ee
  They just correspond to the polarizations of graviton, Kalb-Ramond (KR) 2-form field and dilaton, respectively:
    \beqn\label{gPolarization}
  \varepsilon_{(ab)}^{\mu\nu}&= &
  \frac12 \Big(  \varepsilon_a^\mu  \varepsilon_b^\nu
  +  \varepsilon_a^\nu  \varepsilon_b^\mu\Big) -\frac{1}{D} \delta_{ab} \Pi^{\mu\nu}~, \qquad
\delta^{ab}    \varepsilon_{ab}^{\mu\nu}=\eta_{\mu\nu}    \varepsilon_{ab}^{\mu\nu}=0~, \qquad
  \varepsilon_{(ab)}^{\mu\nu}=  \varepsilon_{(ba)}^{\mu\nu}=  \varepsilon_{(ab)}^{\nu\mu}~, 
\\ \label{BPolarization}
  \varepsilon_{[ab]}^{\mu\nu}&= &
  \frac12 \Big(  \varepsilon_a^\mu  \varepsilon_b^\nu
  -  \varepsilon_a^\nu  \varepsilon_b^\mu\Big)  
  =-  \varepsilon_{[ab]}^{\nu\mu}=  -\varepsilon_{[ba]}^{\mu\nu} ~,
\\
    \varepsilon ^{\mu\nu} &= & \frac{1}{D}\varepsilon_a^\mu  \varepsilon_a^\nu= \frac{1}{D} \Pi^{\mu\nu} ~, \label{dilatonpolss}
  \eeqn
  where we also introduced
     \be\label{Pimunu}
 \Pi^{\mu\nu}(x) \equiv  \delta^{ab}\varepsilon_a^\mu (x) \varepsilon^\nu_b (x)
 =\eta^{\mu\nu}-n^\mu \hat k^\nu(x) -n^\nu\hat k^\mu(x)~.
  \ee 
  
     We will also  frequently use the abstract polarization tensors which satisfy various properties. 
   In particular, the polarization vector  of gluon  $e^\mu$  satisfies
   \be\label{gluonPolarization}
   e  \cdot e  =1~, \qquad k\cdot e=0~, \qquad e^\mu\simeq e^\mu+\lambda k^\mu~,
   \ee
   where the equivalence in the last equation is guaranteed by the gauge invariance.  The last property allows us to choose a gauge where 
   \be\label{enzero}
   e\cdot n =0 ~.
   \ee
   We can then  decompose any gluon polarization vector as
   \be
   e^\mu\simeq\sum_a c_a \varepsilon_a^\mu~,
   \ee
   up to a pure gauge. 
   
%
%
   Similarly, for  closed string massless spectra, we will   use $e^{\mu\nu}$ to denote  the polarization tensors. Let us first consider graviton and  KR 2-form field, whose polarization tensors satisfy
 \be\label{eGB}
 e^{\mu\nu} e_{\mu\nu} =1~, \qquad k_\mu e^{\mu\nu}=k_\nu e^{\mu\nu}=\eta_{\mu\nu}e^{\mu\nu} =0~, \qquad 
 e^{\mu\nu}=s e^{ \nu\mu}~,\qquad
 e^{\mu\nu}\simeq e^{\mu\nu}+   k^\mu \zeta^\nu
+  s \zeta^\mu k^\nu~,\qquad
 \ee 
 where    $k\cdot \zeta=0$, and $s=\pm 1$ for graviton and 2-form field, respectively. They can be expanded in terms of basis   \eqref{gPolarization}\eqref{BPolarization}  that we constructed before:
 \be\label{polarizationexpansion}
 e^{\mu\nu} \simeq \sum^D_{a,b=1\atop a<b \text{ or } a=b<D} c_{ab} \varepsilon_{(ab)}^{\mu\nu}~,
 \qquad\qquad
  e^{\mu\nu} \simeq \sum^D_{a,b=1\atop a<b} c_{ab} \varepsilon_{[ab]}^{\mu\nu}~,
 \ee
 for graviton and Kalb-Ramond 2-form field, respectively.
 
 For dilaton, we have
 \be
  e^{\mu\nu} e_{\mu\nu} =1~,\qquad  e^{\mu\nu}=e^{ \nu\mu}~,
   \qquad k_\mu e^{\mu\nu}=n_\mu e^{\mu\nu}=0~,
 \ee
where $n$ is  another null direction defined such that  $n\cdot n=0, n \cdot   \hat k = 1$. This fixes the dilaton polarization to be  \eqref{dilatonpolss}
\be\label{dilatonPol}
e^{\mu\nu}=\frac{1}{D}\Big( \eta^{\mu\nu}-n^\mu \hat k^\nu  -n^\nu\hat k^\mu  \Big) =\varepsilon^{\mu\nu}
= \sum_a \frac{1}{D}\varepsilon_{aa}^{\mu\nu}~.
\ee
One easily check that  $e^{\mu\nu}\eta_{\mu\nu}=1$.

In four dimensional spacetime with $D=2$,  it is convenient to introduce complex coordinates on the celestial sphere
  \be\label{xzin4D}
  x^1=\frac{\sfz+\sfzb}{2}~,\qquad   x^2=\frac{-i(\sfz+\sfzb)}{2}~, \qquad \sfz=x^1+ix^2~, \qquad \sfzb=x^1-i x^2~.
  \ee
Then we can represent the momentum as $k=\eta \omega_k \hat k^\mu $ where
  \be
  \hat k^\mu=\frac12\Big(1+(x)^2, 2x^1, 2x^2,1-(x)^2\Big)=\frac12 \Big( 1+\sfz\sfzb, \sfz+\sfzb, -i(\sfz-\sfzb), 1-\sfz\sfzb \Big)~.
  \ee
The two polarization vectors are
$
  \varepsilon_1^\mu=(x^1,1,0,-x^1), \;
    \varepsilon_2^\mu=(x^2,  0,1,-x^2)
$. 
 It is more convenient to define the   polarization vectors in the  helicity basis:
  \be\label{gluonPol4D}
   \varepsilon_+^\mu=   \frac{1}{\sqrt2}\Big( \varepsilon_1^\mu- i \varepsilon_2^\mu \Big)
    =\frac{1}{\sqrt2} (\sfzb,1,-i, -\sfzb)~, \qquad
    \varepsilon_-^\mu=  \frac{1}{\sqrt2} \Big( \varepsilon_1^\mu+i  \varepsilon_2^\mu \Big)
    =\frac{1}{\sqrt2} (\sfz,1, i, -\sfz)~,
  \ee
  satisfying $    \varepsilon_+\cdot     \varepsilon_+= \varepsilon_-\cdot     \varepsilon_-=0, \;\varepsilon_+\cdot     \varepsilon_-=1$.
 
 For graviton, the    polarization tensors for two helicities are
 \be\label{gravitonPol4D}
   \varepsilon_\pm^{\mu\nu}=    \varepsilon_\pm^\mu    \varepsilon_\pm^\nu
   = \frac12\Big(\varepsilon_{11}^{\mu\nu} - \varepsilon_{22}^{\mu\nu} 
   \mp  i\varepsilon_{12}^{\mu\nu} \mp i  \varepsilon_{21}^{\mu\nu} 
       \Big)
   =    \varepsilon_{(11)}^{\mu\nu}   \mp    i \varepsilon_{(12)}^{\mu\nu}~.
 \ee
  
  \subsection{Vertex operator  in bosonic string}
  In this subsection, we review the vertex operators in open and closed bosonic string theory.
 \subsubsection{Open string vertex operator}
 At tree level, open string amplitudes are computed by the correlator of  vertex operators inserted at the boundary of the disk, or equivalently the   boundary of upper half plane, which we parametrize  by  $y\in \mathbb R$.
   
  There are two types of vertex operators.  
 The integrated vertex operators  have the following general structure
 \footnote{ Note that the product of operators at coincident points should  always  be understood as normal ordered product.}
 \footnote{ We will sometimes abuse the terminology of vertex operators and call  both $\mathcal V$ and $V$   vertex operators. The exact   meaning should be clear from the context.}
\be\label{IntVertexopen}
\mathcal V^{A,\ell}(k) =\int dy\;   V^{A,\ell}_k(y)= \int dy\;  \mathscr V^{A,\ell} (y) e^{i k\cdot X(y)}~,
\ee
while  the unintegrated vertex operators take  the form
\be\label{UnIntVertexopen}
\mathcal V^{A,\ell}(k) =c(y)  V^{A,\ell}_k(y)=  c(y) \mathscr  V^{A,\ell} (y) e^{i k\cdot X(y)}~.
\ee
where $c$ is the ghost and the position $y$ in the unintegrated vertex operator is arbitrary.  Here $\ell$ is the number of derivatives in $\mathscr V^{A,\ell}_k$ (e.g. $ \p_y X(y) $ has $\ell=1$), and $A$ are the extra quantum numbers labelling the vertex operators, including the polarization  vectors, Chan-Paton factors, etc.  

The spectra of fields in  open string are given by
\be
- k\cdot k=M^2=\frac{1}{\alpha'}\Big(N-\frac{D}{24} \Big) =\frac{1}{\alpha'}\Big(N-1 \Big)~, 
\ee
where $N\in \mathbb Z_{\ge 0}$ is the level and in the last equality we set $D=24$  for the consideration of critical bosonic string.  

When the momentum  becomes on-shell, namely $  k^2\to -M^2= (1- N)/\ap$, the   vertex operator  becomes BRST invariant and should has weight 1, 
\be\label{hVAl}
h(  V^{A,\ell}_k) =h(\mathscr V^{A,\ell}_k )+h( e^{i k\cdot X })
=\ell+\ap k^2=1~,
\ee
Since on-shell momentum satisfies  $k^2 = (1- N)/\ap$, we see that $\ell$ is essentially the level $N$, namely $N=\ell$. 
 

At level 0, we have tachyon whose vertex operator is given by \footnote{ For simplicity we will set all the coefficients in front of the vertex operators to be 1. Meanwhile, in the final formulae of celestial OPEs, we will   normalize the overall coefficient properly to make the result as simple as possible.  All the couplings can be easily restored, as we keep track of the overall factors in the intermediate steps. \label{footnote2}}
\be
  V^{ 0}_k=     e^{i k\cdot X}~, \qquad M^2=-\frac{1}{\ap}~.
\ee

At level 1, we have gluons whose vertex operators are   \cite{Polchinski:1998rr}
\be\label{gluonvertexoperator}
  V^{  1}_k=  
  e_\mu \dot X^\mu e^{i k\cdot X} t^A ~, \qquad M^2=0~.
\ee
where $\dot X\equiv \frac{\p}{\p y}X$. The polarization  $e_\mu$ satisfies  \eqref{gluonPolarization}
\be
k^2=k\cdot e=0~, \qquad e\cdot e=1~.
\ee
For the gluon vertex operator, we also have the Chan-Paton factor  $t^A$, which is just the   gauge algebra generator  with color index $A$. The corresponding structure constant,  denoted as $f^{ABC}$,  is fully anti-symmetric. 

  \subsubsection{Closed string vertex operator }
 At tree level, the closed string amplitudes are computed as the correlator of vertex operators on the sphere, or equivalently   the complex plane, which is parametrized by  coordinates $z,\zb$. 
  
 The integrated vertex operator  has the structure
\be\label{IntVertex}
\mathcal V^{A,\ell}(k) =\int d^2z \;   V^{A,\ell}_k(z,\zb)= \int d^2z\;  \mathscr  V^{A,\ell} (z,\zb) e^{i k\cdot X(z,\zb)}~,
\ee
while  the unintegrated vertex operator takes the form
\be\label{UnIntVertex}
\mathcal V^{A,\ell}(k) =c(z)\tilde c(\zb)  V^{A,\ell}_k(z,\zb)=c(z)\tilde c(\zb)  \mathscr V^{A,\ell} (z,\zb) e^{i k\cdot X(z,\zb)}~,
\ee
where $c,\tilde c$ are the ghosts and the position $z,\zb$ in the unintegrated vertex operator is arbitrary.  Here $\ell$ is the number of holomorphic derivatives in $\mathscr V^{A,\ell}_k$ (which is also the number of anti-holomorphic derivatives), and $A$ are the extra quantum numbers labelling the vertex operators, such as the polarization  tensors.  
 
The closed string spectra are given by
\be\label{km2clsoed}
- k\cdot k=M^2=\frac{2}{\alpha'}\Big(N+\tilde N-2\frac{D}{24} \Big) =\frac{4}{\alpha'}\Big(N-1 \Big) ~,
\ee
where we use the  level matching condition $N=\tilde N\in \mathbb Z_{\ge 0}$, and in the last equality we set $D=24$ for critical  bosonic string. 

In order to be BRST invariant, the on-shell vertex operator should have holomorphic and anti-holomorphic weights 1, namely
\be\label{hVAl}
h(  V^{A,\ell}_k) =h(\mathscr V^{A,\ell}_k )+h( e^{i k\cdot X })
=\ell+\frac \ap 4 k^2=1~,
\ee
and the same  for $\bar h$. Therefore, on-shell momentum should satisfy $ k^2= 4 (1- \ell)/\ap$. Comparing  with the mass-shell condition in  \eqref{km2clsoed}, we again have $N=\ell$.

At level 0, we have closed string tachyon whose vertex operator is given by 
\be
  V^{ 0}_k=
    e^{i k\cdot X}~, \qquad M^2=-\frac{4}{\ap}~.
\ee

At level 1, we have massless fields  whose vertex operators  are given by  \cite{Polchinski:1998rr}
\be
  V^{  1}_k=  
   e_{\mu\nu} \p X^\mu \bar\p X^\nu e^{i k\cdot X}   ~, \qquad M^2=0~.
\ee
Depending on the structure of polarization tensor $  e_{\mu\nu}$, the vertex operator can represent different fields, either graviton, or KR 2-form field, or dilaton. 
 See \eqref{eGB}-\eqref{dilatonPol} for    the properties and explicit forms of polarization tensors  corresponding to these different fields.
 
 \subsection{CFT on the worldsheet and on the celestial sphere}
 Given the vertex operators, one can then compute the string amplitudes.
 In general, the string amplitude is schematically computed by the  following   formula  \cite{Polchinski:1998rr}
\be\label{generalstringAmp}
\mathcal A_n(k_1,k_2, \cdots, k_n) \sim\sum_{\text{topologies}}   e^{-\lambda\chi}
 \int_{\mathcal M_{g,n}} dm\; \EV{ \mathcal V^{A_1,\ell_1}(k_1)\mathcal V^{A_2,\ell_2}(k_2)\cdots \mathcal V^{A_n,\ell_n}(k_n) }_{\text{WS}}~,
\ee
where we need to sum over all the topologies for the string worldsheet, and for each topology, we need to incorporate the contribution from various ghosts properly and integrate over the moduli space of Riemann surface.  
The world-sheet correlator is evaluated through the path integral of Polyakov action with operator  insertions.
The computation of string amplitude is   very hard as one goes to higher loops, but it simplifies dramatically at tree level. 
At tree level, the topology is fixed and there are no moduli to integrate over. 

As a consequence, the string amplitude at tree level is given by 
\be\label{AstringAmp}
\mathcal A_n(k_1,k_2, \cdots, k_n) =\EV{ \mathcal V^{A_1,\ell_1}(k_1)\mathcal V^{A_2,\ell_2}(k_2)\cdots \mathcal V^{A_n,\ell_n}(k_n) }_{\text{WS}}~.
\ee 
In the case of open string, the worldsheet is given by the disk, which is conformally equivalent to the upper half complex plane. The vertex operators are inserted on the boundary of the disk.   In particular, among $n$ vertex operators, three of them should take the unintegrated form \eqref{UnIntVertexopen} and  the rest in the integrated form \eqref{IntVertexopen}, in order to soak up the zero modes. For gluon amplitudes, we should take a trace over the product of Chan-Paton factors $t^A$, which are ordered on the boundary according to their positions. So different ways of inserting the vertex operators on the disk boundary give rise to different Chan-Paton factors, and one needs to include all orderings of insertions.  For closed string amplitude, the worldsheet is given by the sphere, which is conformally equivalent to the whole complex plane. Again to soak up the zero modes, we should choose three vertex operators  in the unintegrated form  \eqref{UnIntVertexopen} and the rest in the integrated form \eqref{IntVertex}.

On the other hand, the celestial amplitudes for massless particles are given by the Mellin transformation of momentum space amplitudes  \cite{Cheung:2016iub,Pasterski:2016qvg,Pasterski:2017kqt} \footnote{ We focus on massless fields in this paper. For   massive fields, the Mellin transformation needs to  be modified.}
\be
\mathcal M_n(x_1, \cdots, x_n)=
\prod_{j=1}^n \int_0^\infty d\omega_j\; \omega_j^{\Delta_j-1}\mathcal A_n(\eta_i \omega_i \hat k_i ) ~, \qquad\quad 
\mathcal A_n(k_i)=A_n(k_i)\;\delta^{D+2}(\sum_{i=1}^n k_i)~.
\ee
In contrast to the momentum space amplitude which has manifest translational invariance, the celestial amplitude is designed to make the   Lorentz symmetry manifest. Indeed, in $D+2$ dimensional Minkowski spacetime we have   Lorentz group $SO(1,D+1)$, which is also  the conformal group of  CFT  in  $D$ dimensions.
The place supporting such a conformal group   is just the   celestial sphere, which sits at the  boundary null infinity of spacetime.  
The celestial amplitude defined above  can  thus be regarded as the correlator of some   celestial operators in a putative celestial conformal field theory living on the celestial sphere
\be\label{celestialcorrelator}
\mathcal M_n(x_1, \cdots, x_n)=\EV{\cO_{\Delta_1}(x_1) \cdots \cO_{\Delta_n}(x_n)}_{\text{CS}}~.
\ee
Note that   for simplicity of notation we have stripped off all the extra labels of the operators 
except for their positions and dimensions. The representation of bulk scattering amplitude in terms of boundary correlator is just reminiscent of  the holographic principle. 

So the string amplitude can be computed in two different ways, either  \eqref{AstringAmp} or \eqref{celestialcorrelator}. To make the connection more precise, we can also  define the  vertex operators in the conformal basis   through a Mellin transformation 
\be\label{VDelta}
\mathcal V^{A,\ell,\eta}_\Delta(x)= \int_0^\infty d\omega\; \omega^{\Delta-1} \mathcal V^{A,\ell}(\eta\omega \hat k) 
~.
\ee
We will   refer to  it  as the \emph{conformal vertex operators} in this paper.
Then the celestial string amplitude is essentially given by the correlator of conformal vertex operators evaluated in worldsheet CFT
\be\label{confVOstring}
\mathcal M_n(x_1, \cdots, x_n)=\EV{\cV_{\Delta_1}(x_1) \cdots \cV_{\Delta_n}(x_n)}_{\text{WS}}~.
\ee

The very similar structure between \eqref{celestialcorrelator} and \eqref{confVOstring} suggests a   map $\mathscr F$  from   world-sheet CFT (WSCFT) to celestial CFT (CCFT), such that the two Hilbert spaces are related as follows
\be
\mathscr F: \qquad H_\text{WSCFT} \to H_\text{CCFT}~.
\ee
In particular, there is a one-to-one map between string  conformal   vertex operators and celestial operators   
\be
\mathscr F: \qquad \cV_{\Delta } \mapsto \cO_{\Delta }~.
\ee
 
 Our goal in this paper is to derive   the celestial OPEs
 \be\label{celestialOPE}
 \cO_{\Delta_1 }(x_1) \cO_{\Delta_2 }(x_2)\sim
 \sum_j 
   C_{\Delta_1\Delta_2}^{\Delta_j}(x_1,x_2) \cO_{\Delta_j }(x_2)~.
 \ee
Due to the map $\mathscr F$, it is sufficient to compute
  \be\label{celestialVOPE}
 \cV_{\Delta_1 }(x_1) \cV_{\Delta_2 }(x_2)\sim
 \sum_j 
 C_{\Delta_1\Delta_2}^{\Delta_j}(x_1,x_2)\cV_{\Delta_j }(x_2)~.
 \ee

We would like to have several remarks here. In spite of the similarity  between   \eqref{celestialcorrelator} and \eqref{confVOstring}, the celestial CFT and worldsheet CFT are very different in many aspects. The worldsheet CFT is always   two dimensional, while the  dimension of celestial CFT depends on the bulk spacetime dimension. Moreover, the worldsheet CFT and thus its correlator \eqref{AstringAmp}  are explicit   and well-defined, while for   celestial CFT which has many unusual features, our understanding remains poor and is mostly about symmetry.  The map $\mathscr F$ suggests that these two are supposed to be related in some way. In particular,   since the integrated conformal vertex operator  $\cV_\Delta(x)$  arises after doing  the worldsheet integration, it is reasonable to speculate that the celestial CFT arises from  worldsheet CFT by some kind of projection of  worldsheet coordinates and going to the labelling space, namely momentum $k$ or $\Delta, x$ space. On the other hand, the similarity between \eqref{celestialcorrelator} and \eqref{confVOstring}  is for tree level amplitude. At loop level, the string amplitude in \eqref{generalstringAmp} is more complicated and only admits perturbative expansion. The      CCFT  correlator \eqref{celestialcorrelator} is however   defined  formally but exactly. To reconcile the difference, the celestial CFT is supposed to also admit some expansion, and then one may compare the two sides order by order. This is indeed the case in AdS/CFT where   the gauge theory in the CFT side admits    large-$N$ expansion and can be compared with the loop expansion of string theory.  Understanding these points may be useful for us to establish a concrete and exact model of celestial holography in string theory.

%
%
 
  \section{Celestial OPE from worldsheet OPE in bosonic string} \label{OPEfromworldsheet}
  
In this section, we will discuss the derivation of celestial OPEs from   worldsheet OPEs following the strategy outlined in the introduction.  The string theory we will consider in  this section  is     bosonic strings,  either open or closed. As a result, we are able to derive the gluon OPEs and  graviton/dilaton/KR field OPEs from open and closed strings, respectively.   The  mixed OPEs involving both gluon and   graviton/dilaton/KR field can be derived from the open-closed string setup, which will be deferred to appendix~\ref{appOpenClosed}. All the OPEs involving gluon, 
 graviton/dilaton/KR  as well as their mixture will also be discussed in the heterotic superstring in the next section. 
 In section~\ref{OPEfromfactorization}, we will also confirm our celestial OPEs through the     collinear factorization of string amplitude.  

  \subsection{OPE in open string }
  Let us now discuss how to derive the celestial  gluon OPEs  from the worldsheet OPEs  in open string theory. As we described before, it is sufficient to compute the  OPE \eqref{celestialVOPE}. For this aim, in principle we need to know  $V_{\Delta_1}(x_1,y_1) V_{\Delta_2}(x_2,y_2)$ for arbitrary $y_1,y_2$,  even if they are very far away from each other on the worldsheet. Here $V_\Delta(x,y) $ is essentially the Mellin transformation of the integrand $V_k(y)$ in  \eqref{IntVertexopen}. Since we are only interested in the collinear limit $x_1\to x_2$, it is reasonable to speculate that the celestial OPE is determined by the worldsheet OPE. \footnote{ To be clear, the worldsheet OPE refers  to the OPE of two operators  approaching each other on the worldsheet, namely $z_1\to z_2$, while the celestial OPE  refers  to the OPE of two celestial operators approaching each other on the celestial sphere, namely $x_1\to x_2$. The celestial coincident limit  $x_1\to x_2$ is equivalent to the collinear limit.  } This will be  our worldsheet approach to deriving the celestial OPEs.  In particular, the celestial gluon OPE  can be derived by computing
  \footnote{ Instead of considering two vertex operators in the integrated forms \eqref{IntVertexopen}, one can also choose one  in the integrated form \eqref{IntVertexopen} and  the other in the unintegrated form \eqref{UnIntVertexopen}. Their OPE leads to another unintegrated vertex operator. The   final results for celestial OPEs  obtained in these two ways are the same.}
      \beqn
\cV^A(p, \zeta)\cV^B(q, \xi )
 &=& 
\int dy_1 dy_2\; \theta(y_1-y_2) \;
    V^A_{p, \zeta}(y_1)    V^B_{q, \xi}(y_2) 
+
\int dy_1 dy_2\;   \theta(y_2-y_1) \;   V^B_{q, \xi}(y_2)     V^A_{p, \zeta}(y_1) 
\qquad\qquad\\&=&\nonumber
\label{VAVBint}
\int dy_1 dy_2\; \theta(y_1-y_2) \;
\zeta\cdot\dot X e^{i p \cdot X  }(y_1)\xi\cdot\dot X e^{i q  \cdot X  }(y_2) t^A t^B
\\& & 
+
\int dy_1 dy_2\; \theta(y_2-y_1) \;
 \xi\cdot\dot X e^{i q  \cdot X  }(y_2)\zeta\cdot\dot X e^{i p \cdot X  }(y_1)   t^Bt^A ~,
   \eeqn
 where we used the explicit form of gluon vertex operators in \eqref{gluonvertexoperator} and $\zeta, \xi$ are the polarization vectors.  We will first compute the worldsheet OPE of two $V$'s in the integrand; then we will exam the behavior of this worldsheet OPE in the collinear limit. As we will see, in the collinear limit, the worldsheet OPE actually localizes to a delta-function, which just gives another vertex operator. Performing the Mellin transformation then leads us to the celestial OPEs.  
 
 Before going to the details, let us add some comments about the formula above \eqref{VAVBint}. 
 Since we are interested in  the worldsheet OPE, the  two   vertex operators should be next to each other and there is no other operator insertion  between them. Nevertheless, we have two different contributions in the formula above, corresponding to two different orderings of vertex operators on the boundary.  On the other hand, the formula above should be understood in the correlator where other operator insertions indeed appear; this gives rise to  the upper and lower limits of integration   for the integrals. As we will see, this brings the subtle issue of boundary contact terms.  
 
Now we can start the derivation of celestial OPEs. As the first step, we need to compute the worldsheet OPE for the integrand in \eqref{VAVBint}. This can be done as the fields $X$ are essentially free scalars. All the detailed computations of worldsheet OPEs are given in appendix~\ref{appOPE}. In particular,  the worldsheet OPE for the integrand in  \eqref{VAVBint} is given by  \eqref{XXexpOPEapdx}:
%
    \beqn\label{openstringOPE}
\zeta\cdot\dot X e^{i p \cdot X  }(y_1)\xi\cdot\dot X e^{i q  \cdot X  }(y_2)
&\sim&
- 2\alpha' y_{12}^{2 \alpha' p\cdot q-2} e^{i (p+q) \cdot X }    
\Bigg[  
 \Big(  \zeta\cdot \xi  -2\ap   \zeta \cdot q\; \xi \cdot p  \Big)  
\\&&\nonumber
+   i   y_{12} \Big(   \zeta \cdot q \;  \xi \cdot \dot X  -\xi \cdot p\; \zeta\cdot \dot X 
+  \big(\zeta\cdot \xi  -2\ap   \zeta \cdot q\; \xi \cdot p \big)p \cdot \dot X
 \Big)  +
 \cdots \Bigg] 
  (y_ 2)~,
  \eeqn
  where we assume $y_1>y_2$.
  
  We will take $p,q$ on-shell, namely $p^2=q^2=0$, then the polarization vectors $\zeta, \xi$ satisfy the  properties in \eqref{gluonPolarization}. We are interested in the collinear limit where $p,q$ are almost parallel and thus $  p \cdot q\to 0$, implying that $p+q$ is   almost on-shell. However, in order to gain information from the OPE,  we can not take the strict collinear limit. Instead,  we will denote 
  $K\equiv p+q$ and  choose a nearby null momentum $  k=\omega_k \hat k$ such that $K=k+\epsilon v$, where 
  $v$ is a generic momentum of order one. Then we have 
  \be
  K^2=2 p \cdot q=k^2+2\epsilon k\cdot v+\epsilon^2 v\cdot  v=2\epsilon k\cdot v+\sO(\epsilon^2)~.
  \ee
 Hence $\epsilon$ has the same order of magnitude with $p\cdot q$ in the collinear limit. Without loss of generality, we will define  $\epsilon=2\ap p \cdot q$.         Then $K=p+q=k+\sO(\epsilon)$.    In the following discussions, we will try to simplify \eqref{openstringOPE} in such a collinear limit.

Given the  null momentum $k$, we can define    the basis for polarization vectors $\varepsilon_a(k)$   \eqref{polarizationBasis}, which satisfy \eqref{Pimunu}
       \be \label{Pimunu2}
       \varepsilon_a^\mu (k) \varepsilon^\nu_a (k)
 =\eta^{\mu\nu}-n^\mu \hat k^\nu -n^\nu\hat k^\mu ~.
  \ee
Contracting with $p^\mu \dot X^\nu$,  we get
       \be 
      p \cdot \varepsilon_a \;  \varepsilon _a \cdot \dot X
 =p \cdot X-p\cdot n\; \hat k \cdot\dot  X - \dot   X \cdot n \; \hat k\cdot p ~.
  \ee
Using  $n \cdot \hat p=1$   \eqref{identitynke} and hence $n \cdot p =\omega_p n \cdot \hat p =\omega_p  $,  \footnote{ We will only consider the celestial OPEs of celestial operators corresponding to    out-going particles in this paper, so $\eta=1$.}
the above equation can be written as
\beqn 
 p \cdot \dot X
 & =&
   p \cdot \varepsilon_a \;  \varepsilon _a \cdot \dot X
  + \frac{1}{\omega_k }         k\cdot p  \; \dot X \cdot n 
 + \frac{\omega_p}{\omega_k} k \cdot\dot  X
  \\& =&
  p \cdot \varepsilon_a \;  \varepsilon _a \cdot \dot X
 + \frac{1}{\omega_k }         p\cdot q  \; \dot X \cdot n 
+ \frac{\omega_p}{\omega_k} k \cdot\dot  X +\sO(\epsilon)~,
\label{pXsimple}
  \eeqn
  where we used $k \cdot p =    (p +q+\sO(\epsilon)) \cdot p=p \cdot q+\sO(\epsilon)$ as $p^2=0$. 
  
  Similarly, we can contract \eqref{Pimunu2} with $\zeta^\mu\dot X^\nu$, yielding
           \be 
      \zeta \cdot \varepsilon_a \;  \varepsilon _a \cdot \dot X
 =\zeta \cdot X-\zeta \cdot n\; \hat k \cdot\dot  X - \dot   X \cdot n \; \hat k\cdot \zeta 
=\zeta \cdot X  -\frac{1}{\omega_k} \dot   X \cdot n \;   q\cdot \zeta 
+\sO(\epsilon)~,
  \ee
where we used $k \cdot \zeta=(p+q+\sO(\epsilon)) \cdot \zeta=q \cdot \zeta+\sO(\epsilon)$ and $\zeta\cdot n=  0$  due to \eqref{gluonPolarization} and  \eqref{enzero}~.  \footnote{ Note that $\zeta\cdot n=0$ can be realized  by choosing a specific gauge. This is inessential as the amplitude is   gauge  invariant  or   BRST invariant.} 
As a result, we have
          \be 
 \zeta \cdot \dot X\; \xi \cdot p =    \zeta \cdot \varepsilon_a \;  \varepsilon _a \cdot \dot X \; \xi \cdot p+
 \frac{1}{\omega_k} \dot   X \cdot n \;   q\cdot \zeta \;     p \cdot \xi
 +\sO(\epsilon)~,
  \ee
 and similarly  
            \be 
 \xi \cdot \dot X\;  \zeta \cdot  q=     \xi \cdot \varepsilon_a \;  \varepsilon _a \cdot \dot X \;  \zeta \cdot  q+
 \frac{1}{\omega_k} \dot   X \cdot n \;   p\cdot  \xi  \;       q\cdot \zeta  
  +\sO(\epsilon)~.
  \ee
  Taking the difference, we find
            \be \label{zetaxiDiff}
 \xi \cdot \dot X\;  \zeta \cdot  q- \zeta \cdot \dot X\; \xi \cdot p
 =
      \xi \cdot \varepsilon_a \;  \varepsilon _a \cdot \dot X \;  \zeta \cdot  q 
        -    \zeta \cdot \varepsilon_a \;  \varepsilon _a \cdot \dot X \; \xi \cdot p 
         +\sO(\epsilon)~.
    \ee
 With \eqref{pXsimple} and  \eqref{zetaxiDiff}, we can simplify the  terms in  the square bracket of \eqref{openstringOPE}
\beqn
&&  \Big(  \zeta\cdot \xi  -2\ap   \zeta \cdot q\; \xi \cdot p  \Big)   
+ i\, y_{12} \Big(   \zeta \cdot q \;  \xi \cdot \dot X  -\xi \cdot p\; \zeta\cdot \dot X 
+  \big(\zeta\cdot \xi  -2\ap   \zeta \cdot q\; \xi \cdot p \big)p \cdot \dot X
 \Big)  \qquad
  \\&=&\nonumber 
   \zeta\cdot h\cdot \xi 
  +i\, y_{12}
   \Big(      \zeta \cdot  q  \; \xi \cdot \varepsilon_a \;  \varepsilon _a \cdot \dot X     
        -    \xi \cdot p  \; \zeta \cdot \varepsilon_a \;  \varepsilon _a \cdot \dot X   
        +  \zeta\cdot h\cdot \xi  \;    p \cdot \varepsilon_a \;  \varepsilon _a \cdot \dot X
        \\ &&\nonumber \qquad\qquad\qquad\qquad
+
   \zeta\cdot h\cdot \xi   \big( 
   \frac{1}{\omega_k }         p\cdot q  \; \dot X \cdot n 
+  \frac{\omega_p}{\omega_k} k \cdot\dot  X 
\big)
 \Big)          +\sO(\epsilon)~,
 \eeqn
 where  we have introduced
 \be
h^{\mu\nu}=   \eta^{\mu\nu}-2\alpha'  q^{\mu} p^{\nu}  ~ , \qquad
\zeta\cdot h\cdot \xi \equiv\zeta_\mu h^{\mu\nu}\xi_\nu=\zeta\cdot \xi  -2\ap   \zeta \cdot q\; \xi \cdot p ~.
\ee
Therefore the worldsheet OPE    \eqref{openstringOPE} can be written as
    \beqn\label{XXexpOPEsimple}
&&\zeta\cdot\dot X e^{i p \cdot X  }(y_1)\xi\cdot\dot X e^{i q  \cdot X  }(y_2)
\\&\sim& \label{tachyonpart}
- 2\alpha' y_{12}^{2 \alpha' p\cdot q-2} e^{i (p+q) \cdot X }     \zeta\cdot h\cdot \xi   
\\&&\label{gluonpart}
- 2i\; \alpha' y_{12}^{2 \alpha' p\cdot q-1} e^{i (p+q) \cdot X }  
 \Big(      \zeta \cdot  q  \; \xi \cdot \varepsilon_c  
        -    \xi \cdot p  \; \zeta \cdot \varepsilon_c 
        +  \zeta\cdot h\cdot \xi  \;    p \cdot \varepsilon_c 
        \Big) \varepsilon _c \cdot \dot X
  \\&&  \label{regularpart}
  - 2i\; \alpha'      \frac{ \zeta\cdot h\cdot \xi  }{\omega_k }    \dot X \cdot n \;  
      p\cdot q  \; y_{12}^{2 \alpha' p\cdot q-1}     e^{i (p+q) \cdot X }  
      \\&&\label{derivativeparft}
- 2 \; \alpha'   \zeta\cdot h\cdot \xi   \;\frac{\omega_p}{\omega_k}    
y_{12}^{2 \alpha' p\cdot q-1}\p e^{i  (p+q) \cdot X } 
    \\&& \label{smallpart}
      +
\sO( y_{12}^{2 \alpha' p\cdot q } )+\sO(2 \alpha' p\cdot q)~,
  \eeqn
where all terms on the right hand side are   evaluated at $y_2$. This is the worldsheet OPE in the collinear limit. To obtain the celestial OPE, we use the following identities to analyze the dominant contributions. For infinitesimal $\epsilon$ and positive $x>0$, we have
\footnote{ They arise  from the generalized function
  \be\nonumber
  \lim_{\epsilon\to 0} \frac{\epsilon}{2} |x|^{\epsilon-1}=\delta(x)
  \ee
 as well as its derivatives.  For this representation of delta function, see \url{https://mathworld.wolfram.com/DeltaFunction.html}.
  }
 \be\label{xepsId}
\epsilon x^{\epsilon -1} =2\delta(x)~, \qquad \epsilon x^{\epsilon -1-n} =\frac{2(-1)^n}{n!} \delta^n(x)~, \qquad 
\epsilon x^{\epsilon +n}=0~, \qquad n=0,1,2,\cdots ~.
\ee
 
 We first observe that in the worldsheet OPE, if we only look at the $X$ factors, \eqref{tachyonpart} and  \eqref{gluonpart} are just the integrand in the vertex operators of tachyon and gluon. 
 Furthermore, if we take the limit  $2 \alpha' p\cdot q\to 1$ and use identities in \eqref{xepsId}, \eqref{tachyonpart}    gives the factor
 $y_{12}^{2 \alpha' p\cdot q-2}\propto \delta(y_{12})/(2\ap p \cdot q-1)$ and the pole  just singles out the tachyon. 
 Therefore  in  the limit  $2 \alpha' p\cdot q\to 1$, the term \eqref{tachyonpart} dominates the OPE.
 Doing the $y$ integral, \eqref{tachyonpart}  indeed gives the tachyon vertex operator, whose mass-shell condition is just $(p+q)^2=2 p\cdot q =1/\ap$.

The more interesting limit for our purpose is the collinear limit $2 \alpha' p\cdot q\to 0$. We will argue that in this limit
only \eqref{gluonpart} survives as a singular term in the OPE at leading order, while the rest are sub-dominant. 

For \eqref{regularpart},  the factor  $p\cdot q  \; y_{12}^{2 \alpha' p\cdot q-1}$  in the collinear limit $p\cdot q\to 0$ becomes   $\delta(y_{12}) $  after using \eqref{xepsId}. This is of order one, hence we will ignore it.

For \eqref{tachyonpart},  the factor  $ y_{12}^{2 \alpha' p\cdot q-2}$ just leads to $\delta'(y_{12}) /p \cdot q$ following \eqref{xepsId}. This looks singular in the collinear limit. For vertex operators, we need to perform the integral as in \eqref{VAVBint},  which schematically gives
\be\label{boundarycontact}
\int dy_1 dy_2 \; \theta(y_1-y_2) \delta'(y_{12}) e^{i (p+q)  \cdot X  }(y_2)
\to   \int dy_1dy_2\; [\delta(y_1-y_2)]^2 e^{i (p+q)  \cdot X  }(y_2)+
 \int dy_2\;  \p e^{i (p+q)  \cdot X  }(y_2)~.
\ee
  The $[\delta(y_1-y_2)]^2$ factor in the first  term  arises from the collision of $y_1,y_2$ and it is   a contact type term. The second type of term with an unrestricted integration range is a total derivative, which normally does not have contributions in amplitudes due to BRST invariance. When we insert   it in the correlator, there are also other operator insertions, then the  second type of term gives rise to the boundary term when the position of  two operators in OPE    coincides with the rest of operator insertions. 
   We will refer to these types  of terms as \emph{boundary contact terms}. 
    Since these boundary contact terms arise  when operator insertions collide and correspond to very singular configurations, we expect that they would finally cancel out and play no role in our OPE analysis.  On the other hand,  \eqref{tachyonpart} corresponds to tachyon whose mass-squared is negative, the boundary contact terms can thus be partially understood as the remnant contribution from fields in string theory with lower masses. Physically, we also expect the collinear limit should   kill the contribution from these fields at leading order.

 For \eqref{derivativeparft}, $\p e^{i(p+q)\cdot X}=\p e^{i k \cdot X}+\sO(\epsilon)=k \cdot \dot X e^{ i k \cdot X} +\sO(\epsilon)$. Since the polarization is along the momentum direction, this is supposed to a pure gauge or BRST-exact term and hence does not contribute. Nevertheless, we can also analyze this term  as before. The factor
 $y_{12}^{2 \alpha' p\cdot q-1}$  becomes $\delta(y_{12}) /p \cdot q$ in the collinear limit. And then
 performing the integral  gives
   \be\label{boundarycontact2}
 \int dy_1 dy_2 \; \theta(y_1-y_2) \delta(y_{12}) \p e^{i  (p+q)  \cdot X }(y_2) 
 \to \int dy_2\;  \p e^{i  (p+q)  \cdot X }(y_2) ~,
 \ee
 which takes the same form we encountered in \eqref{boundarycontact}. 
 
 Actually we can combine \eqref{tachyonpart} and \eqref{derivativeparft} together and get
  \beqn
 &&
  - 2\alpha' y_{12}^{2 \alpha' p\cdot q-2} e^{i (p+q) \cdot X }     \zeta\cdot h\cdot \xi   
   - 2 \; \alpha'   \zeta\cdot h\cdot \xi   \;\frac{\omega_p}{\omega_k}    
y_{12}^{2 \alpha' p\cdot q-1}\p e^{i  (p+q) \cdot X } 
 \\&=&\nonumber
 - 2\alpha' y_{12}^{2 \alpha' p\cdot q-2} e^{i (p+q) \cdot X }     \zeta\cdot h\cdot \xi   
 \frac{  2\omega_p  \alpha' p\cdot q+\omega_k-\omega_p   }{\omega_k} 
   - 2 \; \alpha'   \zeta\cdot h\cdot \xi   \;\frac{\omega_p}{\omega_k}    
\p \Big[ y_{12}^{2 \alpha' p\cdot q-1}  e^{i  (p+q) \cdot X }  \Big]~,
  \eeqn
 where the first term is the same as  \eqref{tachyonpart}  up to a constant, while  the second term is a total derivative and again would lead to at most boundary contact terms.
  
Therefore, up to the subtle boundary contact terms, all terms except for \eqref{gluonpart} in the world-sheet OPE do  not give rise to singular terms in the collinear limit $p\cdot q\to 0$. So   we just need to focus on  \eqref{gluonpart} .
 
To proceed,   we choose the polarization vectors as the basis defined in  \eqref{polarizationBasis}    $\zeta=\varepsilon_a(p), \; \xi =\varepsilon_b(q)$, and furthermore use the notation $p=\omega_p \hat p(x_1),\; q =\omega_q \hat q(x_2), \; k=\omega_k \hat k(x_3)$ where $x_{1,2,3}$ are the coordinates on the celestial sphere.  Then the world-sheet OPE \eqref{openstringOPE} \eqref{gluonpart} becomes
      \beqn\label{XXOPEgluon}
&&\varepsilon_a(p)\cdot\dot X e^{i p \cdot X  }(y_1)\varepsilon_b(q)\cdot\dot X e^{i q  \cdot X  }(y_2)
\\&\sim&\nonumber
- 2i\; \alpha' y_{12}^{2 \alpha' p\cdot q-1} e^{i (p+q) \cdot X }  
 \Big(      \varepsilon_a(p) \cdot  q  \; \varepsilon_b(q) \cdot \varepsilon_c \;    
        -    \varepsilon_b(q) \cdot p  \; \varepsilon_a(p) \cdot \varepsilon_c \;  
                + \varepsilon_a(p) \cdot h\cdot  \varepsilon_b(q) \;    p \cdot \varepsilon_c \;          \Big)
                 \varepsilon _c \cdot \dot X~.
  \eeqn
The next step is to rewrite all the terms on the right hand side   in terms of  celestial   variables $\omega_i, x_i$. 
  Using identities in \eqref{kenDiffpts},
      \be
      \hat k (x_i)\cdot    \hat k (x_j)\equiv  \hat k^\mu_i \cdot    \hat k^\mu_j=-\frac12 ( x_{ij})^2~, \qquad
         \hat k_i \cdot     \varepsilon_a(x_j)= x_{ij}^a~, \qquad
 \varepsilon_a(x_i) \cdot   \varepsilon_b(x_j)=\delta^{ab}~,
  \ee
  we find
  \be\label{epspq1}
  \varepsilon_a(p) \cdot h\cdot  \varepsilon_b(q)
  =  \varepsilon_a(p) \cdot \varepsilon_b(q)  -2\ap     \varepsilon_a(p) \cdot q\; \varepsilon_b(q) \cdot p 
  =\delta^{ab}+2\ap  \omega_p \omega_q x_{12}^a x_{12}^b~,
  \ee
and
  \be\label{epspq2}
   \varepsilon_a(p) \cdot  q  \; \varepsilon_b(q) \cdot \varepsilon_c
   =-\omega_q x_{12}^a \delta^{bc}~, \qquad
      \varepsilon_b(q) \cdot  p  \; \varepsilon_a(p) \cdot \varepsilon_c
   = \omega_p x_{12}^b  \delta^{ac}~.
  \ee 
  To simplify $p\cdot \varepsilon_c$, we need to use the momentum conservation.  Using \eqref{CSmomentum}, the momentum conservation $K=p+q=k+\epsilon v$   can be written as
 \be\label{momentumConse}
 \omega_p \Big( \frac{1+(x_1)^2}{2}, x_1^a,\frac{1-(x_1)^2}{2}  \Big)
 + \omega_q \Big( \frac{1+(x_2)^2}{2}, x_2^a,\frac{1-(x_2)^2}{2}  \Big)
 = \omega_k \Big( \frac{1+(x_3)^2}{2}, x_3^a,\frac{1-(x_3)^2}{2}  \Big)+\epsilon v~,
 \ee 
where the  infinitesimal quantity  $\epsilon$  is now given by
  \be
  \epsilon=2\ap p \cdot q=-\ap \omega_p \omega_q (x_{12})^2~.
  \ee

From \eqref{momentumConse}, it is easy to see that 
\be
\omega_k =\omega_p+\omega_q +\sO(\epsilon)~, \qquad
\omega_k x_3^a =\omega_p x_1^a+\omega_q  x_2^a+\sO(\epsilon)~. \qquad
\ee
As a result, we have 
\be\label{x123a}
x_{13}^a=\frac{\omega_q }{\omega_p+\omega_q} x_{12}^a+\sO(\epsilon)~, \qquad
x_{32}^a=\frac{\omega_p }{\omega_p+\omega_q} x_{12}^a+\sO(\epsilon)~,
\ee
and furthermore
  \be\label{epspq3}
   p \cdot \varepsilon_c =\omega_p \hat p \cdot \varepsilon_c(k) =\omega_p x_{13}^c  =\frac{\omega_p\omega_q}{\omega_p+\omega_q} x_{12}^c+\sO(\epsilon)~.
  \ee
%
  
With these identities \eqref{epspq1} \eqref{epspq2} \eqref{epspq3}, the OPE \eqref{XXOPEgluon}  simplifies to 
       \beqn\label{XXOPEgluon22}
&&\varepsilon_a(p)\cdot\dot X e^{i p \cdot X  }(y_1)\varepsilon_b(q)\cdot\dot X e^{i q  \cdot X  }(y_2)
\\&\sim&\nonumber
   i\; \epsilon   y_{12}^{\epsilon-1} 
\frac{\omega_p x_{12}^b \delta^{ac}+\omega_q x_{12}^a \delta^{bc}
-\frac{\omega_p\omega_q}{\omega_p+\omega_q} x_{12}^c 
\delta_{ab}
-2\ap \frac{\omega_p^2\omega_q^2}{\omega_p+\omega_q}   x_{12}^a x_{12}^bx_{12}^c  
}{p \cdot q} 
 \varepsilon _c \cdot \dot X \; e^{i (p+q) \cdot X }  +\sO(\epsilon)
   \\&\sim&\nonumber
 -4 i\;    \delta( y_{12})\;
I_c\Big(p,\varepsilon_a(p); q,\varepsilon_b(q); \ap\Big)\;
 \varepsilon _c \cdot \dot X \; e^{i (p+q) \cdot X } +\sO(\epsilon) ~, 
  \eeqn
where  used \eqref{xepsId} and have defined  
  \beqn\label{Ipqeps}
   I_c\Big(p,\varepsilon_a(p); q,\varepsilon_b(q); \ap\Big)
 & =&\label{celestI}
\frac{        \varepsilon_a(p) \cdot  q  \; \varepsilon_b(q) \cdot \varepsilon_c \;    
        -    \varepsilon_b(q) \cdot p  \; \varepsilon_a(p) \cdot \varepsilon_c \;  
                + \varepsilon_a(p) \cdot h\cdot  \varepsilon_b(q) \;    p \cdot \varepsilon_c            }{2 p\cdot q}
  \qquad\qquad\\&=&
  \frac{\omega_p x_{12}^b \delta^{ac}+\omega_q x_{12}^a \delta^{bc}
-\frac{\omega_p\omega_q}{\omega_p+\omega_q} x_{12}^c 
\delta^{ab}
-2\ap \frac{\omega_p^2\omega_q^2}{\omega_p+\omega_q}   x_{12}^a x_{12}^bx_{12}^c  
}{  \omega_p \omega_q (x_{12})^2}
  \\& =& 
   -I_c\Big(q,\varepsilon_b(q); p,\varepsilon_a(p); \ap\Big)~,
    \label{antisymIpqeps}
  \eeqn
  which is   anti-symmetric in $p$ and $q$, as shown above.

 Doing the integral as in  \eqref{VAVBint}, we thus get
   \beqn
\cV_a^A(p)\cV_b^B(q  )
 &=&\nonumber
\int dy_1 dy_2\; \theta(y_1-y_2) \;
\Big( -4 i   \,  \delta( y_{12})  
I _c(p,\varepsilon_a(p); q,\varepsilon_b(q); \ap )  
\Big)
 \varepsilon _c \cdot \dot X \; e^{i (p+q) \cdot X } (y_2)  t^A t^B
\\& &\nonumber
+
\int dy_1 dy_2\; \theta(y_2-y_1) \;
\Big( -4i  \,   \delta( y_{21})   I_c(  q,\varepsilon_b(q);p,\varepsilon_a(p);\ap) \Big)  
 \varepsilon _c \cdot \dot X \; e^{i (p+q) \cdot X } (y_1)t^Bt^A 
  \\&\propto&\nonumber
   f^{ABC}    I_c\Big(p,\varepsilon_a(p); q,\varepsilon_b(q); \ap\Big) \times
\int dy_2\;    
 \varepsilon _c \cdot \dot X \; e^{i (p+q) \cdot X } (y_2)   t^C
   \\&\propto &\nonumber
   f^{ABC}     I_c\Big(p,\varepsilon_a(p); q,\varepsilon_b(q); \ap\Big) \times
   \cV_c^C(p+q  )~,
   \eeqn
   where we   used the anti-symmetric property of $I_c$ in \eqref{antisymIpqeps} and   the notation $\cV_a(p)=\cV(p,\varepsilon_a(p))$, $\cV_b(q)=\cV(q,\varepsilon_b(q)) $ and so on. We also used the integral
   \be
   \int dy_1 dy_2\; \theta(y_1-y_2)  \delta( y_{12})  f(y_2)
   =   \int  dy_2\; \theta(0)   f(y_2)=\frac12 \int  dy_2\;      f(y_2)~,
   \ee
   with  $\theta(0)=\frac12$.

Finally, we need to perform the Mellin transformation \eqref{VDelta} in order to go to the conformal basis. Using the following integral,
\beqn\label{collinearBfcn}
&&  
 \int_0^\infty       d\omega_1\; \omega_1^{\Delta_1-1} 
 \int_0^\infty     d\omega_2\;  \omega_2^{\Delta_2-1}\;
\omega_1^\alpha \omega_2^\beta (\omega_1+\omega_2)^\gamma \; F(\omega_1+\omega_2)
\\&=&
 B\Big(\Delta_1+\alpha,\Delta_2+\beta \Big)    
  \int_0^\infty d\omega \; \omega^{ \Delta_P  -1}  \; F(\omega)~,
  \qquad\qquad \Delta_P=\Delta_1+\Delta_2+\alpha+\beta+\gamma~,
  \qquad\nonumber 
\eeqn
 we obtain the final result for celestial gluon OPE
\beqn\label{gluonOPEopen}
&&\cV^A_{\Delta_1,a}(x_1)\cV^B_{\Delta_2,b}(x_2)
\\&\sim&\nonumber
  f^{ABC} 
  \frac{x_{12}^{a}\delta^{bc} B(\Delta_1-1,\Delta_2) 
  +x_{12}^{b}\delta^{ac}B(\Delta_1 ,\Delta_2-1)
 -x_{12}^{c}\delta^{ab} B(\Delta_1 ,\Delta_2)
   }{(x_{12})^2}  
   \cV^C_{\Delta_1+\Delta_2-1,\;c}(x_2)
   \\&&
   -2\ap   f^{ABC} 
  \frac{    x_{12}^a x_{12}^bx_{12}^c    B(\Delta_1+1,\Delta_2+1)
   }{(x_{12})^2}  
     \cV^C_{\Delta_1+\Delta_2+1,\;c}(x_2)~.
   \nonumber
\eeqn
 
\subsubsection{General structure of worldsheet OPE}
Based on our previous OPE analysis of two gluon vertex operators, we   would like now to discuss the general structure of worldsheet OPE. 
From \eqref{openstringOPE} and the derivation above, one expects that in general the world-sheet OPE takes the form 
\be\label{generalOPE}
\mathscr V^{A,\ell_1} e^{ip \cdot X}  (y_1)\mathscr  V^{B,\ell_2} e^{iq \cdot X} (y_2)
\sim \sum_{\ell=0}^\infty y_{12}^{2 \alpha' p\cdot q+\ell-\ell_1-\ell_2 } 
F^{A,\ell_1;B,\ell_2}_{C, \ell} (p,q)  \mathscr V^{C,\ell } e^{i(p+q) \cdot X}(y_2)~,
\ee
where
\be
\mathscr  V^{A,\ell}  e^{i k\cdot X}\sim \prod_i\p^ {\ell_i} X^{\mu_i} e^{i k\cdot X}, \qquad 
\ell=\sum_i \ell_i:
\qquad 
h=\alpha' k^2+\ell  \xrightarrow{\text{on-shell}} \ell-\alpha' M^2= 1~.
\ee
Note that in \eqref{generalOPE} we did not impose the on-shell condition on the momenta and it  holds even off-shell. 

Now we take $p,q$ on-shell, namely  setting $p^2=-m_1^2=(1-\ell_1)/\alpha'$, $q^2=-m_2^2=(1-\ell_2)/\alpha'$,  then we find 
\be
2 \alpha' p\cdot q+\ell-\ell_1-\ell_2=\alpha'(p+q)^2-2+\ell ~.
\ee
For $p+q$, we take it slightly off-shell, namely   $(p+q)^2=-m_3^2+\epsilon/\alpha'=(1-\ell_3)/\alpha'+\epsilon/\alpha'$, then
\be
2 \alpha' p\cdot q+\ell-\ell_1-\ell_2=(1-\ell_3) +\epsilon -2+\ell  
=   \epsilon +\ell-\ell_3 -1~.
\ee
Therefore  \eqref{generalOPE} becomes
\beqn
\mathscr  V^{A,\ell_1} e^{ip \cdot X} \mathscr   V^{B,\ell_2}  e^{iq \cdot X}
& =& \sum_{\ell=0}^\infty y_{12}^{\epsilon +\ell-\ell_3-1 } 
F^{A,\ell_1;B,\ell_2}_{C, \ell} (p,q)  \mathscr V^{C,\ell } e^{i(p+q) \cdot X}(y_2)
\\ &\xrightarrow{\epsilon\to 0}&
 \sum_{\ell=0}^{\ell_3} \frac{1}{\epsilon}\delta^{\ell_3-\ell} (y_{12})
F^{A,\ell_1;B,\ell_2}_{C, \ell} (p,q)  \mathscr V^{C,\ell } e^{i(p+q) \cdot X}(y_2)~.
\eeqn
where we used \eqref{xepsId} and ignored some unimportant factors.

After doing the worldsheet integral, we then  expect to find 
\be
(p+q)^2\to(1-\ell_3)/\ap: \qquad 
\mathcal V^{A,\ell_1}_p  \mathcal V^{B,\ell_2}_q   \sim
   \frac{1}{(p+q)^2-(1-\ell_3)/\ap}
F^{A,\ell_1;B,\ell_2}_{C, \ell_3}\mathcal V^{C ,\ell_3}_{p+q}  +\cdots~,
\ee
where the dots are boundary contact terms of the form \eqref{boundarycontact} arising from $\ell=0, \cdots, \ell_3-1$ as well some terms in $\ell=\ell_3$. The prefactor on the right hand side is just the propagator $1/((p+q)^2+m_3^2)$.

  \subsection{OPE in closed string   }
  Now we switch to the closed string.  As before, we need to first compute 
     \beqn\label{closeOPEstr}
\cV (p, e_1)\cV (q, e_2 )
 &=& 
\int d^2 z_1 d^2 z_2\;
    V _{p, e_1}(z_1,\bar z_1)    V _{q, e_2}(z_2,\bar z_2) 
\\&=& 
\int d^2 z_1 d^2 z_2\; 
e_{1\mu\bar\mu}\p X^\mu\bar\p X^{\bar\mu }e^{i p \cdot X}(z_1,\bz_1) \;
e_{2\nu\bar\nu}  \p X^\mu\bar\p X^{\bar\nu} e^{i p \cdot X}(z_2,\bz_2)  ~,
\label{VVOPEclose}
  \eeqn
   and then perform the Mellin transformation to obtain the celestial OPE. 
  
  As the starting point, we need to know the world-sheet OPE of two operators in the integrand of \eqref{VVOPEclose}, which is computed in \eqref{closestringOPE}: 
  \beqn \label{closestringOPE2}
 &&  \p X^\mu\bar\p X^{\bar\mu }e^{i p \cdot X}(z_1,\bz_1)  \p X^\mu\bar\p X^{\bar\nu} e^{i p \cdot X}(z_2,\bz_2) 
\\&\sim&
  \frac { \ap^2 }{4}|z_{12}|^{  \alpha' p\cdot q-4} e^{i (p+q) \cdot X   }    
\\&&\times
\Bigg[  
  \Big(  \eta^{\mu\nu}  -\frac12\ap   q^\mu \; p^\nu \Big)  
  +   i \; z_{12} \Big(   q^\mu \;  \p X ^\nu  -  p^\nu\;  \p X^\mu
+  \big( \eta^{\mu\nu}  - \frac12\ap   q^\mu \; q^\nu\big)p \cdot \p X  
 \Big)  +
 \cdots \Bigg] 
\\&&\times
\Bigg[  
  \Big(  \eta^{\bar \mu\bar \nu}  -\frac12\ap   q^{\bar \mu} \; p^{\bar \nu} \Big)  
  +   i \; \bz_{12} \Big(   q^{\bar \mu} \;  \p X ^{\bar \nu}  -  p^{\bar \nu}\;  \bar \p X^{\bar \mu}
+  \big( \eta^{\bar\mu\bar\nu}  - \frac12\ap   q^{\bar \mu} \; q^\nu\big)p \cdot \bar\p X  
 \Big)  +
 \cdots \Bigg]  (z_2,\bar z_2)~.
 \qquad\qquad
\eeqn
This OPE is essentially the ``square'' of open string world-sheet OPE \eqref{openstringOPE}, up to the rescaling of $\ap$.  In particular, the terms proportional to $z_{12},\bz_{12}$ can be simplified as in the open string case; in each bracket, we can get  expressions which are similar to  \eqref{XXexpOPEsimple}.

Although many cross product terms from the left and right moving sectors  in \eqref{closestringOPE2} would be generated, we only need to consider the diagonal terms, namely those terms which only depend on the modulus $|z_{12}|$. The non-diagonal terms, like those with factor $z_{12}$, would become  zero after integrating along the angular direction of $z_{12}$.

 Before simplifying this OPE, let us first introduce some useful identities. 
 In the open string case, \eqref{xepsId} plays an important role.  Now we would like to find similar identities for the closed string.  We first parametrize the complex coordinate as
  \be
 z=u+{  i}v=\varrho e^{i \vartheta}~, \qquad \bz=u-{  i}v=\varrho e^{-i \vartheta}~, \qquad
 |z|=\varrho~.
 \ee
 Then we have the following formula
 \footnote{ In this convention, we thus have
 \be\nonumber
 \p_{\bar z}\frac{1}{z}= \pi \delta^{(2)}(z)~, \quad 
 \int d^2  z \;   f(z)  \delta^{(2)}(z)=f(0)~.  
 \ee
 For the representation of delta function in polar coordinates, see \url{https://mathworld.wolfram.com/DeltaFunction.html}. 
 }
 \be\label{deltazclose}
 \delta^{(2)}(z)=\delta(u) \delta(v)=\frac{\delta(\varrho)}{\pi\varrho}=\frac {1}{2\pi} \epsilon \varrho^{\epsilon-2} 
= \frac{ 1}{2\pi} \epsilon |z|^{\epsilon-2}~, \qquad
 d^2z=du dv=\varrho d\vartheta d\varrho ~,
\ee
where $\epsilon$ is again infinitesimal and we used the identity in \eqref{xepsId}. Further taking derivatives of the delta-function gives
\be\label{deltazClose}
\p \delta^{(2)}(z)=-\frac{ 1}{2\pi} \epsilon |z|^{\epsilon-2} z^{-1}~, \quad
\bar\p \delta^{(2)}(z)=-\frac{ 1}{2\pi} \epsilon |z|^{\epsilon-2} \bar z^{-1}~, \quad
\p\bar\p \delta^{(2)}(z)= \frac{ 1}{2\pi} \epsilon |z|^{\epsilon-4}  ~ ,  \quad\cdots~.
\ee
 

We are interested in the behavior of OPE \eqref{closestringOPE2} in the collinear limit $p\cdot q\to 0$. As in the open string case, one can show that most terms in the OPE are not relevant in the collinear limit; they are either regular in the limit $p\cdot q\to 0$, or only contribute boundary contact terms. 
For example, the tachyon   arising from the first term  in two square brackets of \eqref{closestringOPE2}  takes the form 
\be
  |z_{12}|^{  \alpha' p\cdot q-4} e^{i (p+q) \cdot X   }    
\to  \frac{ \p\bar\p \delta^{(2)}(z_{12})}{   p\cdot q }e^{i (p+q) \cdot X   } ~,
\ee
where we used identity in \eqref{deltazClose}. 
Following \eqref{VVOPEclose}, we need to   integrate over $z_1,z_2$. This type of  total derivative  can contribute at most boundary terms after doing the integral. Other terms in the  OPE \eqref{closestringOPE2} can be analyzed similarly. In particular, one can rewrite terms in each square bracket  in \eqref{closestringOPE2} and get expressions similar to the open string case \eqref{XXexpOPEsimple}. 
Then one can show that most terms are not important for our purpose. 

After doing simplification, the only relevant terms in OPE  \eqref{closestringOPE2}     in the collinear limit $p\cdot q\to 0$ are given by
 \beqn\label{closeOPEsimp}
 &&  \p X^\mu\bar\p X^{\bar\mu }e^{i p \cdot X}(z_1,\bz_1)  \p X^\nu\bar\p X^{\bar\nu} e^{i p \cdot X}(z_2,\bz_2) 
\\&\sim&
 - \frac { \ap^2 }{4}|z_{12}|^{  \alpha' p\cdot q-2} e^{i (p+q) \cdot X   }    
  \times \Big(   q^\mu \;  \varepsilon_c^\nu  -  p^\nu\;  \varepsilon_c^\mu
+  \big( \eta^{\mu\nu}  - \frac12\ap   q^\mu \; q^\nu\big)p \cdot  \varepsilon_c 
 \Big) 
 \label{closeOPEsimp112nd}
 \\&&\times
   \Big(   q^{\bar \mu} \;    \varepsilon_{\tilde c} ^{\bar \nu}  -  p^{\bar \nu}\;   \varepsilon_{\tilde c}^{\bar \mu}
+  \big( \eta^{\bar\mu\bar\nu}  - \frac12\ap   q^{\bar \mu} \; q^\nu\big)p \cdot    \varepsilon_\tc 
\label{closeOPEsimp2nd}
 \Big)  
 \times \varepsilon_c\cdot \p X\;   \varepsilon_{\tilde c}\cdot \bar\p X~.
\eeqn
which is just the ``square'' of \eqref{gluonpart}. 

 To simplify further, we choose the polarization tensors as the basis constructed in \eqref{varepsab}:
 \beqn\label{varepsab22}
\varepsilon^{\mu\bar\mu}_{a\tilde a}(p)
= \varepsilon^{\mu }_{a } (p)\varepsilon^{ \bar\mu}_{ \tilde a} (p)~,
 \qquad
\varepsilon^{\nu\bar\nu}_{b\tilde b}(q)
= \varepsilon^{\nu }_{b }(q) \varepsilon^{ \bar\nu}_{ \tilde b} (q)~.
\eeqn
Any polarization for graviton/dilaton/KR field can be expanded in this basis using \eqref{polarizationexpansion} \eqref{dilatonPol}.

Contracting the OPE \eqref{closeOPEsimp} with \eqref{varepsab22}  and taking the limit $p\cdot q\to 0$ gives
\beqn
 && \varepsilon_{a\tilde a\; \mu\bar\mu}(p) \p X^\mu\bar\p X^{\bar\mu }e^{i p \cdot X}(z_1,\bz_1)
 \varepsilon _{b\tilde b\; \nu\bar\nu}(q)  \p X^\nu\bar\p X^{\bar\nu} e^{i p \cdot X}(z_2,\bz_2) 
\\&\sim&
 - \frac { \ap^2 }{4}2\pi\frac{1}{ \alpha' p\cdot q}\delta^{2}(z_{12})  e^{i (p+q) \cdot X   }    
  \times \Big(   q^\mu \;  \varepsilon_c^\nu  -  p^\nu\;  \varepsilon_c^\mu
+  \big( \eta^{\mu\nu}  - \frac12\ap   q^\mu \; q^\nu\big)p \cdot  \varepsilon_c 
 \Big) \varepsilon^{\mu }_{a }  \varepsilon^{\nu }_{b }
\nonumber \\&&\qquad\qquad  \times
   \Big(   q^{\bar \mu} \;    \varepsilon_{\tilde c} ^{\bar \nu}  -  p^{\bar \nu}\;   \varepsilon_{\tilde c}^{\bar \mu}
+  \big( \eta^{\bar\mu\bar\nu}  - \frac12\ap   q^{\bar \mu} \; q^\nu\big)p \cdot   \varepsilon_{\tilde c} 
 \Big) 
  \varepsilon^{ \bar\mu}_{ \tilde a} \varepsilon^{ \bar\nu}_{ \tilde b}\;
  \varepsilon_c\cdot \p X\;   \varepsilon_{\tilde c}\cdot \bar\p X
   \\&\sim& 
 \pi\ap\;     \delta^2(z_{12}) \;      
 \cI_{c\tc}\Big(p,\varepsilon_{a\tilde a}(p); q,\varepsilon_{b\tilde b}(q);\ap,\ap\Big)\;
   \varepsilon_{c\tilde c\; \sigma\bar\sigma} \p X^\sigma     \bar\p X^{\bar\sigma}~,
       \label{closeOPErd2}
\eeqn
where we used the identity in \eqref{deltazclose} and defined  
\beqn
     &&
      \cI_{c\tc}\Big(p,\varepsilon_{a\tilde a}(p); q,\varepsilon_{b\tilde b}(q);\alpha,\beta\Big)
 \\&=&\nonumber
 \frac{1}{-2p\cdot q} \Big[ \varepsilon_a(p) \cdot  q  \; \varepsilon_b(q) \cdot \varepsilon_c \;    
        -    \varepsilon_b(q) \cdot p  \; \varepsilon_a(p) \cdot \varepsilon_c \;  
                +
                \Big(  \varepsilon_a(p) \cdot \varepsilon_b(q)  
                -\frac{\alpha}{2}    \varepsilon_a(p) \cdot  q\; \varepsilon_b(q) \cdot p
                \Big)  p \cdot \varepsilon_c
                \Big]
\qquad\nonumber \\&&\qquad
\times  \Big[ \varepsilon_\ta(p) \cdot  q  \; \varepsilon_\tb(q) \cdot \varepsilon_\tc \;    
        -    \varepsilon_\tb(q) \cdot p  \; \varepsilon_\ta(p) \cdot \varepsilon_\tc \;  
                +
                \Big(  \varepsilon_\ta(p) \cdot \varepsilon_\tb(q)  
                -\frac{\beta}{2}    \varepsilon_\ta(p) \cdot  q\; \varepsilon_\tb(q) \cdot p
                \Big)  p \cdot \varepsilon_\tc
                \Big]
 \nonumber \\&=& 
 \omega_p \omega_q (x_{12}) ^2\; 
       I_c\Big(p,\varepsilon_a(p); q,\varepsilon_b(q); \frac14\alpha\Big)\;
      I_\tc\Big(p,\varepsilon_{\tilde a}(p); q,\varepsilon_{\tilde b}(q); \frac14\beta \Big)  ~.
      \label{curalIclosed}     
\eeqn
The expression for $  I_c\Big(p,\varepsilon_a(p); q,\varepsilon_b(q); \ap\Big)$ is given in \eqref{Ipqeps}. 
  
Following \eqref{VVOPEclose} and using  \eqref{closeOPErd2}, we can now compute the OPE between two vertex operators  of massless fields in closed string theory: 
       \beqn 
\cV_{a\ta}(p)\cV_{b\tb}(q)
 &=&  \nonumber
\int d^2 z_1 d^2 z_2\;
\varepsilon_{a\tilde a\; \mu\bar\mu}(p) \p X^\mu\bar\p X^{\bar\mu }e^{i p \cdot X}(z_1,\bz_1)
 \varepsilon _{b\tilde b\; \nu\bar\nu}(q)  \p X^\nu\bar\p X^{\bar\nu} e^{i p \cdot X}(z_2,\bz_2)
 \\&=& 
\int d^2 z_1 d^2 z_2\; \pi\ap\;     \delta^2(z_{12}) \;      
 \cI_{c\tc}\Big(p,\varepsilon_{a\tilde a}(p); q,\varepsilon_{b\tilde b}(q);\ap,\ap\Big)\;
   \varepsilon_{c\tilde c\; \sigma\bar\sigma} \p X^\sigma     \bar\p X^{\bar\sigma} 
          \\&\propto&  
 \cI_{c\tc}\Big(p,\varepsilon_{a\tilde a}(p); q,\varepsilon_{b\tilde b}(q);\ap,\ap\Big)\;   
      \cV_{c\tc} (p+q  )~,
     \label{VaaVbb}
  \eeqn
 where we used the notation $\cV_{a\ta}(p)\equiv\cV(p, \varepsilon _{a\tilde a}(p)  )$ and similarly for $q$, $p+q$.
 
  Finally, we need to go to the conformal basis of the vertex operators \eqref{VDelta}.
 Using the formula \eqref{collinearBfcn} and performing   the Mellin transformation, we finally obtain the celestial OPE for graviton/dilaton/KR field in closed string theory
   \be\label{closeOPE}
\cV_{\Delta_1,a\ta}(x_1) \cV_{\Delta_2,b\tb}(x_2) \sim   \cK~,
\ee
where \footnote{ Note that the exchange in the parentheses is the simultaneous exchange of all items. }
\beqn\label{closeOPEK}
\nonumber
(x_{12})^2\cK&=&\Bigg[ \Big( \delta^{bc}\delta^{\tb\tc} x_{12}^a  x_{12}^\ta  B(\Delta_1-1,\Delta_2+1)
- \delta^{ac}\delta^{\ta\tb} x_{12}^b  x_{12}^\tc  B(\Delta_1+1,\Delta_2 )
- \delta^{ab}\delta^{\tb\tc} x_{12}^c  x_{12}^\ta  B(\Delta_1 ,\Delta_2+1)
\\&&\nonumber
\qquad
+ \frac12  \delta^{ab}\delta^{\ta\tb} x_{12}^c  x_{12}^\tc  B(\Delta_1+1 ,\Delta_2+1 )
+   \delta^{bc}\delta^{\ta\tc} x_{12}^a  x_{12}^\tb  B(\Delta_1 ,\Delta_2 )
\Big) 
\\&& 
\; +\Big( a\leftrightarrow b,\, \ta\leftrightarrow \tb, \; \Delta_1\leftrightarrow \Delta_2 \Big) \Bigg]
\cV_{\Delta_1+\Delta_2, c\tc}(x_2)
\label{GGOPE1st}
\\&&\nonumber
+\frac{\ap}{2} \Bigg[ 
\Big( \frac12 \delta^{\ta\tb}x_{12}^ax_{12}^b x_{12}^c x_{12}^\tc  B(\Delta_1+2,\Delta_2+2)
-
\delta^{ac}x_{12}^b x_{12}^\tb x_{12}^\ta x_{12}^\tc  B(\Delta_1+2,\Delta_2+1)
\\&&\qquad\qquad
+ (\; a\leftrightarrow \ta,\, b\leftrightarrow \tb, \; c\leftrightarrow \tc \; )
\Big)
\; +\Big( a\leftrightarrow b,\, \ta\leftrightarrow \tb, \; \Delta_1\leftrightarrow \Delta_2 \Big) \Bigg]
\cV_{\Delta_1+\Delta_2+2, c\tc}(x_2)
\label{GGOPE2nd}
\\&&
+\frac{\ap^2}{4}\; \Bigg[x_{12}^ax_{12}^b x_{12}^cx_{12}^\ta x_{12}^\tb x_{12}^\tc B(\Delta_1+3,\Delta_2+3) \Bigg]\cV_{\Delta_1+\Delta_2+4, c\tc}(x_2)~.
\label{GGOPE3rd}
\eeqn


\section{Celestial OPE  from worldsheet OPE  in superstring  }\label{superstringOPE}

In this section, we will generalize the previous discussions on OPEs in bosonic string to superstring.  The computations      are similar, and especially we can also simplify the results in the same way. The main difference is that in superstring, the worldsheet has supersymmetry. As a result, the bosonic and fermionic contributions can   cancel each other, which leads to a simpler result. In particular, the $\ap$ corrections to the worldsheet OPEs and thus also to the celestial OPEs are absent in the supersymmetric sector. This behavior  agrees with  the three-point amplitudes in superstring theory where the $\ap$ corrections also don't appear in the supersymmetric sector. 

We will be mainly focusing on  the heterotic string case. The heterotic string is a good playground for our studies of OPEs because we can realize both graviton and gluon easily in terms of closed string. In this setup, we compute all the OPEs involving gluon and graviton/dilaton/KR field, including their mixed OPEs.   Up to $\ap$ corrections, the OPEs derived in heterotic string agree with those derived in bosonic string. 
We will also briefly discuss the case of  type I and IIA/IIB string, which are very similar to heterotic string. 

\subsection{Heterotic string}

In this subsection, we will first review the vertex operators of gluon and graviton/dilaton/KR field in  heterotic string, and then derive the celestial OPEs from worldsheet OPEs.  

\subsubsection{Vertex operator  in heterotic string}
  
In bosonic closed string, the vertex operators take  two forms, either integrated form with worldsheet integration or unintegrated form with $c\tilde c$ attachment. In superstring theory, the vertex operators further take  a variety of forms, called pictures \cite{Friedan:1985ey,Polchinski:1998rq}. In each supersymmetric left or right moving sector, we can label the vertex operator with the ghost charges $q$ or $\tilde q$. The vertex operator in the $q$ or $\tilde q$ picture contains a factor $e^{q\phi}$ or $e^{\tilde q\tilde \phi}$, where $\phi$ is the bosonized field of $\beta\gamma $ ghost system 
and similarly for $\tilde\phi$. The vertex operators in different pictures can be related via the picture changing operator.  The total ghost charges in each supersymmetric sector is determined by the worldsheet genus. In particular, at tree level  of interest in this paper, the total ghost charge should be -2 in order to soak up the fermionic zero modes.  

More specifically, in heterotic string, the gluon vertex operator in the picture -1 and 0 are respectively given by \cite{Polchinski:1998rq}
 \beqn\label{Vm1}
  V^{(-1)}(z,\bar z) &=&   e^{-\tilde \phi } J^A  \tilde \psi^\nu  e^{i k \cdot X}
  =V _L(z)V^{(0)}_R(\bz)~,
   \\
  V^{(0)}(z,\bar z) &=&   J^A\Big( i \bar\p X^\nu+\frac12 \ap k \cdot {\tilde\psi} {\tilde \psi}^\nu \Big)  e^{i k \cdot X} 
  =V _L(z)V^{(-1)}_R(\bz)~,
 \eeqn
 where $J^A(z)$ are the Kac-Moody currents in the left-moving sector, while $\tilde \phi$ and $\tilde \psi$ are the ghost and world-sheet fermion in the right-moving sector, respectively. Similarly, for the massless graviton/dilaton/KR   fields in heterotic string, their vertex operators in the -1 and 0 picture are given   by
   \beqn\label{Wm1}
W^{(-1)}(z,\bar z) &=&   e^{-\tilde \phi} \p X^\mu \tilde \psi^\nu e^{i k \cdot X}=
  W _L(z)W^{(-1)}_R(\bz)~,
\\
W^{(0)}(z,\bar z) &=&   \p X^\mu \Big( i \bar\p X^\nu+\frac12 \ap k \cdot {\tilde\psi} {\tilde \psi}^\nu \Big)e^{i k \cdot X}=  W _L(z)W^{(0)}_R(\bz)~.
  \eeqn
  
 In these vertex  operators, we also write them in  the product form of left and right moving parts:
 \beqn
&& V _L(z)=  J^A(z) e^{i k \cdot X_L }~, \qquad
W _L(z)=  \p X^\mu e^{i k \cdot X_L }~,
\\ &&
V^{(0)}_R(\bz)= W^{(0)}_R(\bz)=\Big( i \bar\p X^\mu+\frac12 \ap k \cdot {\tilde\psi} {\tilde \psi}^\mu \Big)e^{i k \cdot X_R}~, 
\\ &&
V^{(-1)}_R(\bz)= W^{(-1)}_R(\bz)=e^{-\tilde \phi} {\tilde \psi}^\mu \; e^{i k \cdot X_R}~, 
 \eeqn
 where $X(z,\bz)=X_L(z)+X_R(\bz)$.
 This decomposition enables us to compute the OPE in the left and right moving sectors independently.  The Kac-Moody currents $J^A(z)$ satisfy the OPE
 \be
 J^A(z_1)J^B(z_2) \sim \frac{\lk\delta^{AB}}{z_{12}^2}+\frac{f^{ABC} J^C(z_2)}{z_{12}}~.
 \ee
 Without loss of generality, one can set the level $\lk$ to one, 
 which can always be realized by rescaling the  Kac-Moody currents and structure constants.
 
 \subsubsection{Celestial OPE from heterotic worldsheet}
 Now we derive the celestial OPEs in heterotic string from worldsheet. 
 
 \subsubsection*{Gluon-gluon OPE }
 We want to compute the OPE between gluons.  Although vertex operators in different pictures are physically equivalent, it turns out to be simpler to consider the OPE of two vertex operators in the $-1$  and 0 picture, namely $  V^{(-1)}(z_1,\bar z_1)   V^{(0)}(z_2,\bar z_2)$, which gives another vertex operator  in  $-1$ picture. 
 
 We first consider the left-moving OPE $ V _L(z_1) V _L(z_2)$:
 \beqn
   J^A e^{i p \cdot X_L }(z_1)    J^B e^{i q \cdot X_L }(z_2) 
 &=&
   J^A (z_1) J^B (z_2) \times e^{i p \cdot X_L }(z_1)e^{i q \cdot X_L }(z_2) 
 \nonumber   \\&=&
  \Big(  \frac{ \lk\delta^{AB} }{z_{12}^2}+\frac{f^{ABC} J^C(z_2)}{z_{12}}+\cdots\Big) \times 
z_{12}^{\frac12\ap p \cdot q} \Big( 1+ i\, z_{12} p \cdot \p X+\cdots
\Big) e^{i(p+q) \cdot X_L}(z_2)
  \nonumber  \\&=&
z_{12}^{\frac12\ap p \cdot q-1}  \Big(\frac{ \lk\delta^{AB} }{z_{12}}+   {f^{ABC} J^C }   
+ i\,   \lk\delta^{AB} \, p \cdot \p X+\cdots
\Big) e^{i(p+q) \cdot X_L}(z_2)~.
\label{JAJBOPE2}
 \eeqn
 Following the derivation of \eqref{pXsimple}, we have 
  \beqn \label{pdoxXsimle}
 p \cdot \p X
 & =& 
  p \cdot \varepsilon_c \;  \varepsilon _c \cdot  \p  X
 + \frac{1}{\omega_k }         p\cdot q  \;  \p  X \cdot n 
+ \frac{\omega_p}{\omega_k} k \cdot \p   X +\sO(\epsilon)~.
  \eeqn
  Due to the $p \cdot q$ factor, the second term on the right hand side is regular in the collinear limit  $p \cdot q\to 0$.  For the third term, we can combine it with extra factors in \eqref{JAJBOPE2}, leading to  
  \be
z_{12}^{\frac12\ap p \cdot q-1}\; k \cdot \p   X\; e^{i (p+q)\cdot X_L}
=-i\,z_{12}^{\frac12\ap p \cdot q-1}\;  \p  \; e^{ik\cdot X_L} +\sO(\epsilon)~,
  \ee
  where we used $e^{i(p+q) \cdot X_L}=e^{i k \cdot X_L}+\sO(\epsilon)$ as $p+q=k +\sO(\epsilon)$ in the collinear limit. Since the polarization is along the momentum direction, the  is supposed to be a pure gauge and thus can be ignored.  The more subtle issue of boundary  contact   terms  is the same as that in open bosonic string case, see \eqref{boundarycontact2}. Therefore, we can just keep the first term in \eqref{pdoxXsimle}. 
  
  As a result, in the collinear limit  the left-moving OPE simplifies to
  \beqn
   J^A e^{i p \cdot X_L }(z_1)    J^B e^{i q \cdot X_L }(z_2) 
 &=& 
z_{12}^{\frac12\ap p \cdot q-1}  \Big(\frac{ \lk\delta^{AB} }{z_{12}}+   {f^{ABC} J^C }   
+ i\,   \lk\delta^{AB} \,   p \cdot \varepsilon_c \;  \varepsilon _c \cdot  \p  X+\cdots
\Big) e^{i(p+q) \cdot X_L}(z_2)~.
\qquad\qquad
\label{JAJBOPE}
 \eeqn

 For the right-moving   OPE  $V_R^{(-1)}( \bar z_1)   V_R^{(0)}( \bar z_2)$, it  is computed in \eqref{RightVROPE}
    \beqn
&&
\zeta\cdot \tilde \psi  e^{i p \cdot X_R  }(\bz_1)
  \Big( i \xi\cdot\bar\p X +\frac12 \ap q \cdot {\tilde\psi}\xi\cdot {\tilde \psi}  \Big)e^{i q \cdot X_R}(\bz_2)
   \\&\sim& 
    \frac{  \alpha'  }{2}   \bz_{12}^{\frac12\ap p \cdot q-1} 
e^{i (p+ q) \cdot X_R}
\Big( 
\zeta\cdot  q \; \xi\cdot \tilde \psi - \xi\cdot p \; \zeta\cdot\tilde \psi 
 -\zeta\cdot \xi\; q  \cdot \tilde \psi 
   \Big) ~.
   \label{414bracket}
     \eeqn
 Note that terms in the bracket of \eqref{414bracket} are very similar to the terms  proportional $y_{12}$   in the bracket of  \eqref{openstringOPE}: one just needs to replace $\dot X$ with $\psi$, exchange $p,\zeta$ and $q,\xi$, and ignore $\ap$ corrections.  Therefore we can  also simplify the formula  in the same manner.  Following the steps in the derivation of   \eqref{pXsimple} and   \eqref{zetaxiDiff}, we now have \footnote{ Note $  (p+q) \cdot \varepsilon_\tc=k \cdot \varepsilon_\tc+\sO(\epsilon) =\sO(\epsilon) $ as $k \cdot \varepsilon_\tc=0$.}
\beqn 
 q \cdot \tilde \psi 
 & =&
  -p \cdot \varepsilon_\tc \;  \varepsilon _\tc \cdot  \tilde \psi 
 + \frac{1}{\omega_k }         p\cdot q  \; \tilde \psi  \cdot n 
+ \frac{\omega_q}{\omega_k} k \cdot \tilde \psi  +\sO(\epsilon) ~,
\eeqn
and
\beqn
    \zeta \cdot  q \;  \xi \cdot  \tilde \psi   -   \xi \cdot p\; \zeta \cdot  \tilde \psi 
=
      \zeta \cdot  q  \; \xi \cdot \varepsilon_\tc \;  \varepsilon _\tc \cdot  \tilde \psi  
        -   \xi \cdot p  \;  \zeta \cdot \varepsilon_\tc \;  \varepsilon _\tc \cdot  \tilde \psi   
          +\sO(\epsilon)~.
    \eeqn 
  These two formulae enable  us simplify  \eqref{414bracket}
     \beqn
&&
\zeta\cdot \tilde \psi  e^{i p \cdot X_R  }(\bz_1)
  \Big( i \xi\cdot\bar\p X +\frac12 \ap q \cdot {\tilde\psi}\xi\cdot {\tilde \psi}  \Big)e^{i q \cdot X_R}(\bz_2) 
        \\&\sim& 
    \frac{  \alpha'  }{2}   \bz_{12}^{\frac12\ap p \cdot q-1} 
e^{i (p+ q) \cdot X _R}
\Big( 
\zeta\cdot  q \; \xi\cdot \varepsilon_\tc 
-\xi\cdot p \; \zeta\cdot \varepsilon_\tc
 +\zeta\cdot \xi\; p  \cdot \varepsilon_\tc 
   \Big)\; \varepsilon_\tc \cdot \tilde \psi 
       \label{firstline}
           \\& & 
        +  
          \frac{  \alpha'  }{2}   \bz_{12}^{\frac12\ap p \cdot q-1} 
e^{i (p+ q) \cdot X _R} 
\Big( -\zeta\cdot \xi\;  \frac{1}{\omega_k }         p\cdot q  \; \tilde \psi  \cdot n 
 -\zeta\cdot \xi\;  \frac{\omega_q}{\omega_k} k \cdot \tilde \psi 
   \Big)~.
   \label{secondTermBRST}
  \eeqn
  In the collinear limit $p \cdot q\to 0$, the first term in \eqref{secondTermBRST} is regular. For   the second term, we can rewrite it as: \footnote{ Recall that that $p+q=k+\epsilon v$ where $k$ is null while $p+q$ is slightly off-shell.}
  \be   \label{fssecondTermBRST}
  e^{i (p+ q) \cdot X _R}  k \cdot \tilde \psi 
  =  e^{i k \cdot X _R}  k  \cdot \tilde \psi +\sO(\epsilon)~.
  \ee
  If we combine  \eqref{fssecondTermBRST} with the factor    $  z_{12}^{\frac12\ap p \cdot q-1} $ from the left-moving sector,
  \footnote{\label{ftn10} The left-moving sector may also contribute   $ z_{12}^{\frac12\ap p \cdot q-2} $, then in total we have $ |z_{12}|^{ \ap p \cdot q-2 } z_{12}^{-1}$. This gives zero after integrating along the angular direction of $z_{12}$.
  
 }
   we get
  \beqn\label{form430}
    |z_{12}|^{ \ap p \cdot q-2 }  e^{i k \cdot X _R}  k  \cdot \tilde \psi 
\xrightarrow{p \cdot q\to 0} \frac{\delta^2( z_{12})}{p \cdot q}
 k  \cdot \tilde \psi   e^{i k \cdot X _R}  
 \xrightarrow{\int d^2 z_1\, d^2 z_2}
 \frac{1}{p \cdot q}\int d^2 z_2\;  k  \cdot \tilde \psi   e^{i k \cdot X _R}  ~.
  \eeqn
Since the polarization is along the momentum direction,  this is a pure gauge.   More specifically it is a BRST exact term and thus  plays no role in string amplitude.  Therefore, we can drop all the terms in \eqref{secondTermBRST}. 

Combining the left-moving  OPE \eqref{JAJBOPE} and right-moving  OPE  \eqref{firstline}, we then find
 \beqn
 && 
 e^{-\tilde \phi } J^A  \tilde \psi^\mu  e^{i p \cdot X}(z_1,\bz_1)
 J^B\Big( i \bar\p X^\nu+\frac12 \ap q \cdot {\tilde\psi} {\tilde \psi}^\nu \Big)  
 e^{i k \cdot X}  (z_2,\bz_2)
 \\&\sim&  
    \frac{  \alpha'  }{2}    |z_{12}|^{ \ap p \cdot q-2}  e^{-\tilde \phi }
   \Big(\frac{ \lk\delta^{AB} }{z_{12}}+   {f^{ABC} J^C }   
+ i\,   \lk\delta^{AB} \,   p \cdot \varepsilon_c \;  \varepsilon _c \cdot  \p  X+\cdots
\Big) 
\nonumber\\&&\qquad\qquad\qquad\qquad\times
\Big( 
  q^\mu \;  \varepsilon_\tc ^\nu
- p^\nu \;   \varepsilon_\tc^\mu
 +\eta^{\mu\nu} \; p  \cdot \varepsilon_\tc 
   \Big)  \; \varepsilon_\tc \cdot \tilde \psi  \, e^{i (p+q) \cdot X} ~.
  \qquad\qquad \label{ggOPEsimp22}
 \eeqn
  First, we have the leading term $\lk\delta^{AB}$ coming with a factor $  |z_{12}|^{ \ap p \cdot q-2}z_{12}^{-1} $. This leads to zero after performing the $z_1,z_2$ integral due to the cancellation from  the angular direction of $z_{12}$.  After contracting  with the basis of polarization,  the rest of terms in \eqref{ggOPEsimp22} become 
  
   \beqn
 && 
 e^{-\tilde \phi } J^A  \varepsilon_\ta(p) \cdot \tilde \psi   e^{i p \cdot X}(z_1,\bz_1)
 J^A\Big( i \varepsilon_\tb(q) \cdot \bar\p X+
 \frac12 \ap q \cdot {\tilde\psi}\; \varepsilon_\tb(q) \cdot {\tilde \psi} \Big)  
 e^{i k \cdot X}  (z_2,\bz_2)
 \\&\sim&  \nonumber
     \frac{  \alpha'  }{2}  |z_{12}|^{ \ap p \cdot q-2} 
    \Big( 
  \varepsilon_\ta(p) \cdot q  \;  \varepsilon_\tb(q) \cdot \varepsilon_\tc  
-   \varepsilon_\tb(q) \cdot p \;  \varepsilon_\ta(p) \cdot \varepsilon_\tc
 +\varepsilon_\ta(p) \cdot   \varepsilon_\tb(q)  \; p  \cdot \varepsilon_\tc 
   \Big) 
      \\& & \qquad\qquad  
\times   \Big[ {f^{ABC}  }\; e^{-\tilde \phi }       J^C \; \varepsilon_\tc \cdot \tilde \psi  \, e^{i (p+q) \cdot X}   
       +
   i\,   \lk\delta^{AB} \,   p \cdot \varepsilon_c \;   e^{-\tilde \phi }  \;  \varepsilon _c \cdot  \p  X  \varepsilon_\tc \cdot \tilde \psi  \, e^{i (p+q) \cdot X} \Big]
 \\& \sim& \nonumber
2\pi\;\delta^2(z_{12})\;
           I_\tc\Big(p,\varepsilon_\ta(p); q,\varepsilon_\tb(q); 0\Big) 
 \\& &  \qquad\qquad  
\times   \Big[ {f^{ABC}  }\; e^{-\tilde \phi }       J^C \; \varepsilon_\tc \cdot \tilde \psi  \, e^{i (p+q) \cdot X}   
       +
   i\,   \lk\delta^{AB} \,   p \cdot \varepsilon_c \;   e^{-\tilde \phi }  \;  \varepsilon _c \cdot  \p  X  \varepsilon_\tc \cdot \tilde \psi  \, e^{i (p+q) \cdot X} \Big]~,
 \eeqn
 where we used identity \eqref{deltazclose} and the expression for  $I _\tc( p, \varepsilon_\ta(p)   ; q, \varepsilon_\tb(q);0  )$ in \eqref{Ipqeps} but without $\ap$ corrections. 
 Compared  with  \eqref{Vm1}  \eqref{Wm1}, it is easy to recognize that terms in the square bracket are exactly     the vertex operators  of gluon and graviton/dilaton/KR field. After performing the integral over $z_1,z_2$,  we get 
 \beqn
   \cV_\ta^{(-1) A}(p ) \cV_\tb^{(0) B}(q )
&\sim&
  f^{ABC} \; I_\tc \Big(p,\varepsilon_\ta(p); q,\varepsilon_\tb(q); 0\Big) 
   \cV_\tc^{(-1) C}(p +q)  
  \\&&
  + 
  \delta^{AB} \,   p \cdot \varepsilon_c \; 
 I_\tc \Big(p,\varepsilon_\ta(p); q,\varepsilon_\tb(q); 0\Big) 
    \cW_{c\tc}^{(-1) C}(p +q ) ~, \qquad
    \nonumber
 \eeqn
 up to some coefficients   in front of gluon    and graviton contributions, respectively. 
 
We can further write   in terms of celestial variables. In particular, 
\be
 p \cdot \varepsilon_c=\omega_p \hat p    \cdot \varepsilon_c
 =\omega_p (x_{13})^c =\frac{\omega_p\omega_q}{\omega_p+\omega_q} x_{12} ^c~.
 \ee
 where we used \eqref{kenDiffpts} and \eqref{x123a}. 
 
 Finally performing the Mellin transformation   yields the celestial OPE of two gluons 
 \beqn\label{twogluonOPEheterotic}
&&\cV^A_{\Delta_1,\ta}(x_1)\cV^B_{\Delta_2,\tb}(x_2)
\\&\sim&\nonumber
  f^{ABC} 
  \frac{x_{12}^{\ta}\delta^{\tb\tc} B(\Delta_1-1,\Delta_2) 
  +x_{12}^{\tb}\delta^{\ta\tc}B(\Delta_1 ,\Delta_2-1)
 -x_{12}^{\tc}\delta^{\ta\tb} B(\Delta_1 ,\Delta_2)
   }{(x_{12})^2}  
   \cV^C_{\Delta_1+\Delta_2-1,\;\tc}(x_2)
\\&& \nonumber
  +\delta^{AB}
 x_{12}^c \frac{x_{12}^{\ta}\delta^{\tb\tc} B(\Delta_1 ,\Delta_2+1) 
  +x_{12}^{\tb}\delta^{\ta\tc}B(\Delta_1+1 ,\Delta_2 )
 -x_{12}^{\tc}\delta^{\ta\tb} B(\Delta_1+1 ,\Delta_2+1)
   }{(x_{12})^2}  
   \cW _{\Delta_1+\Delta_2 ,\;c\tc}(x_2)~,  
   \eeqn
 where the first and second line just correspond to the gluon and graviton, respectively. 
The     OPE of gluons  among   themselves is the same as that in bosonic string case \eqref{gluonOPEopen}, except for the absence of $\ap$ corrections. 
 
  \subsubsection*{Gluon-graviton/dilaton/KR   OPE} 
     Now we  consider the  OPE   between gluon and graviton/dilaton/KR field, namely $W^{(-1)}(z_1,\bar z_1)   V^{(0)}(z_2,\bar z_2)$. 
     The left-moving OPE   $ W _L(z_1)V _L(z_2)$ is simply given by 
     \be
     \p X^\rho e^{i p \cdot X_L }(z_1)   J^A   e^{i q \cdot X_L }(z_2)
  =  z_{12}^{\frac12 \ap p \cdot q-1} \Big( -\frac{i}{2} \ap q^\rho+\cdots\Big)  J^A   e^{i (p+q) \cdot X_L }(z_2)~,
   \ee
 which follows from the same derivation of \eqref{OPEoneXdot}. 
    The right-moving OPE $W_R^{(-1)}( \bar z_1)   V_R^{(0)}( \bar z_2)=V_R^{(-1)}( \bar z_1)   V_R^{(0)}( \bar z_2)$ has been discussed above and the final result is given in \eqref{firstline}. 
    
   Combining the left and right moving sectors,  we get 
    \beqn\label{heteroticgluongraviton}
     &&
         e^{-\tilde \phi }    \varepsilon_a \cdot \p X   \varepsilon_\ta  \cdot \tilde \psi  e^{i p \cdot X}(z_1,\bz_1) \;
    J^A   \Big( i \varepsilon_\tb\cdot \bar\p X^\nu+\frac12 \ap p \cdot {\tilde\psi}\; \varepsilon_\tb\cdot {\tilde \psi}^\nu \Big)  e^{i q \cdot X}(z_2,\bz_2)
       \\&\sim&
  -\frac{i}{4}\ap^2 |z_{12}|^{ \ap p \cdot q-2 } 
  q \cdot \varepsilon_a   
\Big( 
\varepsilon_\ta \cdot  q \; \varepsilon_\tb\cdot \varepsilon_\tc 
-\varepsilon_\tb\cdot p \; \varepsilon_\ta \cdot \varepsilon_\tc
 +\varepsilon_\ta \cdot \varepsilon_\tb\; p  \cdot \varepsilon_\tc 
   \Big)\;
          e^{-\tilde \phi }   J^A \varepsilon_\tc \cdot \tilde \psi \;  e^{i (p+ q) \cdot X  }
         \qquad \\&\sim&
-i\, \pi \ap \delta^2(z_{12})\;
 q \cdot \varepsilon_a   \,
 I_\tc\Big(p,\varepsilon_\ta(p); q,\varepsilon_\tb(q); 0\Big) \; 
           e^{-\tilde \phi }    J^A \varepsilon_\tc \cdot \tilde \psi \;  e^{i (p+ q) \cdot X  }~,
    \eeqn
    Integrating over $z_1,z_2$  gives
    \be
    \cW^{(-1)}_{a\ta}(p)  \cV^{(0)A}_\tb(q) \propto
   -  q \cdot \varepsilon_a   \,
 I_\tc\Big(p,\varepsilon_\ta(p); q,\varepsilon_\tb(q); 0\Big)
 \cV^{(-1)A}_\tc(p+q)~,
    \ee
 where we ignore the overall constant for simplicity.   
    
    Further writing in terms of celestial variables and   performing the Mellin transformation, we get 
     \beqn\label{gluongravtionOPEheterotic}
&&\cW _{\Delta_1,a\ta}(x_1)\cV^A_{\Delta_2,\tb}(x_2)
\\&\sim&\nonumber 
 x_{12}^a \frac{x_{12}^{\ta}\delta^{\tb\tc} B(\Delta_1-1 ,\Delta_2+1) 
  +x_{12}^{\tb}\delta^{\ta\tc}B(\Delta_1  ,\Delta_2 )
 -x_{12}^{\tc}\delta^{\ta\tb} B(\Delta_1  ,\Delta_2+1)
   }{(x_{12})^2}  
   \cV^A _{\Delta_1+\Delta_2 ,\tc}(x_2)~.
   \eeqn 
   
    \subsubsection*{Graviton/dilaton/KR OPE} 
Finally we want to discuss  the  OPE  of graviton/dilaton/KR field, namely $W^{(-1)}(z_1,\bar z_1)   W^{(0)}(z_2,\bar z_2)$. 
The OPE in the left moving sector $W _L(z_1)W _L(z_2)$ is given in \eqref{closedLeft} and it is the same as the left moving sector of closed bosonic string. Up to the boundary contact terms, we can simplify this sector in the same way as that in the bosonic string. 
 
 The right-moving OPE $W_R^{(-1)}( \bar z_1)   W_R^{(0)}( \bar z_2)=V_R^{(-1)}( \bar z_1)   V_R^{(0)}( \bar z_2)$ has been discussed above and the final result is given in \eqref{firstline}. 
 Combining the left and right moving sectors, we arrive at almost   the same OPE as that  in the closed bosonic string   \eqref{closeOPEsimp} except that  we need to remove the $\ap$ corrections in the right-moving sector \eqref{closeOPEsimp2nd}, which are absent in heterotic string due to supersymmetry \eqref{firstline}.  As a result, in heterotic string the OPE \eqref{VaaVbb}   becomes 
   \beqn 
\cW^{(-1)}_{a\ta}(p)\cW^{(0)}_{b\tb}(q)
 &\sim&  
 \cI_{c\tc}\Big(p,\varepsilon_{a\tilde a}(p); q,\varepsilon_{b\tilde b}(q);\ap,0\Big)\;   
      \cW^{(-1)}_{c\tc} (p+q  )~.
  \eeqn
  
Then one can derive the celestial OPE in the same way as bosonic string. The final result is given by \eqref{closeOPE}:   ignoring the $\ap^2$ terms in \eqref{GGOPE3rd}, and   keeping all the terms in \eqref{GGOPE1st} as well as the $\ap$ terms in \eqref{GGOPE2nd}  which are  proportional to $x_{12}^ax_{12}^bx_{12}^c$.    \footnote{ More specifically, the  terms in the square bracket of \eqref{GGOPE2nd}  is now given by: 
\be\nonumber
\Bigg[x_{12}^ax_{12}^bx_{12}^c\Big(B(\Delta_1+2,\Delta_2+2) \delta^{\ta\tb}x_{12}^\tc
-B(\Delta_1+1,\Delta_2+2) \delta^{\tb\tc}x_{12}^\ta
-B(\Delta_1+2,\Delta_2+1) \delta^{\ta\tc}x_{12}^\tb\Big)\Bigg]~.
\ee 
}
%
%
%
  
\subsection{Type I, IIA, IIB string}
Now we  briefly discuss the OPEs of closed string massless fields in type I and IIA/IIB superstring. 
 The  vertex operators for massless NS-NS fields in type I and  IIA/IIB string are \cite{Polchinski:1998rq}
      \beqn\label{IIAIIBOp}
W^{(-1,-1)}(z,\bar z) &=&   e^{-\phi-\tilde \phi}  \psi^\mu \tilde \psi^\nu e^{i k \cdot X} ~,
\\ \label{IIAIIBOp2}
W^{(0,0)}(z,\bar z) &=&   \Big( i  \p X^\mu+\frac12 \ap k \cdot { \psi} {  \psi}^\mu \Big)  \Big( i \bar\p X^\mu+\frac12 \ap k \cdot {\tilde\psi} {\tilde \psi}^\mu \Big)e^{i k \cdot X} ~. 
  \eeqn
  Now both the left and right-moving sectors are supersymmetric. Note that for type I superstring,   the KR 2-form field is projected out and we are only left with graviton and dilaton. 
  
It is simpler to study OPE of two operators in $-1$ and $0$ picture, namely $W^{(-1,-1)}(z_1,\bar z_1)W^{(0,0)}(z_2,\bar z_2)$.  
  The    vertex operators  \eqref{IIAIIBOp} \eqref{IIAIIBOp2} can again be decomposed into  the product of left and right moving parts. For both the left and right moving sectors, the OPE  is given   by \eqref{firstline} or its conjugate. 
Combining the  two sectors together, the OPE of  $W^{(-1,-1)}(z_1,\bar z_1)W^{(0,0)}(z_2,\bar z_2)$ is similar to that  in  closed bosonic string 
\eqref{closeOPEsimp112nd}\eqref{closeOPEsimp2nd} except that  we need to replace $\p X,\bar\p X\to \psi,\tilde\psi$ and remove all the $\ap$ corrections, which are absent    now due to greater supersymmetry. It is amusing that the tachyon also does not appear in the OPE, although we have not performed the GSO projection. The final  celestial OPE is given by \eqref{closeOPE} with  all $\ap$ terms removed.   
The absence of $\ap$ correction is consistent with the string amplitude, as we will see in the next section.  

%

\section{Celestial OPE from collinear factorization} \label{OPEfromfactorization}
In the previous two sections, we have derived the celestial OPEs from  the string worldsheet   perspective.  In this   section, we will compute the celestial OPEs using  a different method based on the collinear factorization   of scattering amplitudes.  It turns out that the two approaches give  the same result all the time.

Before going to the computational details, let us first describe the general strategy of deriving celestial OPEs based on the collinear factorization. 
We are interested in the     scattering amplitude of massless fields. 
In the collinear limit  two momenta become  parallel $p_1/\! /p_2$, and the total momentum $ P\equiv p_1+p_2$ also becomes almost on-shell, namely $P^2=2p_1\cdot p_2\to 0$. The amplitude then factorizes in the collinear limit as \footnote{  Since the two collinear particles are massless, the collinear limit thus singles out the massless field propagator connecting $A_3$ and $A_n$.   More generally, to identify the contribution from the field with mass $M$, one can  instead use the propagator $1/(P^2+M^2)$  in \eqref{ampfactorization} and take the   limit $P^2\to -M^2$. }
\be 
A_{n+1}(p_1^{s_1},p_2^{s_2}, \cdots) \xrightarrow   {p_1/\! /p_2} \sum_sA_{3 }(p_1^{s_1},p_2^{s_2}, -P^{ s}) \frac{1}{P^2}A_{n}(P^{s}, \cdots) ~,
\ee
where $s_i$ are the extra quantum numbers labelling the particles.   The prefactor in front of $A_{n}$ is essentially the split function characterizing the collinear behavior:
\be\label{ampfactorization}
 \Split(p_1^{s_1}+p_2^{s_2}\to P^s)=  \frac{1}{P^2}A_{3 }(p_1^{s_1},p_2^{s_2}, -P^{ s}) 
=  \frac{1}{2p_1\cdot p_2 }A_{3 }(p_1^{s_1},p_2^{s_2}, -P^{ s}) ~.
\ee
Therefore, once we know the three-point amplitude, we  also know the split function. Performing the Mellin transformation  then gives the corresponding celestial OPEs. 
  
 \subsection{Celestial OPE for gluon}
In bosonic open string theory,  the  (color-ordered) amplitude  for three  gluons is given by  \cite{Polchinski:1998rr}
\footnote{ Like footnote~\ref{footnote2}, we set the overall factors in  all the string  amplitudes to be one. In the final results of celestial OPEs, we will   choose proper overall factors to make the formulae as simple as possible. 
 We also use the subscripts $g$ and $G$ for gluon and graviton/dilaton/KR field, respectively.}
%
\be\label{bosonicGluonAmp}
A^o_{ggg}=e_{1\mu}e_{2\alpha}e_{3\rho} T^{\mu\alpha\rho}(4\ap)~,
\ee
 while  in heterotic string, the three gluon amplitude is \cite{Polchinski:1998rq}
   \be
 A^H_ {ggg}=  e_{1 \mu }  e_{2\alpha} e_{3\rho } T^{\mu\alpha\rho}(0) ~,
  \ee
where we introduced the tensor  
  \beqn\label{Ttensor}
 T^{\mu\alpha\rho}(\ap)&=&p_{23}^\mu \; \eta^{\alpha\rho}+p_{31}^\alpha\; \eta^{\rho\mu} +p_{12}^\rho\; \eta^{\mu\alpha}
 +\frac{\alpha'}{8} p_{23}^\mu \; p_{31}^\alpha \; p_{12}^\rho~, \qquad\qquad
 p_{ij}=p_i-p_j~.
\eeqn

So up to $\ap$ corrections, the three gluon amplitude is the same in   bosonic and heterotic string. This is not surprising as the low energy effective field actions of  both theories contain  the Yang-Mills theory, which is responsible for the leading non-$\ap$ amplitude above. The absence of $\ap$ correction in the heterotic string is due to supersymmetry, and it is consistent with the absence of $\ap$ corrections in the celestial OPEs in heterotic string that we derived before from worldsheet. 

Since the heterotic gluon amplitude can be regarded as the special limit of  the bosonic one without $\ap$ correction, we will just focus on the bosonic string gluon amplitude \eqref{bosonicGluonAmp}.   More explicitly, the three gluon amplitude  \eqref{bosonicGluonAmp} can be written as
\beqn 
A^o_{ggg} &=&e_1 \cdot p_{23} \; e_2\cdot e_3
+e_2 \cdot p_{31}\; e_3\cdot e_1
+e_3 \cdot p_{12} \; e_1\cdot e_2
+\frac{\alpha'}{2} e_1 \cdot p_{23} \; e_2\cdot p_{31}\; e_3 \cdot p_{12}~ 
\qquad
\\&=&
 2\Big[ e_1\cdot p_{2} e_2\cdot e_3 - e_2\cdot p_{1} e_1\cdot e_3-e_3\cdot p_{2} e_1\cdot e_2
+2\ap e_1 \cdot p_2\; e_2\cdot p_1\; e_3 \cdot p_2
\Big]~,
\label{ooosimpleGluon}
\eeqn
where  in the second equality  we used
 momentum conservation $p_1+p_2+p_3=0$ and $e_i\cdot p_i=0$ to simplify.

We choose $p_1,p_2$   out-going, namely   $\eta_1=\eta_2=1 $, then $p_3$ is incoming $\eta_3=-1$. We also choose the polarization vectors as $e_i=\varepsilon_{a_i}(p_i)$.  Using \eqref{kenDiffpts}, we have
  \be 
p_i \cdot e_j\equiv  p_i \cdot     \varepsilon_{a_j}(x_j)=\eta_i\omega_i   \hat p_i \cdot     \varepsilon_{a_j}(x_j)
  =\eta_i\, \omega_i\, x_{ij}^{a_j}~, \qquad
 \varepsilon_{a_i} (x_i) \cdot   \varepsilon_{a_j}(x_j)=\delta^{a_ia_j}~,
  \ee
which enables us to rewrite \eqref{ooosimpleGluon} as
\beqn
\frac12 A^o_{ggg} (p_i,\varepsilon_{a_i})&=&  e_1 \cdot p_{2} \; e_2\cdot e_3
 -e_2\cdot p_{1} \; e_3\cdot e_1
 -e_3 \cdot p_{2} \; e_1\cdot  e_2
 +2\ap e_1 \cdot p_2\; e_2\cdot p_1\; e_3 \cdot p_2
 \\&=&
 \omega_2 x_{21}^{a_1}\delta^{a_2a_3}
- \omega_1 x_{12} ^{a_2}\delta^{a_1a_3}
- \omega_2 x_{23}^{a_3}\delta^{a_1a_2}
 +2\ap\omega_1\omega_2^2\; x_{21}^{a_1}   x_{12}^{a_2}   x_{23}^{a_3}  ~.
 \label{Agggsimple}
\eeqn
 
   To have sensible results for collinear factorization, we   take $p_1,p_2$ on-shell but $P\equiv p_1+p_2$ slightly off-shell. Then the momentum conservation $P=p_1+p_2=-p_3+\epsilon v$ is 
 \be\label{momcnonservation}
 \omega_1 \Big( \frac{1+(x_1)^2}{2}, x_1^a,\frac{1-(x_1)^2}{2}  \Big)
 + \omega_2 \Big( \frac{1+(x_2)^2}{2}, x_2^a,\frac{1-(x_2)^2}{2}  \Big)
 = \omega_3 \Big( \frac{1+(x_3)^2}{2}, x_3^a,\frac{1-(x_3)^2}{2}  \Big)+\epsilon v~,
 \ee 
 where $\epsilon $ characterizes the deviation from the strict collinear limit and $v$ is an order one vector. Note $P^2=2p_1\cdot p_2=-\omega_1\omega_2 (x_{12})^2 =-\epsilon p_3\cdot v+\sO(\epsilon)$, hence $\epsilon \sim  (x_{12})^2$. This also introduces $\sO(\epsilon)$ uncertainty in the numerator of \eqref{ampfactorization}, and thus an order one $ (x_{12})^0$ uncertainty in the split function. Nevertheless, it would not affect the singular terms in the split function which we are really interested in. 
 
From \eqref{momcnonservation}, we have
  \be 
\omega_1 x_1^a+\omega_2 x_2^a= \omega_3x_3^a+\sO(\epsilon)~, \qquad
\omega_1  +\omega_2 = \omega_3 +\sO(\epsilon)~,
\ee 
and thus
\be
x_{13}^a=\frac{\omega_2}{\omega_1+\omega_2}x_{12}^a+\sO(\epsilon)~, \qquad
x_{32}^a=\frac{\omega_1}{\omega_1+\omega_2}x_{12}^a+\sO(\epsilon)~.
\ee
Then \eqref{Agggsimple} can be simplified further as
\be
  \widehat A_3 (\omega_i, \varepsilon_{a_i}, \alpha')=
  2\Bigg[
 \omega_2 x_{21}^{a_1}\delta^{a_2a_3}
- \omega_1 x_{12} ^{a_2}\delta^{a_1a_3}
+\frac{\omega_1\omega_2}{\omega_1+\omega_2}x_{12}^{a_3}\delta^{a_1a_2}
 +2\ap \frac{\omega_1^2\omega_2^2}{\omega_1+\omega_2}\; x_{12}^{a_1}   x_{12}^{a_2}   x_{12}^{a_3} 
 \Bigg]~.
\ee
Substituting into  \eqref{ampfactorization}, we get the split function
\beqn
  \Split(p_1^{a_1}+p_2^{a_2}\to P^{a_3})
 &=& \frac{    \widehat A_3 (\omega_i, \varepsilon_{a_i}, \alpha')+\sO(\epsilon)}{2p_1\cdot p_2} 
 \\ &=&2  \frac{
  \omega_2 x_{12}^{a_1}\delta^{a_2a_3}
+ \omega_1 x_{12} ^{a_2}\delta^{a_1a_3}
-\frac{\omega_1\omega_2}{\omega_1+\omega_2}x_{12}^{a_3}\delta^{a_1a_2}
-2\ap \frac{\omega_1^2\omega_2^2}{\omega_1+\omega_2}\; x_{12}^{a_1}   x_{12}^{a_2}   x_{12}^{a_3} 
 }{\omega_1\omega_2\; (x_{12})^2   }
 +\sO(\epsilon^0)  ~.\qquad\qquad
\eeqn
Up to an overall factor, this is     the same as the $I$ introduced in \eqref{Ipqeps}. 
 
Further performing the Mellin transformation leads to the following celestial OPE for gluons: 
\beqn\label{OABgluon}
&&\cO^A_{\Delta_1,a_1}(x_1)\cO^B_{\Delta_2,a_2}(x_2)
\nonumber\\&\sim&  
   f^{ABC} 
   \frac{x_{12}^{a_1}\delta^{a_2a_3} B(\Delta_1-1,\Delta_2) 
  +x_{12}^{a_2}\delta^{a_1a_3}B(\Delta_1 ,\Delta_2-1)
 -x_{12}^{a_3}\delta^{a_1a_2} B(\Delta_1 ,\Delta_2)
     }{(x_{12})^2}    \cO^C_{\Delta_1+\Delta_2-1,\;a_3}(x_2) 
  \nonumber \\&&
-  2\ap f^{ABC}   \frac{   x_{12}^{a_1} x_{12}^{a_2}x_{12}^{a_3} B(\Delta_1+1 ,\Delta_2+1)
   }{(x_{12})^2}    
   \cO^C_{\Delta_1+\Delta_2+1,\;a_3}(x_2)   ~,
\eeqn
where we restore  the color factor $f^{ABC}$ and  choose a proper overall normalziation to make the formula as simple as possible.  Here   $\cO^A_{\Delta, a}$ denotes the celestial gluon operator with the polarization vector $\varepsilon_a$ and  color index $A$.  This agrees with \eqref{gluonOPEopen} derived from worldsheet.

 \subsection{Celestial OPE for graviton/dilaton/KR field}
 
In bosonic closed string theory,  the   amplitude  for the closed string massless fields  is given by   \cite{Polchinski:1998rr}
\be\label{cloedgraviton}
A^c_{GGG}= e_{1 \mu\nu}  e_{2\alpha\beta} e_{3\rho\sigma} T^{\mu\alpha\rho}(\ap)T^{\nu\beta\sigma}( \ap)~.
\ee
 So up to the rescaling of $\ap$, the  closed string amplitude is essentially the ``square'' of the open string amplitude \eqref{bosonicGluonAmp}. This is   the simplest example of the famous KLT relation \cite{Kawai:1985xq}. 
 
 In  heterotic string,  the amplitude for massless  fields is \cite{Polchinski:1998rq}
 \be
 A^H_ {GGG}=
  e_{1 \mu\nu}  e_{2\alpha\beta} e_{3\rho\sigma} T^{\mu\alpha\rho}(\ap)T^{\nu\beta\sigma}(0)~,
  \ee
 where the $\ap$ corrections in the right moving sector is absent due to supersymmetry. 
 For type I and IIA/IIB string, they have supersymmetry in both left and right moving sectors, and the  corresponding  massless NS-NS amplitude  has no $\ap$ corrections at all: 
  \be
 A ^{I/II}_ {GGG}=
  e_{1 \mu\nu}  e_{2\alpha\beta} e_{3\rho\sigma} T^{\mu\alpha\rho}(0)T^{\nu\beta\sigma}(0)~.
  \ee
  The absence of $\ap$ in three-point amplitude agrees with the absence of $\ap$ correction in celestial OPEs we derived before from worldsheet. 
  
  Let us just focus on the  amplitude  of  massless fields in    closed bosonic  string \eqref{cloedgraviton}.
 As before, we choose  $p_1,p_2$   out-going and $p_3$ in-coming, namely   $\eta_1=\eta_2=-\eta_3=1 $.  We also choose the polarization tensors  $e_i^{\mu\nu}=\varepsilon^\mu_{a_i}\varepsilon^\nu_{\ta_i}$, which are the basis  for polarizations. 
Then we can simplify the three-point amplitude and write it in terms of celestial variables. The steps are almost identical to the open string case in the previous subsection, except for the doubling. The final result is 
\be
A^c_{GGG}=\widehat A_3(\omega_i, \varepsilon_{a_i},\frac14 \alpha')
\widehat A_3(\omega_i, \varepsilon_{\ta_i},\frac14 \alpha')~.
\ee

Substituting into  \eqref{ampfactorization}, we get the split function
\beqn
  \Split(p_1^{a_1\ta_1}+p_2^{a_2\ta_2}\to P^{a_3\ta_3})
=\frac{A^c_{GGG}}{2p_1\cdot p_2}
= -\frac{\widehat A_3(\omega_i, \varepsilon_{a_i},\frac14 \alpha')
\widehat A_3(\omega_i, \varepsilon_{\ta_i},\frac14 \alpha')}{ (x_{12})^2 \omega_1\omega_2}~.
\eeqn
 Up to an overall factor, the split function  here is   the same as that in \eqref{curalIclosed}. We then need to perform Mellin transformation to obtain the celestial OPEs. The steps are identical, and the results are also just given by \eqref{closeOPE}-\eqref{GGOPE3rd}  except for the replacement $a\to a_1, b\to a_2, c\to a_3, \cV \to \cO$.

 \subsection{Celestial OPE for  gluon and graviton/dilaton/KR field}
 In heterotic string, the three-point amplitude involving two gluons and one graviton/dilaton/KR field is   \cite{Polchinski:1998rq}
  \be\label{heteroticGluonGraviton}
 A^H_ {ggG}=e_{1\mu\nu} e_{2\rho} e_{3\sigma}p_{23}^\mu 
 T^{\nu\rho \sigma}(  0)\delta^{AB}~.
 \ee
 Note that there is no $\ap$ correction  due to supersymmetry. In the case of graviton, the   interaction responsible for this amplitude is just the minimal coupling between gravitons and gluons. For bosonic string, one can also compute the two gluons and one graviton amplitude from the open-closed string setup. This  will be discussed in appendix~\ref{appOpenClosed} and the amplitude is given by \eqref{ampOOCbos}, which suffers $\ap$ corrections as we expected. Here we will just focus on the heterotic case \eqref{heteroticGluonGraviton}.

Following the same procedure as before, one can derive the celestial OPEs from this three-point amplitude.  

We first derive the celestial OPE between gluon and graviton/dilaton/KR field. So we take  $p_1,p_2$ out-going and also choose the polarizations as $e_1^{\mu\nu}=  \varepsilon_ {a_1}^\mu \varepsilon_{\ta_1}^\nu $, $e_2=\varepsilon_{\ta_2}(p_2 ), e_3=\varepsilon_{\ta_3}(p_3 )$. Then \eqref{heteroticGluonGraviton} becomes
\be
 A^H_ {ggG}=2 \varepsilon_{a_1}\cdot p_2\; \widehat A_3(\omega_i, \varepsilon_{\ta_i},0)\delta^{AB}~.
\ee
Substituting into  \eqref{ampfactorization}, and performing the
Mellin transformation, we get the   celestial OPE between gluon and graviton/dilaton/KR field 
 \beqn\label{gluonGravitonOPE}
&&\cO _{\Delta_1,a\ta}(x_1)\cO^A_{\Delta_2,\tb}(x_2)
\\&\sim&\nonumber 
 x_{12}^a \frac{x_{12}^{\ta}\delta^{\tb\tc} B(\Delta_1-1 ,\Delta_2+1) 
  +x_{12}^{\tb}\delta^{\ta\tc}B(\Delta_1  ,\Delta_2 )
 -x_{12}^{\tc}\delta^{\ta\tb} B(\Delta_1  ,\Delta_2+1)
   }{(x_{12})^2}  
   \cO^A _{\Delta_1+\Delta_2 ,\tc}(x_2)~, \qquad
   \eeqn  
   which agrees with  \eqref{gluongravtionOPEheterotic}.

The celestial OPE corresponding to the fusion of two gluons   into one   graviton/dilaton/KR field can be similarly derived. To make the formula  more standard, we exchange the label of 1 and 3 in  \eqref{heteroticGluonGraviton}. Repeating the same steps above leads us  to the following celestial OPE between two gluons
\beqn\label{gluonTogravitonOPE}
&&\cO^A_{\Delta_1,\ta}(x_1)\cO^B_{\Delta_2,\tb}(x_2)
\\&\sim&\nonumber 
 \delta^{AB} x_{12}^c \frac{x_{12}^{\ta}\delta^{\tb\tc} B(\Delta_1 ,\Delta_2+1) 
  +x_{12}^{\tb}\delta^{\ta\tc}B(\Delta_1+1 ,\Delta_2 )
 -x_{12}^{\tc}\delta^{\ta\tb} B(\Delta_1+1 ,\Delta_2+1)
   }{(x_{12})^2}  
   \cO _{\Delta_1+\Delta_2 ,\;c\tc}(x_2)~,\qquad
   \eeqn
which agrees with the second line of \eqref{twogluonOPEheterotic}.   

 \subsection{Celestial OPE  in four dimensions}
 So far, we have derived the celestial OPEs in two different ways. The final formula is a bit complicated.
 Now we specialize to 4D as a consistency check of our results. 
 In 4D, it is  very convenient to use   the helicity basis for gluon and graviton. They are related to the previous polarization basis through some linear   combinations. 
 
Following \eqref{gluonPol4D} and \eqref{gravitonPol4D}, we define the celestial gluon and  graviton operators in the helicity basis   as  
\beqn\label{D4helicitybasis}
\cO_{\Delta,\pm}(\sfz,\sfzb)&=&\frac{1}{\sqrt2}\Big(\cO_{\Delta,1}(x ) \mp i \cO_{\Delta,2}(x) \Big)~,
\\
\cO_{\Delta,\pm 2}(\sfz,\sfzb)&=&
\frac12 \Big(\cO_{\Delta,11}(x)- \cO_{\Delta,22}(x) \mp i \cO_{\Delta,12} (x)\mp i \cO_{\Delta,21}(x)\Big)~,
\eeqn
where the coordinates $\sfz,\sfzb$ and $x$ are related through \eqref{xzin4D}. 

 \subsubsection*{Gluon OPE}
 The general form of celestial gluon OPEs is given in \eqref{OABgluon}. Specializing to 4D and transforming to the helicity basis \eqref{D4helicitybasis}, we find that the first line of  \eqref{OABgluon} without $\ap$   reduces
\beqn
\cO^A_{\Delta_1,+}(\sfz_1,\sfzb_1)\cO^B_{\Delta_2,+}(\sfz_2,\sfzb_2)
&\sim&
\frac{ f^{ABC}}{\sfz_{12}}B(\Delta_1-1,\Delta_2 -1) \cO^c_{\Delta_1+\Delta_2-1,+}(\sfz_2,\sfzb_2)~,
\\
\cO^A_{\Delta_1,-}(\sfz_1,\sfzb_1)\cO^B_{\Delta_2,-}(\sfz_2,\sfzb_2)
&\sim&
\frac{ f^{ABC}}{\sfzb_{12}}B(\Delta_1-1,\Delta_2 -1) \cO^c_{\Delta_1+\Delta_2-1,-}(\sfz_2,\sfzb_2)~,
\eeqn
and
\beqn
 O^A_{\Delta_1,+}(\sfz_1,\sfzb_1)O^B_{\Delta_2,-}(\sfz_2,\sfzb_2)
 &\sim&\nonumber
\frac{ f^{ABC}}{\sfzb_{12}}B(\Delta_1+1,\Delta_2 -1) O^c_{\Delta_1+\Delta_2-1,+}(\sfz_2,\sfzb_2)
\\&&+
\frac{ f^{ABC}}{  \sfz_{12}}B(\Delta_1-1,\Delta_2 +1) O^c_{\Delta_1+\Delta_2-1,-}(\sfz_2,\sfzb_2)~.
\eeqn
 These are indeed the celestial OPEs for gluon in Yang-Mills theory \cite{Fan:2019emx,Pate:2019lpp}. 

  The second line  of \eqref{OABgluon} with $\ap$ coefficient arises from the higher derivative interactions. We can similarly transform it into  the helicity basis. In particular, we find that only the following two OPEs have  singular  terms 
\beqn
&&\cO^A_{\Delta_1,+}(\sfz_1,\sfzb_1)\cO^B_{\Delta_2,+}(\sfz_2,\sfzb_2)\sim
- \ap \; f^{ABC}
\frac{ \sfzb_{12}^2 }{\sfz_{12}}B(\Delta_1+1,\Delta_2 +1) \cO^C_{\Delta_1+\Delta_2+1,-}(\sfz_2,\sfzb_2)~,
\\
&&
\cO^A_{\Delta_1,-}(\sfz_1,\sfzb_1)\cO^B_{\Delta_2,-}(\sfz_2,\sfzb_2)\sim - \ap \; f^{ABC}
\frac{  \sfz_{12}^2 }{\sfzb_{12}}B(\Delta_1+1,\Delta_2 +1) \cO^C_{\Delta_1+\Delta_2+1,+}(\sfz_2,\sfzb_2)~,
\eeqn
while the rest  of OPEs are all regular. The two OPEs above with singular terms have exactly the same structure as that  predicted by the general formula of OPE in \cite{Jiang:2021ovh,Himwich:2021dau}. The rest of helicity configurations only give rise to regular OPEs, as their corresponding amplitudes vanish on-shell. Indeed in 4D, the three-point on-shell amplitudes are fully determined by the helicities due to little group  scaling and locality.  Each singular OPE above is in one-to-one correspondence with the on-shell three-point amplitudes of gluons   arising from either YM theory or higher derivative interactions.

%
%
%
 
\subsubsection*{Graviton  OPE}

Similarly, one can transform the celestial graviton OPE in \eqref{closeOPEK} into the helicity basis in 4D. The final result, up to an overall constant,   is given by
\beqn\label{gravitonOPEsamehelicity}
 \cO _{\Delta_1, +2}(\sfz_1, \bar \sfz_1)\cO _{\Delta_2, +2}(\sfz_2, \bar \sfz_2) &\sim&
 - \frac{\bar \sfz_{12}}{\sfz_{12}}  B(\Delta_1-1, \Delta_2-1) \cO_{\Delta_1+\Delta_2, +2}(\sfz_2,\bar \sfz_2)
  \\&&
-\frac{\ap^2}{16}\; \frac{\bar \sfz_{12}^5}{\sfz_{12}}B(\Delta_1+3, \Delta_2+3) \cO_{\Delta_1+\Delta_2+4, -2}(\sfz_2,\bar \sfz_2)~,
\\
 \cO _{\Delta_1, -2}(\sfz_1, \bar \sfz_1)\cO _{\Delta_2, -2}(\sfz_2, \bar \sfz_2) &\sim&
 - \frac{  \sfz_{12}}{\bar \sfz_{12}}  B(\Delta_1-1, \Delta_2-1) \cO_{\Delta_1+\Delta_2, -2}(\sfz_2,\bar \sfz_2)
    \\&&
-\frac{\ap^2}{16}\; \frac{ \sfz_{12}^5}{\bar \sfz_{12}}B(\Delta_1+3, \Delta_2+3) \cO_{\Delta_1+\Delta_2+4, +2}(\sfz_2,\bar \sfz_2)~,
\\ 
\cO _{\Delta_1, +2}(\sfz_1, \bar \sfz_1)\cO _{\Delta_2, -2}(\sfz_2, \bar \sfz_2) &\sim&
 - \frac{\bar \sfz_{12}}{\sfz_{12}}  B(\Delta_1-1, \Delta_2+3) \cO_{\Delta_1+\Delta_2, -2}(\sfz_2,\bar \sfz_2)
\\&&
- \frac{  \sfz_{12}}{\bar \sfz_{12}}  B(\Delta_1+3, \Delta_2-1) \cO_{\Delta_1+\Delta_2, + 2}(\sfz_2,\bar \sfz_2)~,
\eeqn
where we have only shown the singular terms. 
The structures of OPEs are again in perfect  agreement with    \cite{Pate:2019lpp,Jiang:2021ovh,Himwich:2021dau}.  Here those terms in OPE without $\ap$ correspond to the Einstein gravity, while the terms   with $\ap^2$ coefficient come  from higher derivative interactions of graviton. Note the $\ap $ pieces  in \eqref{closeOPEK}  are absent in the above formulae because they are regular. Each singular term in the OPEs above is again in one-to-one correspondence with the on-shell three-point  amplitudes of gravitons.

 
 \subsubsection*{Mixed OPE  for gluon  and graviton}
Now we consider the mixed OPEs involving both gluons and gravitons. The general formula is given by \eqref{gluonGravitonOPE} and \eqref{gluonTogravitonOPE}. 
Specializing to 4D and writing in the helicity basis, they become
   \beqn
&&\cO _{\Delta_1,\pm 2}(\sfz_1,\sfzb_1)\cO^A_{\Delta_2,\pm }(\sfz_2,\sfzb_2)\sim   
-\Big( \frac{ \sfzb_{12}  }{\sfz_{12}}\Big) ^{\pm 1}
B(\Delta_1-1,\Delta_2  )
 \cO^A_{\Delta_1+\Delta_2 ,\pm}(\sfz_2,\sfzb_2)~,
\\ 
&&\cO _{\Delta_1,\pm 2}(\sfz_1,\sfzb_1)\cO^A_{\Delta_2,\mp }(\sfz_2,\sfzb_2)\sim   
-\Big( \frac{ \sfzb_{12}  }{\sfz_{12}}\Big) ^{\pm 1}
B(\Delta_1-1,\Delta_2+2  )
 \cO^A_{\Delta_1+\Delta_2 ,\mp }(\sfz_2,\sfzb_2)~,
\eeqn
%
   and
\beqn
 \cO^A_{\Delta_1,+}(\sfz_1,\sfzb_1)\cO^B_{\Delta_2,-}(\sfz_2,\sfzb_2)\sim  &&
\delta^{AB }
 \frac{\sfzb_{12}}{\sfz_{12}} B(\Delta_1  ,\Delta_2+2)\cO _{\Delta_1+\Delta_2 ,-2}
 \nonumber\\&&+
 \delta^{AB }
 \frac{\sfz _{12}}{\sfzb_{12}} B(\Delta_1+2  ,\Delta_2 )\cO _{\Delta_1+\Delta_2 ,+2}~,
 \qquad\qquad
\eeqn
which are again consistent with \cite{Pate:2019lpp}  and \cite{Jiang:2021ovh,Himwich:2021dau}.

\section{Celestial OPE in $\cN=2$ string and $w_{1+\infty}$ algebra} \label{N2stringOPE}

In this section, we will further generalize previous OPE discussions to $\cN=2$ string theory \cite{Ooguri:1991fp,Ooguri:1990ww,Marcus:1992wi}. The $\cN=2$ string theory has four dimensional target spacetime with $(2,2) $ signature. So we will first discuss the kinematics and celestial variables in $ (2,2) $ signature. Then  as before, we can compute the worldsheet OPE and celestial OPE.
An interesting feature is that we can even compute all the $\overline{SL(2,\mathbb R)}$
\footnote{ This is just half of the Lorentz group. More precisely  $ {SL(2,\mathbb R)}\times \overline{SL(2,\mathbb R)}$  is the double cover of   $SO(2,2)$ which is the Lorentz group in $(2,2)$ signature. }   descendants in the OPE, and this essentially comes from the momentum conservation in $(2,2)$ signature. The soft sector of such OPE with descendants  just gives the $w_{1+\infty}$ algebra, after   rewriting in terms of chiral modes.  Therefore, we give  an indirect derivation of $w_{1+\infty}$ from worldsheet in  $\cN=2$ string theory.


\subsection{Kinematics in $(2,2) $ signature }

Now we consider the spacetime in $ (2,2)$ signature.  In this case, it is very  convenient to introduce complex coordinates $\cX^1=(X_1+i X_2)/\sqrt{2}, \cX^2=(X_3+i X_4)/\sqrt{2} $ as well as their complex conjugates. The metric is then
\be
ds^2=(dX^1)^2+(dX^2)^2-(dX^3)^2-(dX^4)^2
=\eta_{\mu\nu}  d\cX^\mu d\cX^\nu
=2d\cX^1 d\bar \cX^1-2d\cX^2 d\bar \cX^2~.
\ee

We will write the vector as 
\be\label{Avec}
\bm A=(A^1, A^2, A^{\bar 1}, A^{\bar 2})~.
\ee
We will also use the  Greek indices   $\mu,\nu, \cdots=1,\bar 1,2,\bar 2 $,   unbarred  indices $   i,j,\cdots=1,2$, and  barred indices $\bi,\bj,\cdots=\bar 1,\bar 2$. 
Note we will identify  $\bar A^i \equiv \bar A^\bi \equiv A^\bi $. For real vector,  $A^\bi$ is just the complex conjugate of $A^i$. 

The metric  $\eta_{\mu\nu} $ is given by
\be\label{etamunumetric}
\eta_{i\bj} =\eta_{ \bj i}=\eta^{i\bj}=\eta^{\bj i}=(-1)^{i+1} \delta^{ij}~, \qquad \eta_{i j}=\eta_{\bi\bj}=\eta^{i j}=\eta^{\bi\bj}=0~.
\ee

We introduce the following notation
 \be
A\cdot \bar B=\bar B\cdot   A= A^1 \bar B^1-A^2 \bar B^2~,
\ee 
 where $A=(A^1,A^2), \bar B=(B^{\bar 1},B^{\bar 2})$.
 
Given two vectors $\bm A, \bm B$, we can define their inner product as
\be
\bm B\cdot\bm A=\bm A\cdot\bm B=\eta_{\mu\nu} A^\mu B^\nu=A\cdot \bar B +B\cdot \bar A
=A^1 \bar B^1-A^2 \bar B^2
+B^1 \bar A^1-B^2 \bar A^2~.
\ee
This is real  for  real vectors $A,B$.

Given the vector in \eqref{Avec}, we also define its dual $\bm A^\vee$ as
\be\label{AandAdual}
\bm A=(A^1, A^2, A^{\bar 1}, A^{\bar 2})~, \qquad
\bm A^\vee=(A^1, A^2, - A^{\bar 1}, -A^{\bar 2})~.
\ee
Note that $\bm A^\vee$ is  an imaginary vector as $(A^\vee)^\bi=-((A^\vee)^i)^*$, if $\bm A$ is a real vector. And   we have the inner product 
\be
\bm A^\vee\cdot\bm B
= A\cdot \bar B -B\cdot \bar A
=A^1 \bar B^1-A^2 \bar B^2
-B^1 \bar A^1+B^2 \bar A^2~,
\ee
which is purely imaginary for   real vectors $A,B$. They satisfy $
\bm A \cdot\bm B=-\bm A^\vee\cdot\bm B^\vee, \bm A^\vee\cdot\bm B=-\bm B^\vee\cdot\bm A$.

We are particularly interested in  the null momentum satisfying   $\bm k \cdot \bm k =0$. It can then be parametrized as  
\be\label{nullmom22}
\bm k =\omega_k \; \hat{\bm k} ~, \qquad \omega_k\in \mathbb R~.
\ee
Here the null vector   is 
\be\label{nullVec22}
\hat{\bm k}(\sfz,\sfzb)=\Big(1+\sfz\sfzb+i(\sfz-\sfzb),\; 1-\sfz\sfzb+i(\sfz+\sfzb),\; 
1+\sfz\sfzb-i(\sfz-\sfzb),\; 
1-\sfz\sfzb-i(\sfz+\sfzb)   \Big) ~,
\ee
where $\sfz, \sfzb$ are the coordinates on the celestial torus \cite{Atanasov:2021oyu},  instead of the celestial sphere.
It is worth emphasizing that   the two variables $\sfz, \sfzb$ are    real and  {independent}, instead of  the complex conjugate of each other. 

Then the polarizations are given by 
\beqn\label{varepsilonPls}
\bm\varepsilon_+&=&-\frac{i}{2}\p_\sfz \hat{\bm k}=\frac12(1-i  \sfzb,   1+ i\sfzb,-1-i\sfzb, -1+i\sfzb)~, \qquad\\
 {\bm\varepsilon}_-&=&-\frac{i}{2}  \p_{\sfzb} \hat{\bm k}=\frac12(-1 -i  \sfz ,  1+i\sfz , 1-i\sfz , - 1+i\sfz )~. 
\eeqn
They satisfy 
\be\label{polmomId}
\bm\varepsilon_+\cdot \bm\varepsilon_+
= {\bm\varepsilon}_- \cdot {\bm\varepsilon}_-
=\hat{\bm k}\cdot \hat{\bm k}
=\bm\varepsilon_+ \cdot \hat{\bm k}
= {\bm\varepsilon}_- \cdot \hat{\bm k}
=0~, \qquad 
 {\bm\varepsilon}_+\cdot \bm\varepsilon_-
=1~.
\ee

For  different momenta  with  $\hat{\bm k}_i=\hat{\bm k}(\sfz_i,\sfzb_i)$, we have the identities
\be\label{kpolrelation}
\hat{\bm k}_i\cdot \hat{\bm k}_j =4 \,\sfz_{ij} \,\sfzb_{ij}~, \qquad
\hat{\bm k}_i\cdot \bm\varepsilon_{+j}= 2i\,\sfzb_{ij}~, \qquad
\hat{\bm k}_i\cdot  {\bm\varepsilon }_{-j}=2i\, \sfz_{ij}~,
\ee
and
\be\label{kpolrelation2}  
\bm\varepsilon_{\pm i} \cdot \bm\varepsilon_{\pm j}=0~, \qquad
\bm\varepsilon_{\pm i} \cdot \bm\varepsilon_{\mp j}=1~. \qquad
\ee

For null momentum in \eqref{nullmom22} \eqref{nullVec22}, its dual satisfies 
\be\label{kveePolrelation} 
  \bm k^\vee=\omega_k \hat{\bm k}^\vee
 =\omega_k \Big( 2(1+\sfz^2) \bm\varepsilon_+ +i \, \sfz\,\hat {\bm k }\Big)~,
\ee
so up to a gauge transformation and an overall factor, $\bm k^\vee$ is essentially the positive helicity polarization vector.

We also have 
\footnote{ It is worth mentioning that although $\hat{\bm k}^\vee $ is  essentially the polarization vector, their inner products are non-vanishing, in contrast to the vanishing inner product of two    polarizations with the same helicity in \eqref{kpolrelation2}. The non-vanishing $\hat{ \bm k}_i^\vee\cdot \hat{ \bm k}_j^\vee$ is just due to the gauge transformation \eqref{kveePolrelation} and \eqref{kpolrelation}.
}
\be\label{kijN=2id}
\hat{ \bm k}_i\cdot \hat{\bm k}_j= -\hat{ \bm k}_i^\vee\cdot \hat{ \bm k}_j^\vee
=4 \sfz_{ij} \sfzb_{ij}~, 
 \qquad
\hat{ \bm k}_i^\vee\cdot \hat{ \bm k}_j=  -\hat{ \bm k}_j^\vee\cdot \hat{ \bm k}_i=-4i (1+\sfz_{i}\, \sfz_j) \sfzb_{ij}~.
\ee

\subsection{Vertex operator  in $\cN=2$ string theory}

The $\cN=2$   string  theory is constructed from the $\cN=2$  non-linear $\sigma$-model \cite{Ooguri:1991fp}
\be
S=\frac{1}{\pi}\int d^2 z\, d^2\theta\, d^2\bar\theta\; K(\mathscr X, \bar {\mathscr X})~,
\ee
where $\mathscr X^i$ are   $\cN=2$ chiral superfields  
\be
\mathscr X^i(Z,\bar Z; \theta^-, \bar\theta^-) 
=\cX^i(Z,\bar Z)+\psi_L^i(Z,\bar Z) \theta^-
+\psi_R^i(Z,\bar Z) \bar\theta^- 
+F^i(Z,\bar Z)\bar\theta^- \theta^-, \quad Z=z-\theta^+\theta^-~,
\ee
and $K(\mathscr X,\bar{\mathscr X})=\mathscr X\cdot \bar{\mathscr X}=\mathscr X^1\bar{\mathscr X}^{\bar 1}-\mathscr X^2 \bar{\mathscr X}^{\bar 2}$ is the K{\"a}hler potential for flat metric. More explicitly in terms of component fields, the action reads
\be
S=\frac{1}{\pi}\int d^2z\; \Big( \p \cX\cdot \bar\p \bar\cX+\bar\p \cX \cdot \p \bar \cX
+\bar\psi_L\cdot\bar\p \psi_L+\bar\psi_R\cdot \p \psi_R+\bar F\cdot F\Big)~.
\ee
Thus we have $F=0$ and the following OPEs \footnote{ Note we set $\ap=1$ here. }
\be
\cX^i(z_1, \bz_1) \bar \cX^\bj (z_2, \bz_2)\sim  -\eta^{i\bj} \ln |z_{12}|~, \qquad  
\psi_L^i(z_1) \bar \psi_L^\bj(z_2) \sim    \frac{\eta^{i\bj} }{z_{12}}~, \qquad
\psi_R^i(\bz_1) \bar \psi_R^\bj(\bz_2) \sim  \frac{\eta^{i\bj} }{\bz_{12}}~.
\ee   

The critical dimension of $\cN = 2$ string is four \cite{Fradkin:1981dd,Ooguri:1991fp}.
 In order to have $\cN = 2$ supersymmetry on the world-sheet,    
 the target spacetime should be endowed with a  {complex structure}, implying that the  signature of spacetime can be either (4,0)   or (2,2).
 In the former  (4,0)  case, $\cN = 2$ string only has    ground
state as the physical state in the first-quantized string, and is thus not interesting.
 For the latter (2,2) case and in the simplest version of $\cN=2$ string,   there is a massless  
field in the spectrum,    and it obeys a non-linear differential
equation. The actual Lorentz group of $\cN=2$ string is $U(1,1)\simeq U(1)\times SU(1,1)$, instead of $SO(2,2)$ \cite{Ooguri:1991fp}.  
Note that   $ SU(1,1)$  is also isomorphic to $SL(2,\mathbb R)$.

The vertex operator for the massless   field in $\cN=2$ string theory  is given by \cite{Ooguri:1991fp} 
\beqn 
\widehat V(k) &=&
\int d^2\theta\, d^2\bar\theta\;e^{i(k \cdot \bar{ \mathscr  X}+ \bar k \cdot \mathscr X)}
\\&=&
\Big( ik \cdot \p \bar \cX -i \bar k \cdot \p   \cX  - k \cdot  {\bar\psi_L} \; \bar k \cdot \psi_L \Big) 
\Big( ik   \cdot \bar\p \cX -i \bar k  \cdot\bar \p   \cX  - k \cdot  {\bar\psi_R} \; \bar k \cdot \psi_R \Big) e^{i(k \cdot \bar \cX+ \bar k \cdot \cX)}
~.\qquad
\label{Vkn2}
\eeqn
This looks similar to the   vertex operator  \eqref{IIAIIBOp2} in  type II string,     up to the choice of polarization. Indeed we can write \eqref{Vkn2} as
\beqn\label{N2Vophat}
\widehat V(\bm k) &=&
\Big( i \bm k^\vee \cdot \p  {\bm \cX}    
-\frac12  ( \bm k^\vee \cdot \bm \psi_L )  (\bm k \cdot \bm \psi_L ) \Big)
\Big(   i \bm k^\vee \cdot \bar\p  {\bm \cX}   
-\frac12 ( \bm k^\vee \cdot \bm \psi_R ) (\bm k \cdot \bm \psi_R )
\Big) e^{i \bm k \cdot   {\bm \cX} }~,
\eeqn
where $\bm k^\vee$ is  the dual of $\bm k$ defined in \eqref{AandAdual}. As shown in \eqref{kveePolrelation},   the vector $\bm k^\vee$  is essentially the positive helicity polarization vector,  up to   a gauge transformation and   an overall rescaling.

 Using this vertex operator,   the three-point  amplitude was computed, while higher point amplitudes vanish \cite{Ooguri:1991fp}.  See the review   \cite{Marcus:1992wi}  for other aspects of $\cN=2$ string theory.

\subsection{OPE in $\cN=2$ string theory }
We would like to compute the OPE of vertex operators \eqref{Vkn2}, or equivalently  \eqref{N2Vophat}. The steps are similar to the previous computations, and the final result is given in \eqref{N2OPE}:
 \beqn
\widehat V(\bm k_1) \widehat V(\bm k_2) &\sim& 
\frac14  \Big(\bm k_1^\vee \cdot \bm k_2\Big)^2
 |z_{12}|^{    \bm k_1\cdot \bm k_2 -2}   
 e^{  {i \bm K_3 \cdot   {\bm \cX}   }   }
\\&&\nonumber
\times \Bigg[   
 i\bm K_3^\vee  \cdot \p \bm\cX 
-  \frac{1}{2    }   
(\bm K_3^\vee \cdot \bm \psi_L \; \bm K_3 \cdot \bm \psi_L ) 
 +\sO(\bm k_1\cdot \bm k_2) +\sO(\bm k_1^\vee\cdot \bm k_2^\vee)
 +\sO( z_{12})
   \Bigg] 
 \\  &&\nonumber
 \times
    \Bigg[   
 i\bm K_3^\vee  \cdot \bar\p \bm\cX 
-  \frac{1}{2    }   
(\bm K_3^\vee \cdot \bm \psi_R \; \bm K_3 \cdot \bm \psi_R ) 
 +\sO(\bm k_1\cdot \bm k_2) +\sO(\bm k_1^\vee\cdot \bm k_2^\vee)
 +\sO( \bz_{12})
   \Bigg] ~,
 \eeqn
  where $\bm K_3=\bm k_1+\bm k_2$. It is easy to recognize that we just get $\widehat V(\bm K_3)$ on the right-hand side. Taking the collinear limit $\bm k_1\cdot \bm k_2\to 0$ and using the identity \eqref{deltazclose}  \eqref{kijN=2id},   we arrive at
 \beqn
\widehat V(\bm k_1)\widehat V(\bm k_2) 
 &\sim&
\frac{\pi}{2} \delta^2(z_{12})   e^{  {i \bm k_3 \cdot   {\bm \cX}_L  }   }
 \frac{ \Big(\bm k_1^\vee \cdot \bm k_2\Big)^2}{\bm k_1\cdot \bm k_2}
 \Big[ \widehat V(\bm K_3)+  \sO(\bm k_1\cdot \bm k_2)
 +\sO( z_{12},\bz_{12})\Big]
\\
 &\sim&
-2\pi \delta^2(z_{12})   \omega_1\omega_2
 \frac{ (1+\sfz_1\sfz_2)^2(\sfzb_{12})^2}{\sfz_{12}\sfzb_{12}}
 \Big[\widehat V(\bm K_3)+  \sO(z_{12},\bz_{12})
 +\sO( (\sfz_{12}\sfzb_{12})^1)\Big]~,\qquad
 \eeqn
 where   $  \sO( \bm k_1\cdot \bm k_2)=\sO( (\sfz_{12}\sfzb_{12})^1)$ with $\sfz_{12}$ 
and $\sfzb_{12}$ coming together in a product form.
 
 Performing the integration over $z_1,z_2$ on the worldsheet, we obtain
 \be\label{Vhatomg12}
\widehat \cV(\omega_1, \sfz_1, \sfzb_1) \widehat\cV(\omega_2, \sfz_2, \sfzb_2)
 \sim
 -2\pi\omega_1\omega_2 (1+\sfz_1\sfz_2)^2  \;
  \frac{   \sfzb_{12} }{\sfz_{12} }
\Big[  \widehat\cV(\omega_3, \sfz_3, \sfzb_3)  +\sO( (\sfz_{12}\sfzb_{12})^1)\Big]~,
 \ee
 where we used $\bm k_3=\omega_3 \hat {\bm k}(\sfz_3,\sfzb_3) $ is approximately the same as $ \bm K_3=\bm k_1+\bm k_2$ in the collinear limit, up to order $\sO( (\sfz_{12}\sfzb_{12})^1)$ corrections. 
 The coefficient here is a little awkward. This is because we were using the vertex operators  where the polarization   has an extra factor \eqref{kveePolrelation}: $\bm k^\vee\simeq 2\omega_k (1+\sfz^2)\bm\varepsilon_+$, up to gauge transformation.    To make the equation nicer and make contact with the standard convention, we redefine the vertex operators as follows 
 \be
 \widehat \cV(\omega , \sfz , \sfzb )=2\pi\,\omega^2 (1+\sfz^2)^2\;
  \cV(\omega , \sfz , \sfzb )~.
 \ee
 %
%
Then \eqref{Vhatomg12} takes a nicer form 
  \be\label{VVOPE}
  \cV(\omega_1, \sfz_1, \sfzb_1)  \cV(\omega_2, \sfz_2, \sfzb_2)
 \sim
- \frac{(\omega_1+\omega_2)^2} { \omega_1\omega_2}  \frac{   \sfzb_{12} }{\sfz_{12} }
\Big[   \cV(\omega_3, \sfz_3, \sfzb_3) 
    +\sO( (\sfz_{12}\sfzb_{12})^1)+\sO(\sfz_{12})
    \Big]~.
 \ee
 In the collinear limit, we can just set $\sfz_3, \sfzb_3\to \sfz_2, \sfzb_2,\omega_3\to \omega_1+\omega_2$ as they are close on the celestial sphere and the energy is additive. Performing the Mellin transformation, we get   the celestial OPE: 
   \be
   \cV_{\Delta_1}( \sfz_1, \sfzb_1)  \cV_{\Delta_1} (  \sfz_2, \sfzb_2)
 \sim
  - \frac{  \sfzb_{12} }  {\sfz_{12} } B(\Delta_1-1, \Delta_2-1)
\;  \cV_{\Delta_1+\Delta_2}( \sfz_2, \sfzb_2) ~. 
  \ee
  This just unsurprisingly reproduces the  celestial    OPE of two gravitons both  with positive helicity \eqref{gravitonOPEsamehelicity}. 
  
  Actually, in the current situation, we can go further. In the previous sections, we were considering the spacetime in  Minkowski signature and correspondingly the celestial sphere is in Euclidean  signature.  This is different from the current   $(2,2)$  split signature in a subtle but interesting way. 
 In 4D Minkowski spacetime with $(1,3)$ signature, three particles satisfying momentum conservation can not become  on-shell simultaneously, unless all momenta are strictly parallel and point along the same direction. This constraint is relaxed in (2,2) signature. 
 Given two null momenta $\bm k_1, \bm k_2$ and $\bm K_3=\bm k_1+\bm k_2$, the equation \eqref{kpolrelation} tells us $\bm K_3^2=2\bm k_1\cdot\bm k_2\propto \sfz_{12}\;\sfzb_{12}$. So we can make $\bm K_3$ null   by just setting $\sfz_{12}=0$ or $\sfzb_{12}=0$, without forcing the strict alignment of $\bm k_1$ and $ \bm k_2$.
 In particular, in (2,2) signature, $\sfz_{12} $ and $\sfzb_{12}$ are independent.    We will approach the collinear limit $\bm k_1\cdot\bm k_2\to 0$ by setting $\sfz_1\to \sfz_2$, while keeping $\sfzb_{12}$ arbitrary.   In particular, using \eqref{nullVec22}, the momentum conservation $\bm k_3=\bm k_1+\bm k_2+\sO(\sfz_{12})$ implies 
 \be\label{omeag123z12}
 \omega_3=\omega_1+\omega_2 +\sO(\sfz_{12})~, \qquad \sfz_3=\sfz_1+ \sO(\sfz_{12})=\sfz_2 +\sO(\sfz_{12})~, \qquad 
 \sfzb_3=\sfzb_2+\frac{\omega_1}{\omega_1+\omega_2}\sfzb_{12}+\sO(\sfz_{12} )~.
 \ee
 Now let us reconsider \eqref{VVOPE}. Since we are only interested in terms   at leading  order in $\sfz_{12}$,  this means we just need to consider $\sfz_{12}^0$ terms inside the square bracket of  \eqref{VVOPE}. Therefore we can just set $\sfz_3=\sfz_2$   in $ \cV(\omega_3, \sfz_3, \sfzb_3)$ and furthermore  use the relation in \eqref{omeag123z12}. As a result,  \eqref{VVOPE} becomes 
   \beqn
  \cV(\omega_1, \sfz_1, \sfzb_1)  \cV(\omega_2, \sfz_2, \sfzb_2)
&\sim& 
- \frac{(\omega_1+\omega_2)^2} { \omega_1\omega_2}  \frac{   \sfzb_{12} }{\sfz_{12} }
\Big[   \cV(\omega_1+\omega_2, \sfz_2, \sfzb_2+\frac{\omega_1}{\omega_1+\omega_2}\sfzb_{12}) 
 +\sO(\sfz_{12})
    \Big]
    \\&\sim& 
 -   \sum_{n=0}^\infty \frac{1}{n!}
 { \omega_1^{-1+n}\omega_2^{-1}}{(\omega_1+\omega_2)^{  2-n}}
  \frac{   \sfzb_{12}^{n+1} }{\sfz_{12} }
  \bar\p^n  \cV(\omega_3, \sfz_2, \sfzb_2) +\sO((\sfz_{12})^0)~,
  \qquad\qquad
 \eeqn
 where we essentially performed a Taylor expansion in the second line.
 Performing the Mellin transformation gives
    \be\label{VVOPEdes}
  \cV_{\Delta_1}( \sfz_1, \sfzb_1)  \cV_{\Delta_2} (  \sfz_2, \sfzb_2)
 \sim
-\sum_{n=0}^\infty \frac{1}{n!} 
  \frac{   \sfzb_{12}^{n+1} }{\sfz_{12} }
  \bar\p^n  \cV_{\Delta_1+\Delta_2}( \sfz_2, \sfzb_2)  
    B(\Delta_1+n-1, \Delta_2-1) ~,
 \ee
which is exact to all orders in  $\sfzb_{12}$  but to   leading order in $ \sfz_{12}$. This   coincides with the celestial OPE of two positive helicity gravitons including all the $\overline{SL (2,\mathbb R)}$ descendant contributions \cite{Guevara:2021abz,Jiang:2021ovh}. 

\subsection{Descendant  in OPE from momentum conservation }

In the previous discussions, we derive the celestial OPE for the massless field in $\cN=2$ string   with all  $\overline{SL(2,\mathbb R)}$ descendant contributions included. In this subsection, we want to show that this is a general feature and can be easily  generalized to all celestial OPEs in (2,2) signature.  
%

As explained, in (2,2) signature we can have three momenta conserved and null without fully pointing along the same direction. In particular we can vary  $\sfz_{12},\sfzb_{12} $ independently.  In the collinear limit, the OPE of two operators has the following general structure
\beqn\label{OPEstr}
O(\omega_1, \sfz_1, \sfzb_1) O(\omega_2, \sfz_2, \sfzb_2)
&\sim& 
S(\omega_1,\omega_2,\sfz_{12},\sfzb_{12})\Big[   O(\omega_3, \sfz_3, \sfzb_3)
 +\sO (\sfz _{12}\sfzb_{12}) 
    \Big]~,
\eeqn
where $\omega_3, \sfz_3, \sfzb_3$ parametrize the null momentum $\bm k_3$ which is approximately  the total momentum $\bm k_1+\bm k_2$, The difference  between $\bm k_1+\bm k_2$ and $\bm k_3$ vanishes in the strict collinear limit and accounts for the $\sO (\sfz _{12}\sfzb_{12}) $ uncertainty  in the bracket. We would like to realize the collinear limit such that $\sfz_{12}\to 0$ while keeping  $\sfzb_{12}$ arbitrary. And in the bracket of \eqref{OPEstr}, we only keep the $(\sfz_{12})^0$ term. This enables us to set $\sfz_3 \to \sfz_2$ at  this order. The momentum conservation in the collinear limit \eqref{omeag123z12} further enables us to write $\omega_3,\sfzb_3$ in terms of $\omega_1,\omega_2,\sfzb_1,\sfzb_2$ exactly in $\sfzb $ direction but at leading order $\sO((\sfz_{12})^0)$ in $\sfz$ direction.  As a consequence, \eqref{OPEstr} finally reduces to 
\beqn\label{OPEstrsimple}
O(\omega_1, \sfz_1, \sfzb_1) O(\omega_2, \sfz_2, \sfzb_2)
&\sim& 
S(\omega_1,\omega_2,\sfz_{12},\sfzb_{12})\Big[   O(\omega_1+\omega_2, \sfz_2, \sfzb_2+\frac{\omega_1}{\omega_1+\omega_2}\sfzb_{12}) 
 +\sO(\sfz_{12})
    \Big]~,\qquad
\eeqn
where $S$ only depends on $\sfz_{12},\sfzb_{12}$   through their  differences due to the translational invariance on the celestial sphere/torus. In general  we can assume that $S(\omega_1,\omega_2,\sfz_{12},\sfzb_{12})$ has power law dependence on energy
\be
S(\omega_1,\omega_2,\sfz_{12},\sfzb_{12})=\omega_1^\alpha \omega_2^\beta(\omega_1+\omega_2)^\gamma T(\sfz_{12},\sfzb_{12})~,
\ee
and thus
\be
 S(t\omega  ,(1-t)\omega ,\sfz_{12},\sfzb_{12})
 =
t^\alpha(1-t)^\beta \; \omega^{\alpha+\beta+\gamma} T(\sfz_{12},\sfzb_{12})~.
\ee
Performing the Mellin transformation and using the following identity  
\be
\int d\omega_1\,d\omega_2 \; \omega_1^{\Delta_1-1}  \omega_2^{\Delta_2-1}
= \int_0^1 dt\int d\omega \;\omega^{\Delta_1+\Delta_2-1} 
t^{\Delta_1-1}  (1-t)^{\Delta_2-1}~, \qquad
   t=\frac{\omega_1}{\omega_1+\omega_2}~, \quad \omega=\omega_1+\omega_2~,
\ee
respectively on two sides of \eqref{OPEstrsimple}, we obtain
   \beqn\label{OPEdesfs}
 \cO_{\Delta_1 }(\sfz_1, \sfzb_1)\cO_{\Delta_2 }(\sfz_2, \sfzb_2)  
 & \sim&\nonumber
    \int_0^1 dt  \int d\omega\;
     t^{\Delta_1+\alpha-1}(1-t)^{\Delta_2+\beta-1} \; 
  \omega^{ \Delta_1+\Delta_2+\alpha+\beta +\gamma -1} T(\sfz_{12},\sfzb_{12})  
  \cO(\omega,\sfz_2, \sfzb_2+t \sfzb_{12}) 
  \\ &\sim&
T(\sfz_{12},\sfzb_{12})     \int_0^1 dt\; \cO_{ \Delta_1+\Delta_2+\alpha+\beta +\gamma }(\sfz_2, \sfzb_2+t \sfzb_{12})       t^{\Delta_1+\alpha-1}(1-t)^{\Delta_2+\beta-1}
\nonumber \\ &\sim&
T(\sfz_{12},\sfzb_{12})   \sum_{n=0}^\infty
\frac{( \sfzb_{12})^n}{n!}
\bar\p^n\cO_{ \Delta_1+\Delta_2+\alpha+\beta +\gamma }(\sfz_2, \sfzb_2  )       B(\Delta_1+\alpha+n ,\Delta_2+\beta )~.
\label{OPEwithalldescendant}
 \qquad\quad 
 \eeqn
 So once we work out the primary operators in the OPEs, namely  the leading $n=0$ term in the above expansion, then we may read  off $\alpha,\beta,\gamma$. The formula above   enables us to include all $n>0$ contributions, and thus compute all the  $\overline{SL(2,\mathbb R)}$   descendants.

Our derivation above essentially relies on the momentum conservation in (2,2) signature. The same type of formula was derived before from the conformal symmetry in celestial CFT.  There the   formula for the OPE with $\overline{SL(2,\mathbb R)}$    descendants is given by \cite{Guevara:2021abz,Jiang:2021ovh} 
   \beqn\label{OPEdes}
 && \cO_{\Delta_1,J_1 }(\sfz_1,\sfzb_1)\cO_{\Delta_2 ,J_2}(\sfz_2,\sfzb_2)    
\nonumber   \\ &\sim&
  \mathcal N_{\cO_1\cO_2}^{\cO_3}   \frac{\sfzb_{12}^{N-M}}{\sfz_{12}^{M+N}} 
\int_0^1  dt \; \cO_{ \Delta_3,J_3 } (\sfz_2, \sfzb_2+t \sfzb_{12}) 
\; t^{\Delta_1 -J_1-M+N -1 }(1-t)^{ \Delta_2 -J_2-M+N-1 }~,
\qquad
 \eeqn
 where
  \be
 M=\frac{\Delta_1+\Delta_2-\Delta_3}{2}~, \qquad N=\frac{J_1+J_2-J_3}{2}~,
 \ee
One easily sees that they are very similar. It is also easy to verify that the descendant contributions in \eqref{OPEwithalldescendant} are consistent with the general OPE formula   with descendants in \cite{Jiang:2021ovh}.
 
\subsection{$w_{1+\infty}$ algebra from OPE  }
We have  now derived the celestial OPE \eqref{VVOPEdes} including all  the $\overline{SL(2,\mathbb R)}$ descendants in $\cN=2$ string theory.  Now we focus on the   soft sector,  namely the set of operators with special integral dimensions. More specifically, we define the soft current as  \cite{Guevara:2021abz,Jiang:2021ovh} 
\be\label{softcurrentHk}
H^l(\sfz,\sfzb) =\lim_{\Delta\to l}(\Delta-l) \cO_\Delta(\sfz,\sfzb) ~, \qquad 
l=2, 1, 0, \cdots~.
\ee
The soft currents can be  further decomposed into chiral currents    \cite{Guevara:2021abz,Jiang:2021ovh} 
\be\label{Hksfzb}
H^l(\sfz,\sfzb) =\sum_{n=1-i}^{i-1} \sfzb^{i-n-1}\frac{\cH^i_n(\sfz)}{ (i-n-1)! (i+n-1)!}~, \qquad
l=4-2i~, 
\ee
where the range of indices are 
\be
 \quad i=1,\; \frac32, \; 2,\; \cdots~, \qquad n=1-i, \;2-i,\; \cdots ,\;  i-1~.
\ee
There are thus $2i-1$ chiral currents $\cH_n^i(\sfz)$ which transform under the $(2i-1)$-dimensional representation of $\overline{SL(2,\mathbb R)}$.  After doing some algebraic manipulations, one can show that \eqref{VVOPEdes}  gives rise to  the following chiral OPEs \cite{Jiang:2021ovh} 
\be
\cH^i_n(\sfz) \; \cH^j_m   (0) \sim -\frac{ 2  }{\sfz} \Big(   m(i-1) -n(j-1) \Big) 
 \cH^{i+j-2 }_{n+m}(0)~.
\ee
In terms of commutators, they are
\be\label{w1pinf}
[\cH^i_n , \; \cH^j_m    ]=  - 2 \Big(   m(i-1) -n(j-1) \Big)  
 \cH^{i+j-2 }_{n+m} ~. 
\ee
This  just gives the $w_{1+\infty}$ algebra, or  more precisely, the loop algebra of the wedge algebra of $w_{1+\infty}$ algebra \cite{Guevara:2021abz,Strominger:2021lvk}. 

Therefore, we have derived the $w_{1+\infty}$ algebra in $\cN=2$ string theory. Note that the appearance of $w_{1+\infty}$ in  $\cN=2$ string  is not accidental. At classical leve, $w_{1+\infty}$  appears as the symmetry group of   self-dual gravity in (2,2) signature   \cite{Boyer:1985aj}, where the only degree of freedom is  Kahler potential. The quantization of this theory is just given by   the $\cN=2$ string \cite{Ooguri:1991fp}.  

 Here we obtain the $w_{1+\infty}$ algebra by first deriving the celestial OPE \eqref{VVOPEdes} with all  $\overline{SL(2,\mathbb R)}$  descendants from worldsheet OPE in $\cN=2$ string theory, and then performing the mode expansion into chiral currents. However, our construction is indirect. It would be desirable to construct directly  the generators $\cH^i_n$   from the worldsheet, and then show that they satisfy the $w_{1+\infty}$ algebra \eqref{w1pinf}. Let us add some comments here.  One can indeed perform the Mellin transformation \eqref{VDelta} directly on the vertex operator \eqref{N2Vophat} and obtain the conformal vertex operator. It contains several terms each with   factor   $\Gamma(\Delta+s_i) (-i\hat{\bm k}\cdot \bm \cX)^{-\Delta-s_i}$, $s_i\in \mathbb Z$. 
  Since $\Gamma(\Delta+s_i) $ has a pole when $\Delta+s_i \in \mathbb Z_{\le 0}$, the definition of soft current in \eqref{softcurrentHk} indeed gives meaningful result when $l=-s_i,-s_i-1, \cdots$.  The soft currents are then essentially some polynomials of  $(-i\hat{\bm k}\cdot \bm \cX)$.   However, the range of  index $l$ seems to not work exactly  as we expected.
Moreover, it is not clear how to decompose the resulting soft currents into chiral currents \eqref{Hksfzb}. \footnote{ Due to the relation \eqref{nullVec22}\eqref{varepsilonPls}, the soft currents are also  polynomials in both $\sfz$ and $\sfzb$. However, \eqref{Hksfzb}   shows that the soft currents are supposed to be   Laurent series   in $\sfz$. 
}
A detailed and better understanding is needed to solve these confusions, and we leave the direct construction to the future. 

%
%

 \section{Conclusion and outlook}\label{Conclusion}
 
 To summarize, in this paper we provide  an approach to deriving celestial OPEs from   the worldsheet    in   string theory. Our results are corroborated by the collinear  factorization of string amplitudes, and are applicable to   general dimensions, corresponding to  Einstein-Yang-Mills theory with possible higher derivative corrections. For $\cN=2$ string theory, we also obtain  the descendant contributions in the celestial OPE, whose soft sector leads to the $w_{1+\infty} $ symmetry. 
The connection between celestial sphere and string worldsheet initiated  in this paper may finally help us to construct a concrete model for celestial holography in string theory.    Besides this ambitious goal,  various questions about celestial OPEs remain to be further studied.

First of all, our results of  celestial OPEs  include   the $\ap$ corrections but not the quantum correction as we were only focusing  on  the tree level amplitude.   It would be interesting to derive the string loop corrections to the celestial OPEs. The derivation from worldsheet at loop level seems to be much more complicated; in particular, one needs to take into account the integration over the moduli space of Riemann surface.  
Nevertheless, a simpler approach may   be   considering     the factorization of string amplitudes at loop level and then performing Mellin transformation. 

Even at tree level, it is still not clear how to derive the celestial OPE corresponding to the fusion of two gluon operators from the open string   into a graviton operator from the closed string. 
Although we sidestepped this question by going to heterotic string and obtained the desired celestial OPEs, it is still conceptually very important to derive such an OPE from the open-closed setup. 

Moreover, our derivation in this paper is mostly about the primary operators in the OPEs, although we discussed the descendants in $\cN=2$ string theory. It would be  very interesting to understand how to systematically incorporate  the descendant contribution in  the OPEs. 
Since the descendants are fully determined by  symmetry, the more basic question may be how to    implement various symmetries on the celestial sphere through  some worldsheet generators.
 
 Furthermore, it would be important to study the vertex operators in the conformal basis directly. 
 In our current derivation, we  first derive  the OPEs of  worldsheet vertex operators    in momentum space, and then Mellin transform  to the conformal basis.
Although the momentum space offers a bridge, it  makes the connection between celestial sphere and worldsheet less transparent. So
understanding the  vertex operators and their OPEs directly in conformal basis would be very useful. 
  
Last but not least, it would also be very insightful to have a more direct understanding of $w_{1+\infty}$ symmetry from the worldsheet in $\cN=2$ string theory. 
Although we    derived the $w_{1+\infty}$ symmetry based on the celestial OPEs with descendants purely from string theory,  the origin of $w_{1+\infty}$ symmetry is not clear. A direct construction of  $w_{1+\infty}$ generators would make the symmetry more transparent and also enables us to discover the implications of such an infinite dimensional symmetry algebra.  The study of $\cN=2$ string theory and self-dual gravity is particularly interesting, as the   self-duality of these gravitational theories  suggests the chirality of the dual celestial conformal field theory.  The chiral nature and the infinite dimensional symmetry  bring  lots of simplifications, and may finally lead us to an exact model for celestial holography.

%
%
%
%
%
%
%
%
%
%
%
%
%
%
 
  \acknowledgments
 We would like to thank   Rodolfo Russo  and Arkady Tseytlin for interesting discussions and comments on the draft. 
This work was  supported by the Royal Society grant, \textit{“Relations, Transformations, and Emergence in Quantum Field Theory”}, and 
by the Science and Technology Facilities Council (STFC) Consolidated Grant ST/T000686/1 \textit{``Amplitudes, strings  \& duality''}.

\appendix 
 \section{OPE on the string worldsheet}\label{appOPE}
 In this appendix, we give all the details of computing the worldsheet OPEs in various string theories. 
 
 \subsection{OPE in open bosonic string}
 Let us first consider the OPEs in open bosonic string theory. 
 We will assume that all $X^\mu$ satisfy the Neumann boundary conditions  and the basic OPE is   given by \footnote{ Let us also recall the weights of various operators 
 $
  h(\p_y^\ell X(y))=\ell, \; h(e^{i k\cdot X})=\alpha' k^2 
$.
Although being non-primary, $X $ formally has   weight 0. 
}
\be\label{XXopenOPE}
X^\mu(y_1) X^\nu(y_2) \sim  - 2\alpha' \eta^{\mu\nu} \ln|y_1-y_2|   ~.
\ee From now on, we will assume $y_1> y_2$ and thus  the absolute value symbol in \eqref{XXopenOPE} can be removed. 

Taking derivatives of \eqref{XXopenOPE}, we get
\be
\dot X^\mu(y_1) X^\nu(y_2) \sim  - 2\alpha' \eta^{\mu\nu}  \frac{1}{y_{12}}~, \qquad
\dot X^\mu(y_1) \dot X^\nu(y_2) \sim  - 2\alpha' \eta^{\mu\nu}  \frac{1}{  y_{12}^2}~,
\ee
where $\dot X(y)=\p_y X(y)$.

Now, we can compute the OPEs of various composite operators made out of $X$. 
The most important formula  in our  OPE computations     is the following identity in free field theory
\be\label{ABCformula}
:e^{A}:\; :e^B:=e^{\EV{AB}}:e^{A+B}:~,
\ee
where $A, B$ are   collections of annihilation and creations operators of free fields, and $:\cdots:$ denotes the normal order product.  

\paragraph{{OPE without derivative.}} In particular, we can take $A,B$ in \eqref{ABCformula} as the free fields themselves and then obtain 
 \be\label{pqexpOPE}
:e^{ip \cdot X(y_1) }:\; :e^{i q\cdot X(y_2 ) }:= y_{12}^{2\ap p \cdot q} :e^{ip \cdot X(y_1)+i q\cdot X(y_2 ) }: ~,
\ee
where 
\beqn\label{normalOrdDf}
 : e^{ip \cdot X(y_1)+i q\cdot X(y_2 ) }: &=&
 : e^{ip \cdot X(y_1) }e^{ i q\cdot X(y_2 ) }: 
=\sum_{n=0}^\infty \frac{y_{12}^n}{n!}:[ \p^n e^{ip \cdot X  } ]e^{ i q\cdot X  }(y_2 ): 
\\&=&
 \sum_{n=0}^\infty \frac{y_{12}^n}{n!}:[e^{-ip \cdot X  }  \p^n e^{ip \cdot X  } ]e^{ i (p+q)\cdot X  }(y_2 ): 
 \\&=&
: \Big( 1+y_{12} \; i p \cdot \dot X+\frac{y_{12}^2}{2}\big( -(p \cdot \dot X)^2+i p \cdot \ddot X \big)+\cdots
\Big)  e^{ i (p+q)\cdot X  }(y_2 ): ~.
\qquad
\eeqn
As a result, we have 
 \be
 e^{ip \cdot X(y_1) } e^{i q\cdot X(y_2 ) } = y_{12}^{2\ap p \cdot q} \Big[ 1+y_{12} \; i p \cdot \dot X+\frac{y_{12}^2}{2}\Big( -(p \cdot \dot X)^2+i p \cdot \ddot X \Big)+\cdots
\Big]   e^{ i (p+q)\cdot X  }(y_2 )~.
\ee
where we have removed the normal ordering symbol for simplicity of notation. \footnote{  Rigorously speaking, different  orders of operators inside the normal order product usually give different results. But in our free field OPEs and for the vertex operators under consideration, the difference does not matter. In particular, $: e^{ik \cdot X}\dot X : $  and $:\dot X e^{ik \cdot X}:$ only differ  by $\p  e^{ik \cdot X}$, which is a total derivative and is thus inessential after integration. For this reason, we will not be rigorous about the ordering  of operators.   }

By further taking derivatives in \eqref{pqexpOPE}, we similarly get the following OPE
\beqn\label{ddexpOPE}
\p_{y_1} e^{i p\cdot X(y_1) }  \p_{y_2} e^{i q \cdot X(y_2)}  &\sim&
-  y_{12} ^{2 \alpha' p\cdot q-2}   e^{i(p+q) \cdot X }
\\&&
\times
 \Bigg[2 \alpha' p\cdot q(2 \alpha' p\cdot q-1)+2i \alpha'  y_{12} \, p\cdot q \Big(-q\cdot \dot X  +2\alpha' p\cdot q \; p \cdot \dot X \Big)+\cdots \Bigg] (y_2) ~.
\qquad
\nonumber
\eeqn

\paragraph{{OPE with one derivative.}}
We are also interested in the OPEs of operators involving   derivatives $\dot X$. This can be done by using the following trick. We can regard $\dot X$ as arising from  the Taylor expansion  of the exponentiation of free fields, namely 
\be
\dot X^\mu e^{i p \cdot X} =-i\frac{\p}{\p \zeta_\mu}e^{i p \cdot X+i \zeta\cdot \dot X} | _{\zeta=0}~.
\ee
Since the OPEs of free field exponential   operators can be computed using \eqref{ABCformula}, we can thus also  easily obtain  OPEs of composite operators involving $\dot X$. This can also be easily generalized to composite operators with multiplet $\dot X$ or even higher derivatives $\ddot X$,  etc. 

Using this trick, we can compute the OPE of vertex operators for tachyon and gluon as follows:
 \be\label{XeOPE0}
 e^{i p \cdot X  }(y_1)\dot X^\nu e^{i q  \cdot X  }(y_2)
=- i \frac{\p}{\p \xi_\nu}e^{i p \cdot X }(y_1) 
e^{i q \cdot X+i \xi\cdot \dot X} (y_2) \Big| _{ \xi=0}~.
\ee
The OPE on the right hand side can be evaluated using \eqref{ABCformula}
\beqn
&&e^{i p \cdot X }(y_1) 
e^{i q \cdot X+i \xi\cdot \dot X} (y_2)
\\&=&e^{2 \alpha' p\cdot q \ln y_{12}-2\alpha'  \xi \cdot p /y_{12}  }
:e^{i p \cdot X(y_1)+i q \cdot X (y_2)} e^{  i \xi\cdot \dot X(y_2)}  :
\\&=&
e^{2 \alpha' p\cdot q \ln y_{12}-2\alpha'  \xi \cdot p /y_{12}  }
 : \Big(1+  i \xi \cdot \dot X(y_2 )+\cdots \Big) 
e^{i p \cdot X(y_1)+i q \cdot X (y_2)}:
\nonumber
\\&= &  \nonumber
y_{12}^{2 \alpha' p\cdot q} \Big(1-2\alpha'   \xi \cdot p /y_{12} +\cdots \Big) 
 :  \Big(1+  i \xi \cdot \dot X(y_2 )+\cdots \Big) 
   \Big( 1+y_{12} \; i p \cdot \dot X(y_2)+\cdots \Big) 
e^{i (p+ q) \cdot X (y_2)}:
\\&= &  \nonumber
y_{12}^{2 \alpha' p\cdot q-1} \Big( -2\alpha'   \xi \cdot p   +
  i  y_{12}\xi \cdot \dot X +\cdots \Big)
   \Big( 1+y_{12} \; i p \cdot \dot X+\cdots \Big) 
e^{i (p+ q) \cdot X  }(y_2)
\\&= &  \nonumber
y_{12}^{2 \alpha' p\cdot q-1} \Big( -2\alpha'   \xi \cdot p (1+y_{12} \; i p \cdot \dot X)  +
  i  y_{12}\xi \cdot \dot X(y_2 )+\cdots \Big) 
e^{i (p+ q) \cdot X  }(y_2)
\eeqn
Then the OPE in \eqref{XeOPE0} is given by
\be\label{OPEoneXdot}
e^{i p \cdot X }(y_1)  \dot X^\nu e^{i q \cdot X } (y_2)
\sim
y_{12}^{2 \alpha' p\cdot q-1} 
\Big(  2i\, \alpha'    p^\nu    
+
   y_{12}  (\dot X^\nu
   -2  \alpha'    p^\nu   p \cdot \dot X)
    +\cdots \Big) 
e^{i (p+ q) \cdot X (y_2)}~.
\ee

\paragraph{{OPE with two derivatives.}}
Using the same   trick, we can compute the vertex operators of two gluons as follows
\be\label{XeOPE}
\dot X^\mu e^{i p \cdot X  }(y_1)\dot X^\nu e^{i q  \cdot X  }(y_2)
=- \frac{\p}{\p \zeta_\mu} \frac{\p}{\p \xi_\nu}e^{i p \cdot X+i \zeta\cdot \dot X}(y_1) 
e^{i q \cdot X+i \xi\cdot \dot X} (y_2) \Big| _{\zeta=\xi=0}~.
\ee
%
 Using the identity \eqref{ABCformula}, the OPE on the right-hand-side can be calculated: 
\beqn
&&e^{i p \cdot X+i \zeta\cdot \dot X}(y_1) 
e^{i q \cdot X+i \xi\cdot \dot X} (y_2)
\\&=&e^{2 \alpha' p\cdot q \ln y_{12} +2\alpha' (\zeta \cdot q-\xi \cdot p)/y_{12}+ 2\alpha'\zeta\cdot \xi /y_{12}^2 }
:e^{i p \cdot X(y_1)+i q \cdot X (y_2)}e^{ i \zeta\cdot \dot X(y_1)}
e^{  i \xi\cdot \dot X(y_2)}  :
\\&=&e^{2 \alpha' p\cdot q \ln y_{12} +2\alpha' (\zeta \cdot q-\xi \cdot p)/y_{12}+ 2\alpha'\zeta\cdot \xi /y_{12}^2 }
 :\Big(1+  i \zeta\cdot \dot X(y_1) +\cdots \Big) 
  \Big(1+  i \xi \cdot \dot X(y_2 )+\cdots \Big) 
e^{i p \cdot X(y_1)+i q \cdot X (y_2)}:
\nonumber
\\&= &  \nonumber
y_{12}^{2 \alpha' p\cdot q} \Big(1+2\alpha' (\zeta \cdot q-\xi \cdot p)/y_{12}+ 2\alpha'\zeta\cdot \xi /y_{12}^2+ 2\ap^2 (\zeta \cdot q-\xi \cdot p)^2/y^2_{12}+\cdots \Big) 
\\&  &  \nonumber
\times
:\Big(1+  i \zeta\cdot \dot X(y_2) +y_{12}   i \zeta\cdot \ddot X(y_2) +\cdots \Big) 
  \Big(1+  i \xi \cdot \dot X(y_2 )+\cdots \Big) 
e^{i p \cdot X(y_1)+i q \cdot X (y_2)}:
\\&= &  \nonumber
y_{12}^{2 \alpha' p\cdot q}: \Bigg[\frac{1}{y^2_{12}} 
\Big( 2\alpha'\zeta\cdot \xi  -4\ap^2  \zeta \cdot q\; \xi \cdot p  \Big)   
+\frac{2i\alpha' }{y_{12}}\Big(   \zeta \cdot q \;  \xi \cdot \dot X (y_2) -\xi \cdot p\; \zeta\cdot \dot X(y_2)  \Big)  +
 \cdots \Bigg] 
 \\&  &  \nonumber\qquad\qquad\qquad
\times  \Big(1+iy_{12} p \cdot \dot X(y_2) +\cdots \Big)
e^{i (p+q) \cdot X }:(y_ 2)
\\&= &  \nonumber
y_{12}^{2 \alpha' p\cdot q-2} e^{i (p+q) \cdot X }\Bigg[  
\Big( 2\alpha'\zeta\cdot \xi  -4\ap^2  \zeta \cdot q\; \xi \cdot p  \Big)    \Big(1+iy_{12} p \cdot \dot X\Big)
+  2i\alpha' \; y_{12} \Big(   \zeta \cdot q \;  \xi \cdot \dot X  -\xi \cdot p\; \zeta\cdot \dot X  \Big)  +
 \cdots \Bigg] 
(y_ 2)~.
\eeqn
where the dots represent terms  which are quadratic or higher order  in $\xi$, $ \zeta$ or in  $y_{12}$.  


Combining this OPE and  \eqref{XeOPE}, we then obtain
\beqn\label{XXexpOPEapdx}
\zeta\cdot\dot X e^{i p \cdot X  }(y_1)\xi\cdot\dot X e^{i q  \cdot X  }(y_2)
&\sim&
- y_{12}^{2 \alpha' p\cdot q-2} e^{i (p+q) \cdot X }    
\Bigg[  
2\alpha' \Big(  \zeta\cdot \xi  -2\ap   \zeta \cdot q\; \xi \cdot p  \Big)  
\\&&\nonumber
+  2i\alpha' \; y_{12} \Big(   \zeta \cdot q \;  \xi \cdot \dot X  -\xi \cdot p\; \zeta\cdot \dot X 
+  \big(\zeta\cdot \xi  -2\ap   \zeta \cdot q\; \xi \cdot p \big)p \cdot \dot X
 \Big)  +
 \cdots \Bigg] 
  (y_ 2)~,
\eeqn
or equivalently
\beqn\label{XXexpOPE1233}
 \dot X^\mu e^{i p \cdot X  }(y_1) \dot X^\nu e^{i q  \cdot X  }(y_2)
&\sim&
- y_{12}^{2 \alpha' p\cdot q-2} e^{i (p+q) \cdot X }    
\Bigg[  
2\alpha' \Big(  \eta^{\mu\nu}  -2\ap   q^\mu \; p^\nu \Big)  
\\&&\nonumber
+  2i\alpha' \; y_{12} \Big(   q^\mu \;  \dot X^\nu  -  p^\nu\;  \dot X^\mu
+  \big( \eta^{\mu\nu}  -2\ap   q^\mu \; q^\nu\big)p \cdot \dot X
 \Big)  +
 \cdots \Bigg] 
  (y_ 2)~.
\eeqn
As a consistency check, we can set $\zeta=i  p, \xi =iq$, then \eqref{XXexpOPEapdx} reduces to  \eqref{ddexpOPE} as expected.

  \subsection{OPE in closed  bosonic string}
  Now we switch to the closed bosonic string. It has the basic  OPE
    \be\label{XXbasicOPE}
  X^\mu(z_1,\bz_1)   X^\nu(z_2,\bz_2)  \sim -\frac \ap 2\eta^{\mu\nu}  \ln |z_{12}|^2~.
  \ee
 It is more convenient to separate $X^\mu$ into the  left and right movers:
  \be\label{XLR}
  X^\mu(z,\bar z) =X^\mu_L(z)+X^\mu_R(\bar z)~. 
  \ee
The two sectors are independent and have the following OPEs
      \be\label{XXclosedOPE}
  X_L^\mu(z_1 )   X_L^\nu(z_2 )  \sim -\frac \ap 2\eta^{\mu\nu}  \ln  z_{12}~, \quad
    X_R^\mu(\bz_1 )   X_R^\nu(\bz_2 )  \sim -\frac \ap 2\eta^{\mu\nu} \ln  \bz_{12}~, \quad
      X_L^\mu(z_1 )    X_R^\nu(\bz_2 )  \sim 0~. \quad
  \ee
  In general any vertex operator in closed string theory can be decomposed into the product of the left and right moving pieces. So to compute the OPE of two  vertex operators in closed string theory, one just needs to compute the OPEs in the  left and right   moving  sectors independently,  and then take  their product.  In both the left and right moving sectors, the free OPE \eqref{XXclosedOPE} is almost identical to the open string case \eqref{XXopenOPE}, except for the reduction  of   $\ap$ by a factor of 4.
  \footnote{ In particular, the weights of $e^{i k\cdot X}$ are also reduced by 4, namely
$
  h(e^{i k\cdot X})=  \bar h(e^{i k\cdot X})=\frac\ap 4 k^2
$.
 }
 
We are particularly interested in   massless fields in closed string,  whose   vertex operators are given by
 \be
   V^{\mu\bar\mu}(z,\bz) =\p X^\mu\bar\p X^{\bar\mu }e^{i p \cdot X}(z,\bz)
 \ee
We can decompose it into the   left-moving and right-moving parts: 
  \be
    V^{\mu\bar\mu}(z,\bz) = V_L^\mu(z)V_R^{\bar\mu}(z) , \qquad
    V_L^\mu(z) =\p X_L e^{i p \cdot X_L}=\p X  e^{i p \cdot X_L}, \quad
  V_R^{\bar\mu}(z) =\bar\p X_R e^{i p \cdot X_R} =\bar\p X  e^{i p \cdot X_R}~. \qquad
 \ee
 
   We would like to compute the  OPE of two such vertex operators, namely 
 \beqn
 \p X^\mu\bar\p X^{\bar\mu }e^{i p \cdot X}(z_1,\bz_1)  \p X^\nu\bar\p X^{\bar\nu} e^{i p \cdot X}(z_1,\bz_1) 
 &=&  V^{\mu\bar\mu}(z_1,\bz_1)   V^{\nu\bar\nu}(z_2,\bz_2) 
 \\ &=&
  V_L^\mu(z_1) V_L^\nu(z_2)\times V_R^{\bar\nu}(\bz_1) V_R^{\bar\nu}(\bz_2)~.
  \qquad\qquad  \qquad\qquad
 \eeqn 
 
 
 

As described before, we can compute the OPEs in the left and right moving sectors independently. The computation of  OPE  in the  left moving sector is exactly the same as that in the open string case \eqref{XXexpOPE1233}, and the final result is given by 
  \beqn\label{closedLeft}
 && V_L^\mu(z_1) V_L^\nu(z_2) =
 \p X_L^\mu e^{i p \cdot X_L  }(z_1) \p X_L^\nu e^{i q  \cdot X_L  }(z_2)
\\&\sim&\nonumber
- \frac { \ap }{2}z_{12}^{\frac12 \alpha' p\cdot q-2} e^{i (p+q) \cdot X_L  }    
\Bigg[  
  \Big(  \eta^{\mu\nu}  -\frac12\ap   q^\mu \; p^\nu \Big)  
 \\&&   \qquad\qquad  \qquad\qquad
+   i \, z_{12} \Big(   q^\mu \;  \p X ^\nu  -  p^\nu\;  \p X^\mu
+  \big( \eta^{\mu\nu}  - \frac12\ap   q^\mu \; q^\nu\big)p \cdot \p X  
 \Big)  +
 \cdots \Bigg] 
  (z_ 2)~. \qquad\qquad
\eeqn
The right moving OPE is similar. 
After combining the left and right moving OPEs together, we get the world-sheet OPE of two vertex operators for massless fields in closed string theory: 
  \beqn\label{closestringOPE}
 &&  \p X^\mu\bar\p X^{\bar\mu }e^{i p \cdot X}(z_1,\bz_1)  \p X^\nu\bar\p X^{\bar\nu} e^{i p \cdot X}(z_2,\bz_2) 
\\&\sim&\nonumber
  \frac { \ap^2 }{4}|z_{12}|^{  \alpha' p\cdot q-4} e^{i (p+q) \cdot X   }    
\\&&\nonumber \times
\Bigg[  
  \Big(  \eta^{\mu\nu}  -\frac12\ap   q^\mu \; p^\nu \Big)  
  +   i \; z_{12} \Big(   q^\mu \;  \p X ^\nu  -  p^\nu\;  \p X^\mu
+  \big( \eta^{\mu\nu}  - \frac12\ap   q^\mu \; q^\nu\big)p \cdot \p X  
 \Big)  +
 \cdots \Bigg] 
\\&&\nonumber \times
\Bigg[  
  \Big(  \eta^{\bar \mu\bar \nu}  -\frac12\ap   q^{\bar \mu} \; p^{\bar \nu} \Big)  
  +   i \; \bz_{12} \Big(   q^{\bar \mu} \;  \p X ^{\bar \nu}  -  p^{\bar \nu}\;  \bar \p X^{\bar \mu}
+  \big( \eta^{\bar\mu\bar\nu}  - \frac12\ap   q^{\bar \mu} \; q^\nu\big)p \cdot \bar\p X  
 \Big)  +
 \cdots \Bigg] 
(z_2,\bar z_2)~.
  \nonumber
\eeqn

   \subsection{OPE in heterotic string}
  Now we consider the OPE in heterotic string. The bosonic fields are the same as that in \eqref{XLR} and \eqref{XXclosedOPE}. The new ingredient is the right moving fermions on the worldsheet which have the following OPE
  \be
  \tilde \psi^\mu(\bz_1) \tilde \psi^\nu(\bz_2) \sim \frac{\eta^{\mu\nu}}{\bz_{12}}~.
  \ee

%
  
 We would like to compute the following right moving OPE: 
  \be\label{VRm10}
    V^{(-1)}_R(\bz_1)   V^{(0)}_R(\bz_2)  =e^{-\tilde \phi}
\tilde \psi^\mu e^{i p \cdot X_R  }(\bz_1)
  \Big( i \bar\p X^\nu+\frac12 \ap k \cdot {\tilde\psi} {\tilde \psi}^\nu \Big)e^{i q \cdot X_R}(\bz_2)~.
  \ee
 
For this purpose, let us first present  the following two OPEs
   \be
  \tilde \psi^\mu(\bz_1)  k \cdot {\tilde\psi} {\tilde \psi}^\nu (\bz_2) 
  \sim \frac{k^\mu \tilde \psi^\nu (\bz_2) -\eta^{\mu\nu} k \cdot \tilde \psi  (\bz_2)  }{\bz_{12}}~,
  \ee
  and 
   \be
e^{i p \cdot X_R }(\bz_1)  \bar\p X^\nu e^{i q \cdot X_R } (\bz_2)
\sim
\bz_{12}^{\frac12 \alpha' p\cdot q-1} 
\Big(  \frac{i}{2}  \alpha'    p^\nu    
- \frac12  \alpha'   \bz_{12}  p^\nu   p \cdot \bar\p X
+
   \bz_{12}  \bar\p X^\nu +\cdots \Big) 
e^{i (p+ q) \cdot X } (\bz_2)~,
\ee
which can be obtained as in the open string case \eqref{OPEoneXdot}.
  
Then the OPE in \eqref{VRm10} can be evaluated straightforwardly 
  \beqn
&&
\tilde \psi^\mu e^{i p \cdot X_R  }(\bz_1)
  \Big( i \bar\p X^\nu+\frac12 \ap q \cdot {\tilde\psi} {\tilde \psi}^\nu \Big)e^{i q \cdot X_R}(\bz_2)
  \\&=&
  i \tilde \psi^\mu e^{i p \cdot X_R  }(\bz_1)
   \bar\p X^\nu e^{i q \cdot X_R}(\bz_2)
  +
   \frac12 \ap \tilde \psi^\mu(\bz_1)  q \cdot {\tilde\psi} {\tilde \psi}^\nu(\bz_2)
   \times
    e^{i p \cdot X_R  }(\bz_1)
  e^{i q \cdot X_R}(\bz_2)
    \\&=&
   i \tilde \psi^\mu \bz_{12}^{\frac12 \alpha' p\cdot q-1} 
\Big(  \frac{i}{2}  \alpha'    p^\nu    
- \frac12  \alpha'   \bz_{12}  p^\nu   p \cdot \bar\p X
+
   \bz_{12}  \bar\p X^\nu +\cdots \Big) 
e^{i (p+ q) \cdot X_R (\bz_2)}
\\&&+
\frac12  \alpha'  \Big( \frac{q^\mu \tilde \psi^\nu (\bz_2) -\eta^{\mu\nu} q \cdot \tilde \psi  (\bz_2)  }{\bz_{12}}+\cdots\Big)
\bz_{12}^{\frac12\ap p \cdot q} \Big( 1+\bz_{12} \; i p \cdot \bar\p X+\cdots \Big) 
e^{i (p+ q) \cdot X (\bz_2)}
    \\&=&
    \bz_{12}^{\frac12\ap p \cdot q-1} 
e^{i (p+ q) \cdot X_R  }
\Big( 
 \frac{i}{2}  \alpha'     i \tilde \psi^\mu   p^\nu    
- \frac12  \alpha'   \bz_{12}    i \tilde \psi^\mu   p^\nu   p \cdot \dot X
+
   \bz_{12}    i \tilde \psi^\mu  \dot X^\nu+\cdots
       \\& &\qquad\qquad\qquad\qquad \quad
   +\frac12  \alpha' q^\mu \tilde \psi^\nu  
   -\frac12  \alpha' \eta^{\mu\nu} q  \cdot \tilde \psi  +\cdots
\Big) (\bz_2)
    \\&=&
    \bz_{12}^{\frac12\ap p \cdot q-1} 
e^{i (p+ q) \cdot X_R }
\Big( 
 \frac12  \alpha' \big[  q^\mu \tilde \psi^\nu - p^\nu \tilde \psi^\mu 
 -\eta^{\mu\nu} q  \cdot \tilde \psi 
  \big] 
+\cdots
   \Big) (\bz_2)~.
  \eeqn
 
So the final result for the OPE in \eqref{VRm10} after contracting with polarization vectors is 
\beqn
&&\label{RightVROPE}
\zeta\cdot \tilde \psi  e^{i p \cdot X_R  }(\bz_1)
  \Big( i \xi\cdot\bar\p X +\frac12 \ap q \cdot {\tilde\psi}\xi\cdot {\tilde \psi}  \Big)e^{i q \cdot X_R}(\bz_2)
   \\&\sim& 
    \frac{  \alpha'  }{2}   \bz_{12}^{\frac12\ap p \cdot q-1} 
e^{i (p+ q) \cdot X_R  }
\Big( 
\zeta\cdot  q\; \xi\cdot \tilde \psi - \xi\cdot p \; \zeta\cdot\tilde \psi 
 -\zeta\cdot \xi\; q  \cdot \tilde \psi 
   \Big)(\bz_2)~.
   \nonumber
  \eeqn
%
%

 \subsection{OPE in $\cN=2$ string}
 
For  $\cN=2$ string, we have   bosonic fields $\cX^\mu $ and fermionic fields $\psi_L^\mu,\psi_R^\mu$. We can agin decompose $\cX^\mu $ into left and right movers 
\be
\cX^\mu(z,\bz)=\cX_L^\mu(z)+\cX_R^\mu(\bz) ~,
\ee
which obey the OPEs \cite{Ooguri:1991fp}
\be
\cX_L^i(z_1 ) \bar \cX_L^\bj (z_2 )\sim  - \frac{\eta^{i\bj}}{2} \ln (z_1-z_2)~,\qquad
\cX_R^i(\bz_1 ) \bar \cX_R^\bj (\bz_2 )\sim  - \frac{\eta^{i\bj}}{2} \ln (\bz_1-\bz_2)~.
\ee
 
 The OPEs for fermions are  \cite{Ooguri:1991fp}
\be
\psi_L^i(z_1) \bar \psi_L^\bj(z_2) \sim    \frac{\eta^{i\bj} }{z_1-z_2}~, \qquad
\psi_R^i(\bz_1) \bar \psi_R^\bj(\bz_2) \sim  \frac{\eta^{i\bj} }{\bz_1-\bz_2}~.
\ee
 
It turns out to be more convenient to consider the four component fields, namely $\cX^\mu $ and $\psi_L^\mu,\psi_R^\mu$ where $\mu=i, \bar i=1,2,\bar 1, \bar 2$. 
 Then we have the OPEs
 \be
\cX_L^\mu(z_1 )   \cX_L^\nu (z_2 )\sim  - \frac{\eta^{\mu\nu}}{2} \ln (z_{12})~,\qquad
\cX_R^\mu(\bz_1 )   \cX_R^\nu (\bz_2 )\sim  - \frac{\eta^{\mu\nu}}{2} \ln (\bz_{12})~,
\ee
and
 \be
\psi_L^\mu(z_1)   \psi_L^\nu(z_2) \sim    \frac{\eta^{\mu\nu} }{z_{12}}~, \qquad
\psi_R^\mu(\bz_1)   \psi_R^\nu(\bz_2) \sim  \frac{\eta^{\mu\nu} }{\bz_{12}}~,
\ee
where $\eta^{\mu\nu}$ is given in \eqref{etamunumetric}.
 
We would like to compute the OPE of  two vertex operators given in \eqref{N2Vophat}.  
 As before, we can decompose the vertex operators into the independent left- and right-moving sectors:
\beqn
\widehat V(\bm k,z,\zb) &=&\widehat V_L (\bm k,z) \widehat V_R(\bm k,\zb) ~,
 \\
\widehat V_L(\bm k,z) &=& 
\Big( i \bm k^\vee \cdot \p  {\bm \cX_L}    
-\frac12  ( \bm k^\vee \cdot \bm \psi_L )  (\bm k \cdot \bm \psi_L ) \Big)
e^{i \bm k \cdot   {\bm \cX}_L }~,
 \\
 \widehat V_R(\bm k,\zb) &=& 
\Big(   i \bm k^\vee \cdot \bar\p  {\bm \cX_R}   
-\frac12 ( \bm k^\vee \cdot \bm \psi_R ) (\bm k \cdot \bm \psi_R )
\Big) e^{i \bm k \cdot   {\bm \cX} _R}~.
\eeqn
They can be further written as
\beqn\label{VLBF}
 \widehat V_L(\bm k,z) &=& 
\exp\Big[ i \bm k^\vee \cdot \p  {\bm \cX_L}    
+
{i \bm k \cdot   {\bm \cX}_L }\Big]
 \Big(1-\frac12  ( \bm k^\vee \cdot \bm \psi_L )  (\bm k \cdot \bm \psi_L ) \Big) \Big|_\text{linear term in  $\bm k^\vee$  }~,
\eeqn
and similarly for $V_R$, where we only keep    terms  which are linear in the    $\bm k^\vee$.  

Then bosonic and fermionic parts in \eqref{VLBF} can be considered separately. 
The bosonic contribution in the left moving OPE can be derived as in the open sting case \eqref{XXexpOPEapdx}:  
\beqn
&&e^{  
{i \bm k_1 \cdot   {\bm \cX}_L +i \bm k_1^\vee \cdot \p  {\bm \cX_L}    }
}(z_1) 
e^{  
{i \bm k_2 \cdot   {\bm \cX}_L +i \bm k_2^\vee \cdot \p  {\bm \cX_L}    }
}(z_2)  
\nonumber\\&= &  \nonumber
z_{12}^{\frac12   \bm k_1  \cdot \bm k_2 -2} : \Bigg[  
\Big( \frac12 \bm k_1^\vee\cdot \bm k_2^\vee  -\frac14      \bm k_1^\vee \cdot \bm k_2 \; \bm k_2^\vee \cdot \bm k_1  \Big)
 \nonumber\\&&\qquad\qquad\qquad
 +  \frac{i}{2}  z_{12} \Big(    \bm k_1^\vee \cdot \bm k_2  \;  \bm k_2^\vee \cdot \p \bm \cX(z_2)  -\bm k_2^\vee \cdot \bm k_1 \; \bm k_1^\vee \cdot \p \bm \cX (z_1)
-i \bm k_1^\vee \cdot \bm k_2 +i \bm k_2^\vee \cdot \bm k_1
 \Big) 
 \nonumber\\&&\qquad\qquad\qquad
- z_{12} ^2\Big(   \bm k_1^\vee\cdot \p \bm \cX(z_1) -i\Big)
\Big( \bm k_2^\vee \cdot \p \bm \cX(z_2 )-i\Big)
  \nonumber\\&&\qquad\qquad\qquad
   +\sO(z_{12}^3)
 \cdots \Bigg] 
e^{  
 {i \bm k_1 \cdot   {\bm \cX}_L  }(z_1) 
+{i \bm k_2 \cdot   {\bm \cX}_L     }(z_2)  
 }:~,
\eeqn
where the normal ordering of two operators at different points should be understood in the same way as in \eqref{normalOrdDf}. 

The  fermionic contribution in the left moving OPE  is
\beqn
&& ( \bm k_1^\vee \cdot \bm \psi_L )  (\bm k_1 \cdot \bm \psi_L ) (z_1)
  ( \bm k_2^\vee \cdot \bm \psi_L )  (\bm k_2 \cdot \bm \psi_L ) (z_2)
  \\&=&
   \frac{\bm k_1^\vee \cdot \bm k_2\; \bm k_1 \cdot \bm k_2^\vee
   -\bm k_1^\vee \cdot \bm k_2^\vee \; \bm k_1 \cdot \bm k_2 
    }{z_{12}^2} 
     \\& &
+   \frac{\bm k_1^\vee \cdot \bm k_2 }{z_{12}}:\bm k_1  \cdot \bm \psi(z_1) \bm k_2^\vee \cdot \bm \psi(z_2):
+    \frac{\bm k_1 \cdot \bm k_2^\vee }{z_{12}}:\bm k_1^\vee \cdot \bm \psi(z_1) \bm k_2  \cdot \bm \psi(z_2):
  \\& &
  -     \frac{\bm k_1 \cdot \bm k_2 }{z_{12}}:\bm k_1^\vee \cdot \bm \psi(z_1) \bm k_2^\vee \cdot \bm \psi(z_2):
-     \frac{\bm k_1^\vee \cdot \bm k_2^\vee }{z_{12}}:\bm k_1  \cdot \bm \psi(z_1) \bm k_2  \cdot \bm \psi(z_2):   
  \\& &
+:   \bm k_1^\vee \cdot \bm \psi_L \; \bm k_1 \cdot \bm \psi_L  (z_1)
    \bm k_2^\vee \cdot \bm \psi_L \;\bm k_2 \cdot \bm \psi_L   (z_2):~,
\eeqn
where we used 
\be
\bm p \cdot \bm \psi_L(z_1) \bm q \cdot \bm \psi_L(z_2)
= \frac{\bm p \cdot \bm q}{z_{12}} +:\bm p \cdot \bm \psi_L(z_1) \bm q \cdot \bm \psi_L(z_2):~.
\ee
 
Then the OPE of two operators of the form \eqref{VLBF} is given by
 \beqn
&&
 \widehat V_L(\bm k_1,z_1)  \widehat V_L(\bm k_2,z_2) \Big|_{\text{ linear in } \bm k_1^\vee \text{ and } \bm k_2^\vee}
 \\&\sim&
z_{12}^{\frac12   \bm k_1\cdot \bm k_2 -1} : \Bigg[  
 \frac{1}{2z_{12}} \bm k_1^\vee \cdot  \bm k_2^\vee \Big(    1
   -\frac12 \bm k_1 \cdot \bm k_2 
     \Big)  
\label{KdotVV}\\&& \qquad\qquad\quad
+  \frac{i}{2   }\bm k_1^\vee \cdot \bm k_2 \;
   ( \bm k_1^\vee \cdot \p\bm \cX(z_1) +   \bm k_2^\vee \cdot \p\bm \cX(z_2) )
\\&& \qquad\qquad\quad
+  \frac{1}{4    } \bm k_1^\vee  \cdot  \bm k_2 \Big(   
-   ( \bm k_2^\vee \cdot \bm \psi_L \; \bm k_2 \cdot \bm \psi_L ) (z_2) 
   -
  ( \bm k_1^\vee \cdot \bm \psi_L \; \bm k_1 \cdot \bm \psi_L ) (z_1)  
\\&& \qquad\qquad\quad
 -   :  \bm k_2^\vee \cdot \bm \psi(z_2) \bm k_1  \cdot \bm \psi(z_1):
- :\bm k_1^\vee \cdot \bm \psi(z_1) \bm k_2  \cdot \bm \psi(z_2):
 \Big) 
\\&&\label{k1k2fermi} \qquad\qquad\quad
  -    \frac14  {\bm k_1 \cdot \bm k_2 } :\bm k_1^\vee \cdot \bm \psi(z_1) \bm k_2^\vee \cdot \bm \psi(z_2):
-   \frac14   {\bm k_1^\vee \cdot \bm k_2^\vee  }:\bm k_1  \cdot \bm \psi(z_1) \bm k_2  \cdot \bm \psi(z_2):   
\\&& \qquad\qquad\quad
 +\sO( z_{12})
   \Bigg] e^{  
 {i \bm k_1 \cdot   {\bm \cX}_L  }(z_1) 
+{i \bm k_2 \cdot   {\bm \cX}_L     }(z_2)  
 }:~,
 \eeqn
 where we used $\bm k_1^\vee \cdot \bm k_2=-\bm k_2^\vee \cdot \bm k_1$.
 
 We are interested in the collinear limit $\bm k_1\cdot \bm k_2 \to 0$. In this limit, we also have  $\bm k_1^\vee \cdot  \bm k_2^\vee=-\bm k_1\cdot \bm k_2\to 0$ following \eqref{kijN=2id}. Therefore the  $1/z_{12}$ term \eqref{KdotVV} and fermion bilinear term \eqref{k1k2fermi} can be dropped in the collinear limit. 
 \footnote{  Actually, the $1/z_{12}$ term is BRST exact and can be dropped even not in the collinear limit. Recall that $\bm k^\vee$ is just the positive helicity polarization vector $\bm \varepsilon_+$ \eqref{kveePolrelation}, up to an overall rescaling and a gauge transformation.  For $\bm \varepsilon_+$, we have the property  that 
 $
\bm \varepsilon_+(\sfz_i,\sfzb_i)\cdot\bm\varepsilon_+(\sfz_j,\sfzb_j)=0
$ \eqref{kpolrelation2}. So if we   replace   $\bm k^\vee$ with $\bm \varepsilon_+$, the term $\bm k_1^\vee \cdot \bm k_2^\vee$ should be absent completely. The appearance of the term $\bm k_1^\vee \cdot \bm k_2^\vee$ is due to the gauge transformation, and thus is BRST exact.   }
The remaining terms can be combined into a very simple form:
 \beqn\label{VLVLOPEfinal}
&&
 \widehat V_L(\bm k_1,z_1)  \widehat V_L(\bm k_2,z_2) \Big|_{\text{ linear in } \bm k_1^\vee \text{ and } \bm k_2^\vee}
 \\&\sim&
 \frac{1}{2   }\bm k_1^\vee \cdot \bm k_2 \;  z_{12}^{\frac12   \bm k_1\cdot \bm k_2 -1}    
 e^{  
{i (\bm k_1+\bm k_2) \cdot   {\bm \cX}_L  }    }
\\
  &&
  \times\Bigg[   
 i ( \bm k_1^\vee+  \bm k_2^\vee )\cdot \p\bm \cX 
-  \frac{1}{2    }   
   ( \bm k_1^\vee+ \bm k_2^\vee) \cdot \bm \psi_L \; ( \bm k_1+ \bm k_2) \cdot \bm \psi_L  
 +\sO(\bm k_1\cdot \bm k_2)
 +\sO( z_{12})
 \cdots \Bigg] (z_2)~.\qquad
 \eeqn
 The computation of $ \widehat V_R (\bm k_1,\bz_1)  \widehat V_R(\bm k_2,\bz_2) $
 is identical and the final result is just given by \eqref{VLVLOPEfinal} except for the replacement of $z $ with $\bar z$. 
 Combining the left and right moving OPEs together, we get the final OPE 
 \beqn  \label{N2OPE}
\widehat V(\bm k_1) \widehat V(\bm k_2) &\sim& 
 \frac14 \Big(\bm k_1^\vee \cdot \bm k_2\Big)^2
 |z_{12}|^{    \bm k_1\cdot \bm k_2 -2}   
 e^{  {i \bm K_3 \cdot   {\bm \cX}   }   }
\\&&
\times \Bigg[   
 i\bm K_3^\vee  \cdot \p \bm\cX 
-  \frac{1}{2    }   
(\bm K_3^\vee \cdot \bm \psi_L \; \bm K_3 \cdot \bm \psi_L ) 
 +\sO(\bm k_1\cdot \bm k_2)
 +\sO( z_{12})
   \Bigg] 
 \\  &&
 \times
    \Bigg[   
 i\bm K_3^\vee  \cdot \bar\p \bm\cX 
-  \frac{1}{2    }   
(\bm K_3^\vee \cdot \bm \psi_R \; \bm K_3 \cdot \bm \psi_R ) 
 +\sO(\bm k_1\cdot \bm k_2)
 +\sO( \bz_{12})
   \Bigg] ~,
 \eeqn
  where $\bm K_3=\bm k_1+\bm k_2$.

\section{OPE and amplitude in open-closed string theory} \label{appOpenClosed}

In this section of appendix, we will discuss the celestial OPEs involving both gluons and gravitons from the open-closed string setup.  

\subsection*{Basic OPE in open-closed string }

In the open string case with spacetime filling D-brane, the fields $X^\mu$   satisfy the Neumann boundary condition on the boundary, namely
   \be\label{opneBC}
   \p X^\mu(z,\bar z)  =\bar\p X^\mu(z,\bar z) \qquad \text{ for }\qquad  z=\bar z=y \in\mathbb R~.
   \ee
 
 This can be realized by adding a mirror image contribution to the   OPE in  the closed string   \eqref{XXbasicOPE}, namely  we now have
    \beqn\label{XXopenclosedcase}
 X^\mu(z_1,\bar z_1)   X^\nu(z_2,\bar z_2)  
\sim
  -\frac \ap 2 \eta^{\mu\nu} \ln |z_1-z_2 |^2  -\frac \ap 2 \eta^{\mu\nu} \ln |z_1-  \bar z_2 |^2 ~.
   \eeqn 
 Then it is easy to check that this OPE indeed satisfies the boundary condition in \eqref{opneBC}. On the boundary of the disk, we have   $X(y) \equiv X(z=y,\bar z=y)$ with $y\in \mathbb R$, and we also   recover the OPE \eqref{XXopenOPE} in the open string case.
 
 As before, we can decompose $  X(z,\bz)=X_L(z)+X_R(\bz)   $.
  Then \eqref{XXopenclosedcase} becomes  
     \be\label{XLRopenclosedOPE}
     X_L^\mu(z_1)   X_L^\nu (z_2)\sim   - \frac{\ap}{2} \eta^{\mu\nu} \ln z_{12}  ~, \qquad
     X_R^\mu(\bz_1)   X_R^\nu (\bz_2)\sim   - \frac{\ap}{2} \eta^{\mu\nu} \ln \bz_{12}  ~, 
   \ee
   and furthermore
   \be
        X_L^\mu(z_1)   X_R^\nu (\bz_2)\sim   - \frac{\ap}{2} \eta^{\mu\nu} \ln (z_1-\bz_2)  ~.
   \ee
   As a result, we have  $X(y)=X_L(z=y)+X_R(\bz=y)$ and the following OPEs     \cite{Blumenhagen:2013fgp}
   \beqn\label{XLRopenclosedOPEyy}
 &&
     X^\mu(y)   X^\nu_L(z)\sim   -  \alpha' \eta^{\mu\nu} \ln(y-z) ~, \quad
          X^\mu(y)   X^\nu_R(\bz)\sim   -  \alpha' \eta^{\mu\nu} \ln(y-\bz) ~,
      \\ &&  
      X(y) X(z,\bar z)         \sim   
    -  \alpha' \eta^{\mu\nu} \ln|y-z|^2~   .   
   \eeqn
   
   In particular, we also have $\dot X(y)=2 \p X_L(z)|_{z=y}=2 \bar\p X_R(\bz)|_{\bz=y}$.
    
\subsection*{Gluon and graviton mixed amplitude from open-closed string   }
%
%
%
%

We want to compute the three-point amplitude of two gluons and one graviton/dilaton/KR field. They are labelled by their momenta $p_1,p_2, p_3$ and polarizations $\zeta_{1\mu},\zeta_{2\mu},e_{\mu\nu}$.
They satisfy  the momentum conservation and the polarization transversality conditions
\be\label{p123property}
p_1+p_2+p_3=0~, \qquad \zeta_1\cdot p_1=\zeta_2\cdot p_2=e_{\mu\nu}  p_3^\mu=e_{\mu\nu}  p_3^\nu=0~,
\ee
as well as the on-shell conditions
\be\label{p123Onshell}
  \qquad p_1^2=p_2^2=p_3^2=0~, \qquad \Rightarrow\qquad p_1\cdot p_2=p_2\cdot p_3=p_3\cdot p_1=0~.
\ee
We further assume that the closed string polarization tensor  is traceless $e_{\mu\nu}\eta^{\mu\nu}=0$, so we do not consider dilaton. \footnote{This is to avoid the contraction between the left    and right moving pieces  of closed string vertex operators, which turns out to bring divergence and needs to be treated properly. We thank Rodolfo Russo for discussions on this point.    }

We would like to compute three-point amplitude from the    open-closed string setup. At tree level, this is given by   the   correlator on the disk,   where we  insert the open string vertex operators on the boundary of the disk and closed vertex operators in the interior of the disk. In particular, our three-point amplitude involves   two open   and one closed string fields, and can be computed by 
\be\label{Aooc3}
A^{ooc}_3=\int \frac{dy_1\, dy_2\, d^2 z\; }{V_\text{CKG}}\EV{\zeta_1\cdot \dot X e^{ip_1\cdot X}(y_1)
\zeta_2\cdot \dot X e^{ip_2\cdot X}(y_2)
e_{\mu\nu}  \p X^\mu \bar\p X^\nu e^{i p_3\cdot X}(z,\bar z)} \Tr(t^A t^B)~,
\ee
where we need to divide the volume of the conformal Killing group, which is   $PSL(2,\mathbb R)$.   In practice we need to fix the    $PSL(2,\mathbb R)$ invariance. Following the prescription in \cite{Blumenhagen:2013fgp}, we can set 
\be
y_1=-y_2=-y~, \qquad    z=i~, \qquad  \bz=-i~,
\ee
but we also need to take into account  the non-trivial Jacobian of the transformation from the fixed coordinates to the parameters of $ PSL(2,\mathbb R)$. 
Using the coordinate transformation $y_1=\tilde y- y, y_2=\tilde y+  y$, $z=u+iv, \bar z=u-iv$,  the integration measure becomes  $dy_1\, dy_2\, d^2 z=2dy d\tilde y du dv $. An infinitesimal $  PSL(2, \mathbb R)$ transformation  acts as $\delta z=\alpha+\beta z+\gamma z^2$ where $\alpha,\beta, \gamma $ are real. The Jacobian between them is  \cite{Blumenhagen:2013fgp}
\be
\Big|\frac{\p( u,v,\tilde y)}{\p(\alpha,\beta, \gamma)}\Big|=1+y^2~, \qquad
 \text{ at }\qquad \tilde y=u=0~, \quad v=1~.  
\ee 
Then  \eqref{Aooc3} becomes
\be
A^{ooc}_3=\int  dy\; 2(1+y^2)\;\EV{\zeta_1\cdot \dot X e^{ip_1\cdot X}(-y)
\zeta_2\cdot \dot X e^{ip_2\cdot X}(y)
e_{\mu\nu}  \p X^\mu \bar\p X^\nu e^{i p_3\cdot X}(i, -i)} \Tr(t^A t^B)~.
\ee

To compute the correlator in the integrand, we rewrite it as
\beqn\label{integrandzeta12}
&&
 \EV{\zeta_1\cdot \dot X e^{ip_1\cdot X}(y_1)
\zeta_2\cdot \dot X e^{ip_2\cdot X}(y_2)
e_{\mu\nu}  \p X^\mu \bar\p X^\nu e^{i p_3\cdot X}(z,\bar z)}
\\&=&e_{\mu\nu} \frac{\p^2}{\p \xi_\mu  \p\bar \xi_\nu}
\EV{ e^{ip_1\cdot X+i \zeta_1\cdot \dot X }(y_1)
  e^{ip_2\cdot X+i\zeta_2\cdot \dot X }(y_2)
   e^{i p_3\cdot X_L+i  \xi\cdot\p  X_L}(z) 
     e^{i p_3\cdot X_R+i  \bar\xi\cdot\bar\p    X_R}( \bar z) }\Big|_{\text{linear terms  in }  \zeta_1,\zeta_2, e}~,
     \qquad
     \nonumber
\eeqn
where we only keep the terms which are linear in $\zeta_1, \zeta_2$ and $e$, or equivalently  terms linear in $\zeta_1, \zeta_2$ and $\xi, \bar \xi$ in the correlator.  The correlator can be evaluated using the formula \eqref{ABCformula} and OPE \eqref{XLRopenclosedOPEyy}. The computation is   straightforward and can be simplified by using \eqref{p123property}\eqref{p123Onshell}. After further performing the $y$ integration, we obtain the final result. Up to an overall constant, the three-point amplitude is given by 
\be\label{AoocAmp}
A^{ooc}_3=\delta^{AB}\Big(    \zeta_1\cdot p_2\; \zeta_2\cdot e \cdot p_1
-\zeta_2\cdot p_1\; \zeta_1\cdot e \cdot p_1
+\zeta_1\cdot \zeta_2\; p_1\cdot e \cdot p_1
-\ap \zeta_1\cdot  p_2\; \zeta_2 \cdot p_1\; p_1\cdot e \cdot p_1\Big)~,
\ee
where $A\cdot e\cdot B\equiv\frac12 (A^\mu B^\nu e_{\mu\nu}+A^\mu B^\nu e_{\nu\mu})$.
Using \eqref{p123property}\eqref{p123Onshell},  \eqref{AoocAmp} can be further rewritten as 
\be\label{ampOOCbos}
A^{ooc}_3=\frac18\; e_{\mu\nu}  \zeta_{1\rho}\zeta_{2\sigma}
\Big[ p_{12}^\nu S^{  \rho\sigma\mu}(\ap)+p_{12}^\mu S^{  \rho\sigma\nu}(\ap)\Big]\delta^{AB}~,
\ee
where 
  \beqn
S^{   \rho\sigma \mu}(\ap)&=& 
p_{12}^\mu\; \eta^{\rho\sigma}+p_{23}^\rho \; \eta^{\sigma\mu}
+p_{31}^\sigma\; \eta^{ \mu\rho}  
 +\frac{\alpha'}{4} p_{12}^\mu\;p_{23}^\rho \; p_{31}^\sigma    ~.
\eeqn
The result is similar to the amplitude for two gluon and one graviton/dilaton/KR field   in the heterotic string case \eqref{heteroticGluonGraviton}, except for the symmetrization of closed string polarization tensor   and non-trivial $\ap$ corrections   which are absent in the heterotic string  due to supersymmetry. The tensor $S(\ap)$ here is essentially $T(2\ap)$ in  \eqref{Ttensor}, up to the relabeling.  Note that in the purely open and closed string case, we have $T(4\ap)$ and $T(\ap)$ in \eqref{bosonicGluonAmp} and \eqref{cloedgraviton}, respectively. 

Since the closed string polarization tensor is always symmetrized,   the massless fields from closed string can not be Kalb-Ramond 2-form field  in order to have a non-vanishing amplitude. 

It is worth mentioning that the result in \eqref{AoocAmp}   comes from the $\ap^3$ and $\ap^4$ order terms in the correlator \eqref{integrandzeta12}. But actually the leading   terms in   the correlator \eqref{integrandzeta12} is of    order $\ap^2$, which is non-vanishing  and takes the form
\be\label{vanishAfterInt}
  \ap^2 \Big[ \frac{\zeta_1 \cdot \xi\; \zeta_2 \cdot \bar\xi}{(y+i)^4}
+
\frac{\zeta_2 \cdot \xi\; \zeta_1 \cdot \bar\xi}{(y-i)^4}
\Big]~.
\ee
However, after performing the integration over $y$, this term vanishes.  As a result, the leading interaction between gluon and graviton is indeed the minimal coupling between them, as it should be. 

\subsection*{Celestial OPE of  gluon and graviton }
Now we want to compute the worldsheet OPE between closed string and open string vertex operators. As before, we can decompose the closed string vertex operator into the left and right moving part, and then use the formula \eqref{ABCformula} and \eqref{XLRopenclosedOPEyy}. The steps are similar to the previous cases, so we only write down the final result: \footnote{ Note that inside the square bracket, there are also order one terms which are proportional to $(z-y)/(\zb-y)$ or its inverse; these terms vanish after doing $z$ integral on the upper half plane. This is supposed to be the origin of \eqref{vanishAfterInt}, and both of them disappear only after integration.  }
\beqn
 &&
 e_{\mu\nu} \p X^\mu \bar\p X^\nu e^{i p \cdot X  }(z,\bz ) \zeta\cdot   Xe^{i q  \cdot X  }(y )
 \\&\sim& 
    i \ap^2   | z-y|^{2\ap p\cdot q-2} e^{i (p+q) \cdot X }
 \qquad 
 \\&&\label{firstcontact}
  \qquad
 \times \Big[ 
  \frac{2i( \ap p\cdot \zeta \; q\cdot  e \cdot q-q \cdot e\cdot \zeta)}{z-y}
%
 %
 \\&&\label{secondusefl}
 \qquad\quad 
+  q\cdot  e \cdot q\;  \zeta \cdot \dot X
+q\cdot e\cdot \zeta \; p\cdot \dot X 
 -p \cdot \zeta \; q\cdot e\cdot \dot X
 -\ap  p \cdot \zeta \; q\cdot e\cdot q \;p \cdot \dot X
 \\&&\qquad\quad
 +\sO(z-\bar z)  +\sO(z-y)  +\sO(\bz-y) 
\Big](y)~,\qquad
 \eeqn
 where we use the relation $\dot X=2\p X_L=2\p X_R$ on the boundary. 
 Note that only the symmetric part of the closed string polarization tensor contributes.
 We are interested in the collinear limit $p\cdot q\to 0$. Then the first term \eqref{firstcontact} is supposed to contribute only   the boundary contact terms, so we will ignore it.  The second line \eqref{secondusefl} is the one relevant here. 
 
 To proceed, we choose polarizations as the bases $\zeta=\varepsilon_\tb(q), e^{\mu\nu}=\varepsilon^\mu_a(p) \varepsilon^\nu_\ta(p)$. 
 Then \eqref{secondusefl} can be rewritten as
 \be
 \varepsilon _a\cdot q
 \Bigg[ q\cdot\varepsilon _\ta   \;     \varepsilon _\tb \cdot \dot X
 -p \cdot \varepsilon_\tb \; \varepsilon_\ta\cdot \dot X
 +\Big(  \varepsilon _\ta\cdot \varepsilon_\tb 
 -\ap  p \cdot \varepsilon _\tb \; \varepsilon _\ta\cdot q \Big)\;p \cdot \dot X
 \Bigg]
 +\Big(a\leftrightarrow \ta \Big) ~.
 \ee
 The terms in the bracket are very similar to the terms proportional to  $y_{12}$ in \eqref{openstringOPE}  once replacing $\zeta\to \varepsilon_\ta, \xi \to \varepsilon_\tb$, and $\ap\to \frac12 \ap$. Therefore, the simplification of this formula is also similar to the open string case there. Following the same derivation leading to \eqref{gluonpart}, up to the  boundary contact terms, the final result here is given by 
 \be
 \varepsilon _a\cdot q
 \Bigg[ q\cdot\varepsilon _\ta   \;     \varepsilon _\tb \cdot   \varepsilon _\tc
 -p \cdot \varepsilon_\tb \; \varepsilon_\ta\cdot  \varepsilon _\tc
 +\Big(  \varepsilon _\ta\cdot \varepsilon_\tb 
 -\ap  p \cdot \varepsilon _\tb \; \varepsilon _\ta\cdot q \Big)\;p \cdot  \varepsilon _\tc
 \Bigg]  \varepsilon _\tc \cdot \dot X
 +\Big(a\leftrightarrow \ta \Big) ~.
 \ee 
 Up to symmetrization and $\ap$ correction, this is similar to  the heterotic case \eqref{heteroticgluongraviton}. 
 
In the collinear limit,  the factor $| z-y|^{2\ap p\cdot q-2} $ localizes to a delta-function $\delta^2(y-z)/p\cdot q$, as one can see from \eqref{deltazclose}. Integrating over $y$ and $z$, we get the OPE
\beqn
\cV_{a\ta}(p) \cV^A_\tb(q)
&\sim& 
\frac{i \pi \ap}{4 }
\frac{1}{p\cdot q}
 \varepsilon _a\cdot q
 \Bigg[ q\cdot\varepsilon _\ta   \;     \varepsilon _\tb \cdot   \varepsilon _\tc
 -p \cdot \varepsilon_\tb \; \varepsilon_\ta\cdot  \varepsilon _\tc
 +\Big(  \varepsilon _\ta\cdot \varepsilon_\tb 
 -\ap  p \cdot \varepsilon _\tb \; \varepsilon _\ta\cdot q \Big)\;p \cdot  \varepsilon _\tc
 \Bigg]  \cV^A_\tc(p+q)
 +\Big(a\leftrightarrow \ta \Big) ~
 \nonumber\\&\sim& 
 \frac{i \pi \ap}{2 }
 \varepsilon _a\cdot q
  \; I_\tc\Big(p,\varepsilon_\ta(p); q,\varepsilon_\tb(q); \frac12\ap\Big)
    +\Big(a\leftrightarrow \ta \Big) ~,
\eeqn
 where $I $ is given in \eqref{Ipqeps}.
 
Further performing the Mellin transformation, we finally get the celestial OPE between open and closed string massless fields, up to an overall constant:
     \beqn\label{openclosedgluongraviton}
&&\cV _{\Delta_1,a\ta}(x_1)\cV^A_{\Delta_2,\tb}(x_2)
\\&\sim&\nonumber 
 x_{12}^a \frac{x_{12}^{\ta}\delta^{\tb\tc} B(\Delta_1-1 ,\Delta_2+1) 
  +x_{12}^{\tb}\delta^{\ta\tc}B(\Delta_1  ,\Delta_2 )
 -x_{12}^{\tc}\delta^{\ta\tb} B(\Delta_1  ,\Delta_2+1)
   }{(x_{12})^2}  
   \cV^A _{\Delta_1+\Delta_2 ,\tc}(x_2)   +\Big(a\leftrightarrow \ta \Big) 
   \\&&
-2\ap    \frac{ x_{12}^a x_{12}^\ta x_{12}^\tb x_{12}^\tc  
   }{(x_{12})^2}  B(\Delta_1+1  ,\Delta_2+2)
   \cV^A _{\Delta_1+\Delta_2 +2,\tc}(x_2)~.
   \eeqn 
  Since we have assumed that the polarization tensor is  traceless and furthermore only its symmetric part contributes, the closed string massless field under our consideration can only be graviton. We thus need to subtract the trace part in   \eqref{openclosedgluongraviton}  following \eqref{gPolarization}. 
%
   This then gives the gluon and graviton  OPE from the open-closed string setup. 
 One can check that  the same result can be obtained from  the collinear factorization \eqref{ampfactorization} and the three-point amplitude \eqref{AoocAmp}  derived before. In particular, the relative coefficients of $\ap$ correction are also the  same. The perfect agreement thus justifies our amplitude   and OPE calculations. Therefore, the celestial OPE can also be obtained from the open-closed string setup. However, the emergence of closed string field from open string field, namely the fusion    of two gluons into one graviton, is not clear from the OPE perspective. 

 \bibliographystyle{JHEP} 
 
\bibliography{BMS.bib} 
  
\end{document}